# Competing ferroelectric and smectic order: modulated structures through molecular design.

*Grant J. Strachan, Ewa Górecka, Jadwiga Szydłowska, Damian Pociecha**

Faculty of Chemistry, University of Warsaw, ul. Pasteura 1, 02-093 Warsaw, Poland
* pociu@chem.uw.edu.pl

*Abstract*: We demonstrate that the balance between polar and positional order can be systematically tuned through molecular engineering, providing direct control over the emergence of polar and modulated liquid-crystalline phases, allowing for versatile strategy for the design of functional ferroelectric soft materials. We show that polar orthogonal smectic phases ($SmA_F$ and $SmA_{AF}$), promoted by the self-segregation of aromatic cores and sufficiently long terminal chains, are readily destabilized by strong longitudinal dipolar interactions that energetically penalize parallel alignment of molecular dipoles within a smectic layer. In contrast, the tilted ferroelectric $SmC_F$ phase is remarkably robust across the entire homologous series, indicating that molecular tilt efficiently relieves dipolar frustration within the smectic layers. We further demonstrate that the interplay between microsegregation and electrostatic interactions stabilizes the new modulated $SmC_M$ phase, characterized by incommensurate electron-density waves, particularly for compounds with short terminal chains. For longer homologs controlling the spatial distribution of fluorinated molecular fragments and terminal-chain length enabled the targeted formation of broken-layer-type modulated polar phases (2D or 3D).

**1. Introduction**

The recently discovered proper ferroelectric nematic ($N_F$) phase represents an interesting example of broken inversion symmetry in soft matter[1-5]. While the $N_F$ phase exhibits polar order without translational order, ferroelectric smectic phases additionally develop periodic density modulations associated with layered structures. This adds an additional 'dimension' to the evolution of these phase structures, as the emergence of polar order may occur either before or after the development of layered structure on cooling[6-10]. In most cases, the translational and polar ordering in smectic phases is strongly coupled, leading to commensurate periodicities within the system. However, such an assumption is not generally required by symmetry. The possibility that polarization and density modulations are incommensurate introduces an additional structural degree of freedom, thereby enriching the phase behaviour of ferroelectric matter. In some systems spatial variation of polarization can evolve independently of the translational order parameter associated with density modulation. An example of partially decoupled translational and polar ordering has been discussed in modulated bent-core smectics within the framework of general tilt structures[11]. In these systems, the density wave defines the 2D-modulated layered organization, while the local tilt and associated polar order develop an additional modulation that is not uniquely determined by the basic 2D periodicity. As a consequence, the characteristic length scales associated with density and polar ordering may be incommensurate. Further support for this concept is provided by a number of ferroelectric crystalline systems; in incommensurate ferroelectrics, polarization-related ordering is not locked to the underlying crystallographic periodicity. More complex examples are provided by polar vortex lattices observed in ferroelectric oxide superlattices, where the polarization field forms mesoscale rotational structures with well-defined periodicity emerging from the balance of electrostatic and elastic energies[12]. A similar mechanism may be relevant in soft systems such as the proper ferroelectric smectic phases, offering the potential for a variety of new modulated polar liquid crystalline phases. Understanding the interplay between polar and density modulations in ferroelectric smectic phases is therefore essential both from a fundamental perspective and for the design of responsive soft materials with tunable mesoscale architectures.

Here we report six homologous series of liquid crystalline materials with elongated terminal chains and systematically varied patterns of lateral fluorine substitution on the mesogenic core (Fig. 1). They exhibit unusually rich phase behaviour, including a re-entrant polar nematic phase, broken mirror symmetry helical structures, as well as multiple modulated smectic phases with long and short characteristic periodicities. These materials are given codes $\mathbf{T_n}$**-X-Y**, where n is the number of carbon atoms in the terminal chain, and **X** and **Y** refer to the number of fluorine substituents on rings 2 and 3 in the mesogenic core, respectively (see Fig. 1). For liquid crystals, extending the length of the terminal chain is a standard approach to promote the formation of lamellar structure in general. However, we have previously found that the distribution of fluorine substituents in the mesogenic core can also significantly promote or suppress the formation of ferroelectric smectic phases[13]. Therefore, we decided to explore how these two molecular design strategies interact to influence the stabilization of layer formation and polar ordering. The sometimes-competing nature of these effects has led to the observation of several new polar phase structures.

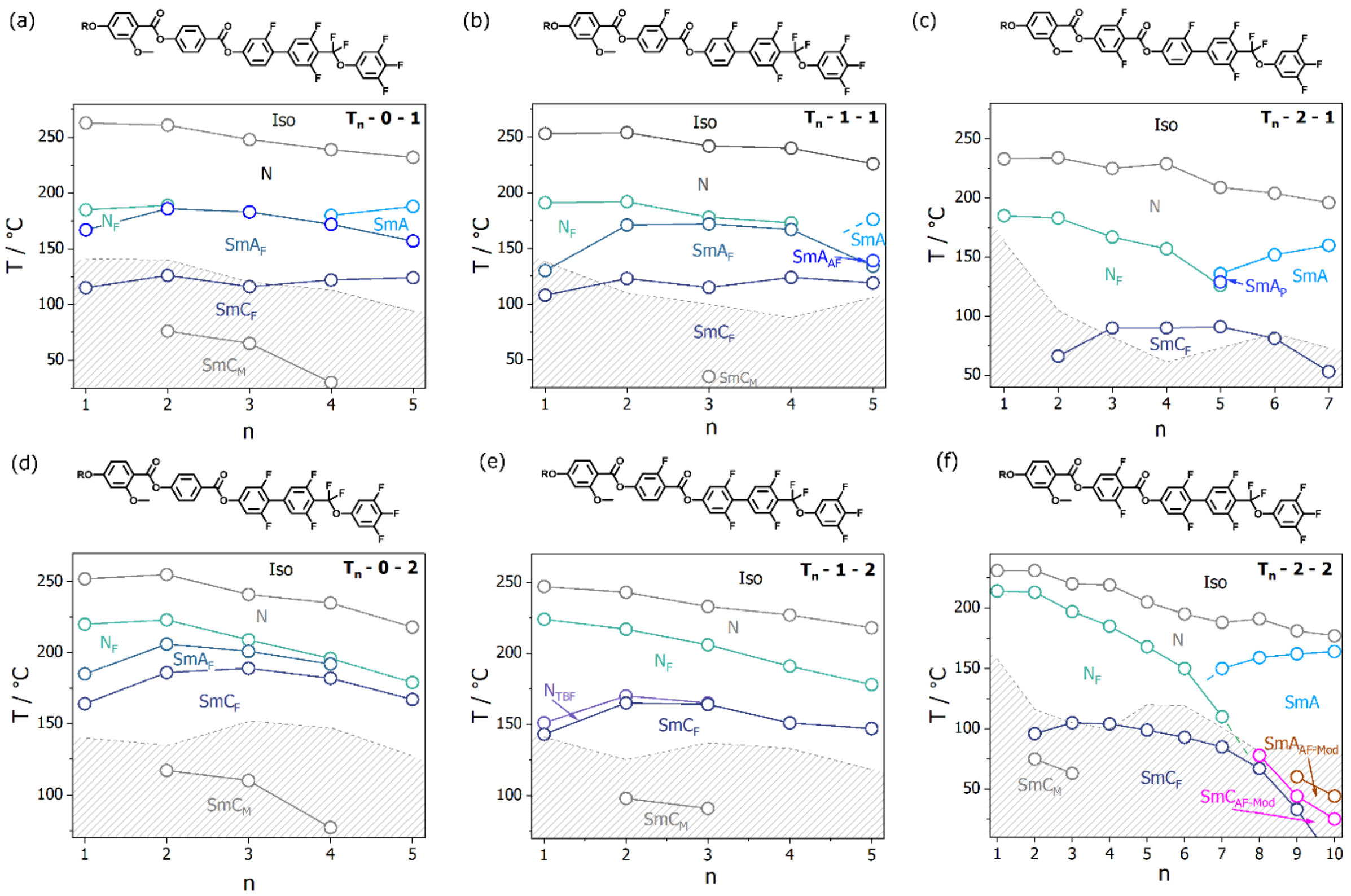

**Figure 1**. Phase diagrams for **T$_n$-X-Y** homolog series, general molecular structures are given above. The shaded areas show the temperature range of solid crystal phase stability.

## 2. Experimental

New compounds were prepared using standard methods and full synthetic details and chemical characterization data are provided in the accompanying Supporting Information, together with description of used experimental techniques.

### Identification of Liquid Crystal Phases

The typical phase sequence in these materials is Iso-N-N$_F$-SmA$_F$-SmC$_F$, although for the **T$_1$-1-2** fluorination pattern the SmA$_F$ phase was replaced with the N$_{TBF}$ phase, in which helical polar structure is spontaneously formed[14]. Preliminary identification of these and other known phases was carried out based on the observation of characteristic textures using polarised optical microscopy and X-ray diffraction studies. In cells treated for planar alignment, the nematic phase gave a planar texture with optical flickering due to director fluctuations, and the N$_F$ phase produced distinct conical defects anchored at the cell spacers. In the N$_{TBF}$ phase, a striped texture developed due to the formation of heliconical structure, with diffraction colours clearly visible by eye.

The non-polar smectic A phase formed elongated fan-like texture and between two untreated glass plates it could be sheared to a homeotropic texture. In contrast, the $SmA_F$ phase formed curved mosaic regions and did not show homeotropic texture, which is in line with its polar nature, as such a texture would lead to a high density of bound charges at the surfaces. The textures produced in the $SmC_F$ phase were highly dependent on the preceding phase (see Tab. S1). The X-ray diffraction experiments revealed only diffuse signals in the small and wide-angle regions for the N, $N_F$ and $N_{TBF}$ phases. For the SmA, $SmA_F$ and $SmC_F$ phases, the small-angle diffraction signal was sharp, and its width was limited only by instrumental broadening. The position of this signal corresponded to the layer thickness, approximately equal to the molecular length. In all these phases the wide-angle signal remained diffuse, consistent with liquid-like molecular order in smectic layers. The phase identification was also supported by measurement of the optical birefringence (Fig. S1). For the phase sequence Iso-N-$N_F$-$SmA_F$-$SmC_F$, represented by material $T_3$-0-2, the birefringence undergoes step-like increases at each transition, although the change at the $SmA_F$-$SmC_F$ transition is very gradual, consistent with the slow textural development observed with POM. For materials forming the $N_{TBF}$ phase, e.g. **$T_2$-1-2**, the $N_F$-$N_{TBF}$ phase transition is accompanied by a decrease in the optical birefringence, as the tilting of the molecules with respect to the helical axis lowers the extraordinary refractive index and increases the ordinary one.

In material **$T_4$-0-2**, at the transition between the N and $N_F$ phases, a transient, chevron-like texture (Fig. S2) was observed, which indicates the formation of an antiferroelectric $N_X$ phase. However, the very short range of this phase (<1 K) prevented its further characterization. Compound **$T_5$-1-1** also formed an antiferroelectric $SmA_{AF}$ phase between the SmA and $SmA_F$ phases, which was identified on a basis of tristable switching behavior in an electric field (Fig. S3). Observations of the optical textures produced in a cell treated for planar alignment showed only very subtle changes across SmA-$SmA_{AF}$-$SmA_F$ phase sequence, and this is consistent with previous reports[15]. The temperature evolution of the optical birefringence showed a small increase at the SmA-$SmA_{AF}$ transition, and a larger jump at the $SmA_{AF}$-$SmA_F$ transition (Fig. S3). The layer spacing gradually increased on cooling in the SmA phase but began to decrease upon transition to the $SmA_{AF}$ phase (Fig. S4), suggesting some splay of the director at the antiferroelectric domain walls.

## 3. Results

### Characterization of new modulated liquid crystal phases

In all the homolog series reported here a new polar liquid crystalline phase, $SmC_M$, was observed for short homologs, below the $SmC_F$ phase. POM observations revealed a ‘smoothing out’ of the $SmC_F$ texture on the transition to $SmC_M$ phase (Fig. S5). The XRD measurements indicated that the layer spacing drastically shortens at the transition from the $SmC_F$ phase. For homologs with shorter terminal chains, this appeared as a jump in the layer spacing, while the change became more continuous as terminal chain length increased (Fig. 2 and Fig. S4).

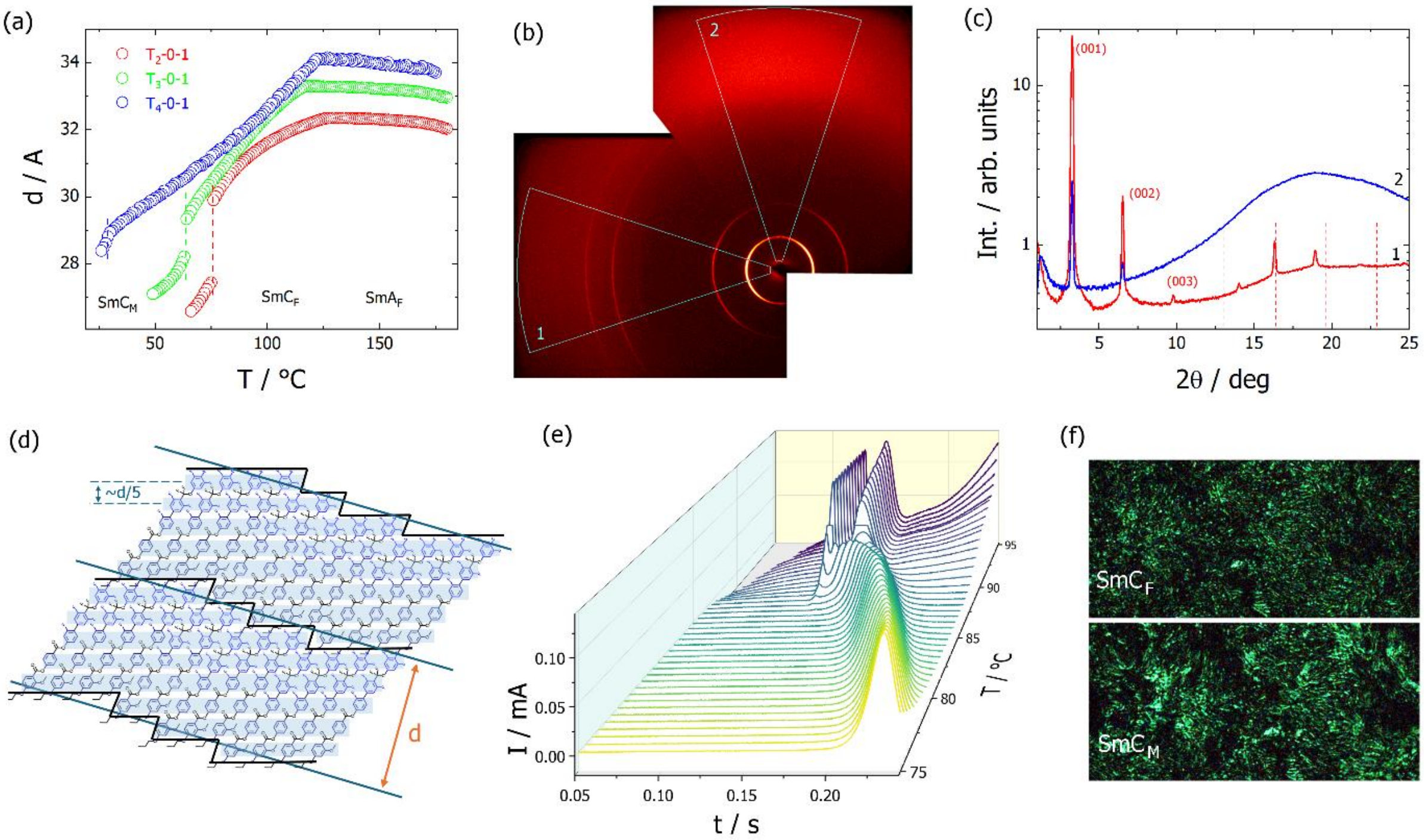

**Figure 2**. $SmC_M$ phase: (a) Smectic layer thickness vs. temperature measured for selected homologs of **$T_n$-0-1** series; (b) Broad-angle 2D XRD pattern for $SmC_M$ phase of compound **$T_3$-0-1** at T=55 °C and (c) diffractograms obtained by integration of the pattern over azimuthal angle, in selected areas indicated in (b): red and blue in 1 and 2, respectively, dashed lines shows expected positions of higher harmonics (00l) smectic layer signal; (d) schematic drawing of molecular arrangement in $SmC_M$ phase, positional correlations between molecular segments (blue lines) give rise to diffraction signals in high angle range; (e) temperature evolution of polarization switching current profile, across $SmC_M$ and $SmC_F$ phases, for compound **$T_3$-1-2**; (f) SHG microscopy images taken in $SmC_F$ and $SmC_M$ phases of compound **$T_3$-1-2**, emission of green light upon irradiation with IR laser (λ=1064 nm) evidences the non-centrosymmetric, polar character of both phases.

The layer spacing changes in the $SmC_M$ phase suggest a tilted phase, although with a different layer contraction coefficient than in the preceding $SmC_F$ phase. Interestingly, in the wide-diffraction angle region, in addition to the diffused signal (corresponding to ~4.5A), reflecting liquid like positional correlation of neighbouring molecules inside the layers, a set of sharp signals also develops. The strongest of these is positioned around the $5^{th}$ harmonic of the layer spacing, and this periodicity seems to correlate with the size of a phenyl unit. However, these signals were not exactly commensurate with the layer spacing. The signals were already present in the preceding $SmC_F$ phase; however, their width reflects only short-range correlations, and on cooling they gradually narrowed to become sharp Bragg-like reflections in the lower temperature $SmC_M$ phase (Fig. S6). The presence of sharp signals associated with much smaller periodicities than the molecular length indicates an additional density wave related to positional correlations between segments of molecular cores having different electrostatic potential[16], while the layer periodicity is still determined by the electron density contrast between the mesogenic core and terminal chains. Most probably, tilted molecules form "blocks" in which their translational positions are correlated along the layer. The width of these blocks is not strictly defined, as no signal corresponding to a regular 2D structure is detected. If the wave vectors associated with the smectic layers and the sublayers of molecular segments are neither commensurate nor collinear, additional reflections are generated, which should be observed in the diffraction pattern. However, the exact density-modulated structure cannot be reconstructed from the present X-ray diffraction data, as this would require precise indexing of reflections in the XRD pattern for a well-aligned, monodomain sample. The $SmC_M$ phase was strongly SHG active (Fig. 2f), however contrary to the $SmC_F$ phase it does not show any relaxation modes in dielectric spectroscopy (Fig. S7). Considering the proposed structure of the phase, it may be that the offset structure of the 'blocks' removes the simple, low energy fluctuations of azimuthal direction of polarization, which provide the mechanism that produces the strong low frequency relaxation mode observed by dielectric spectroscopy in the $SmC_F$ phase. Polarization switching measurement in the $SmC_M$ phase showed a clear single current peak per half a cycle of applied triangular-wave voltage, associated with polarization reversal, in contrast to the characteristic response recorded in the $SmC_F$ phase, with two (asymmetric) current peaks. [8,9,17]. Moreover, the repolarization in the modulated phase appeared at considerably higher electric field (Fig. 2 and Fig. S8). This suggests that the switching process in this $SmC_M$ phase occurs via a different mechanism, which is consistent with the absence of any relaxation mode in the dielectric spectra. The presence of a clear repolarization current peak and SHG signal indicates that the $SmC_M$ phase is ferroelectric and in combination with its complex XRD pattern suggests it is a new modulated type of polar smectic phase.

In the **$T_n$-2-2** series, where homologs up to n = 10 were studied, two additional modulated smectic phases were observed between the non-polar SmA and ferroelectric $SmC_F$ phases. Both phases appeared for long homologs with n > 8. X-ray diffraction studies confirmed that the molecular arrangement within the layers remained liquid-like, as only diffuse scattering was detected in the wide-angle region (Fig. S9). For the higher-temperature phase, denoted $SmA_{AF\text{-}Mod}$, the X-ray diffraction pattern showed, in addition to the (001) reflection corresponding to the layer spacing, sharp satellite reflections indexed as (011) and (0−11) (Fig. 3). These satellite peaks indicate the presence of a density modulation wave propagating along the layers. However, the absence of a measurable (010) reflection suggests that the modulation amplitude is very weak. A similar diffraction pattern has previously been reported for the antiferroelectric $SmA_{AF}$ phase[18], although in the present materials the modulation periodicity is significantly smaller, with blocks containing approximately 9 molecules. The phase did not exhibit SHG activity in the ground state, which is consistent with its antiferroelectric structure. The lower-temperature modulated phase, $SmC_{AF\text{-}Mod}$, produced a considerably more complex X-ray diffraction pattern in the small-angle region (Fig. 3). The observed reflections could not be indexed using any two-dimensional lattice. The best agreement was obtained assuming a 3D structure with $P2_1ma$ symmetry (space group 26). The absence of the (100) and (001) reflections, together with the presence of their second-order harmonics, (200) and (002) signals, indicates that the electron density distribution along the a and c crystallographic directions is centered, while the presence of the (201) reflection suggests a herringbone arrangement of the structural units within the ac plane. These results imply that the smectic layers become distorted and form finite blocks arranged in a zig-zag pattern. In this arrangement, the central block is tilted in the opposite direction to the outer ones and shifted by half of the unit-cell dimension. By comparing both modulated phases, it is reasonable to assume that the antiferroelectric order is preserved in the bc plane of the $SmC_{AF\text{-}Mod}$ unit cell, whereas the block tilt develops in the ac plane. Consequently, the primary density modulation along the b direction is associated with antiferroelectric domain boundaries, while the modulation along the a direction originates from the displacement of the tilted blocks and the electron-density modulation within the ac plane is mainly attributed to the contrast between aromatic and alkyl regions. The XRD data alone do not allow one to determine whether the antiferroelectric ordering is restricted to the bc plane or whether polarity also alternates within the ac (tilt) plane. In the latter case, the arrangement of polar blocks would resemble a double-splay nematic structure[19,20]. No distinct switching-current peak was detected in this phase, which may result from the small size of the ferroelectric blocks forming the complex three-dimensional structure. Upon cooling, the $SmC_{AF\text{-}Mod}$ phase transforms into the $SmC_F$ phase, leading to a significant simplification of the X-ray diffraction pattern. In the $SmC_F$ phase, only reflections associated with the layer spacing remain visible. These reflections appear at azimuthal angles close to the direction of the (201) reflection observed in the preceding phase, which is consistent with a phase transition involving the growth and expansion of ferroelectric blocks or domains.

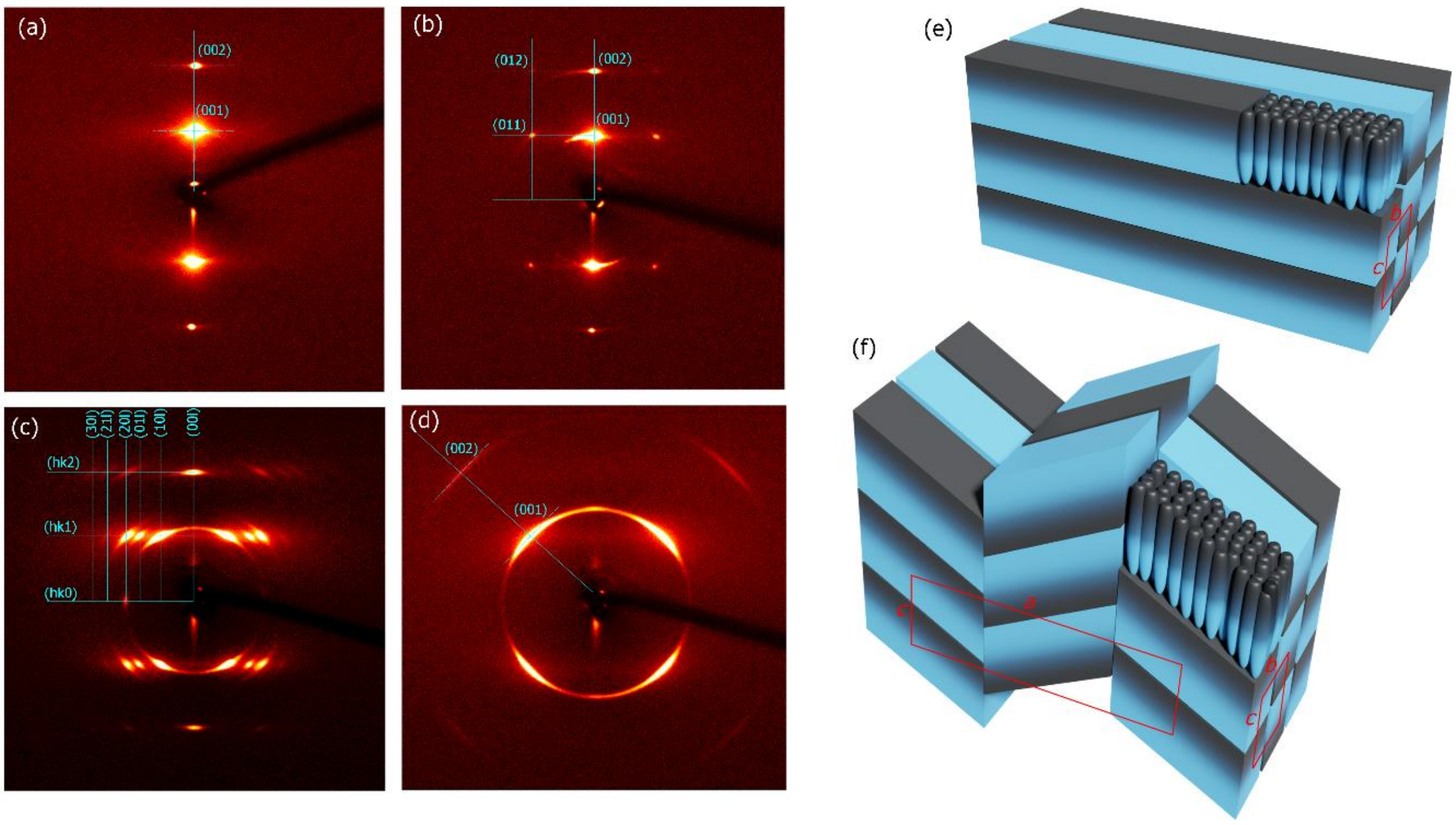

**Figure 3**. 2D small angle XRD patterns recorded in SmA, $SmA_{AF\text{-}Mod}$, $SmC_{AF\text{-}Mod}$ and $SmC_F$ phases (a-d, respectively) of compound **T₉-2-2**, diffraction signals were indexed assuming simple lamellar structure in SmA and $SmC_F$ phases, orthogonal two-dimensional (2D) structure in $SmA_{AF\text{-}Mod}$ phase and three-dimensional (3D) crystallographic lattice of $P2_1ma$ symmetry in $SmC_{AF\text{-}Mod}$ phase. Schematic drawing of modulated 2D and 3D structures of $SmA_{AF\text{-}Mod}$ (e) and $SmC_{AF\text{-}Mod}$ (f) phases, colouring of the ellipsoids (and molecular blocks) signifies the longitudinal dipole moment of molecules, *abc* are unit cell dimensions.

## 4. Discussion

### Trends in phase behavior on increasing terminal chain length and changing fluorination pattern

The studied materials differ in their fluorination pattern, and for all series the elongation of the terminal alkyl chain lowers the onset temperature of polar order, as expected, because the increasing separation between mesogenic cores carrying dipole moments weakens the dipolar interactions. In contrast, the formation of lamellar structures shows a more complex behaviour than the simple increase of transition temperatures with chain length that is typically observed for weakly polar molecules. Comparison of the total N and $N_F$ phases range in the three series with a single fluorine substituent on ring 3, **$T_n$-0-1**, **$T_n$-1-1**, and **$T_n$-2-1**, reveals that the addition of consecutive fluorine atoms to phenyl ring 2 of the mesogenic core progressively destabilize orthogonal SmA and $SmA_F$ smectic phases (Fig. 1). As a result, in the **$T_n$-0-1** series the $N_F$ phase is observed for homologs with n < 3, in **$T_n$-1-1** for n < 4, and in **$T_n$-2-1** for n < 6. These results indicate that increasing the number of fluorine atoms distributed along the mesogenic core weakens the tendency to form orthogonal lamellar structures. This weakening of lamellar order leads to particularly interesting behaviour in the **$T_5$-2-1** homolog. Upon cooling, the phase sequence N–SmA–$SmA_F$–$N_F$–$SmC_F$ is observed, in which the ferroelectric nematic $N_F$ phase appears below more ordered smectic phases[21], an example of re-entrant behaviour[22,23]. It appears that in this material the smectic order is sufficiently weak to melt once strong polar ordering develops in the system[21].

A similar shift of the non-polar SmA phase toward longer-chain homologs is observed when comparing the **$T_n$-0-2**, **$T_n$-1-2**, and **$T_n$-2-2** series, which contain two fluorine substituents on phenyl ring 3. The same overall trend is observed for these materials, and the presence of two fluorine substituents destabilises the smectic A phases still further. Consequently, in the **$T_n$-1-2** series no orthogonal smectic phases are observed (up to n = 5), and the suppression of the $SmA_F$ phase allows for the appearance of the heliconical $N_{TBF}$ phase in a narrow temperature range between $N_F$ and $SmC_F$ phase for homologs up to n=3. In the **$T_n$-2-2** series the SmA phase appears only for molecules with terminal chains above 6 carbon atoms. This series also exhibits re-entrant $N_F$ phase behaviour. Moreover, for long homologs of this series modulated phases composed of smectic layers fragments are observed. It appears that the formation of such "broken-layer" modulated phases is a consequence of the relatively weak lamellar order.

The results clearly demonstrate that polar orthogonal smectic phases, $SmA_F$ and $SmA_{AF}$, while promoted by self-segregation of aromatic cores and sufficiently long terminal chains, can be readily destabilised by polar interactions that strongly energetically disfavour parallel alignment of longitudinal molecular dipoles within a smectic layer. In contrast, the tilted ferroelectric smectic phase $SmC_F$ is observed for virtually all mesogens studied here. This indicates that molecular tilt within the layers reduces the unfavorable dipole–dipole interactions between neighboring molecules. At the same time, tilting enables favorable interactions between electron-density waves formed along the molecular cores due to the uneven distribution of segments with different electronegativity[13]. It should also be noted that spatial segregation of different segments of the mesogenic core appears to be responsible for the formation of the $SmC_M$ phase, which is characterized by incommensurate electron-density modulations. This effect is particularly pronounced for homologs with short terminal chains. Extension of the terminal chains increases the thickness of the alkyl sublayers separating molecular cores bearing dipole moments, thereby reducing positional correlations between the cores in neighboring layers.

## 5. Conclusion

To summarize, we have shown that the balance between polar and smectic orders can be rationally tuned through appropriate molecular design. Frustration between lamellar packing and polar interactions can be intentionally introduced through molecular engineering. One can weaken the tendency toward polar orthogonal lamellar organization by a more uniform distribution of fluorine atoms along the mesogenic core, while elongation of terminal chains promotes lamellar structures. This enables the formation of broken-layer-type modulated polar phases at a certain balance of intermolecular forces. Such an approach opens new possibilities for the programmable design of nanoscale organization.

**6. Acknowledgments**: This research was supported by the National Science Centre (Poland) under the grant no. 2024/53/B/ST5/03275.

*Supplementary material for*

## Competing ferroelectric and smectic order: modulated structures through molecular design.

Grant J. Strachan, Ewa Górecka, Jadwiga Szydłowska, Damian Pociecha*

*Faculty of Chemistry, University of Warsaw, ul. Pasteura 1, 02-093 Warsaw, Poland*

# Contents

## 1. Supplementary Figures

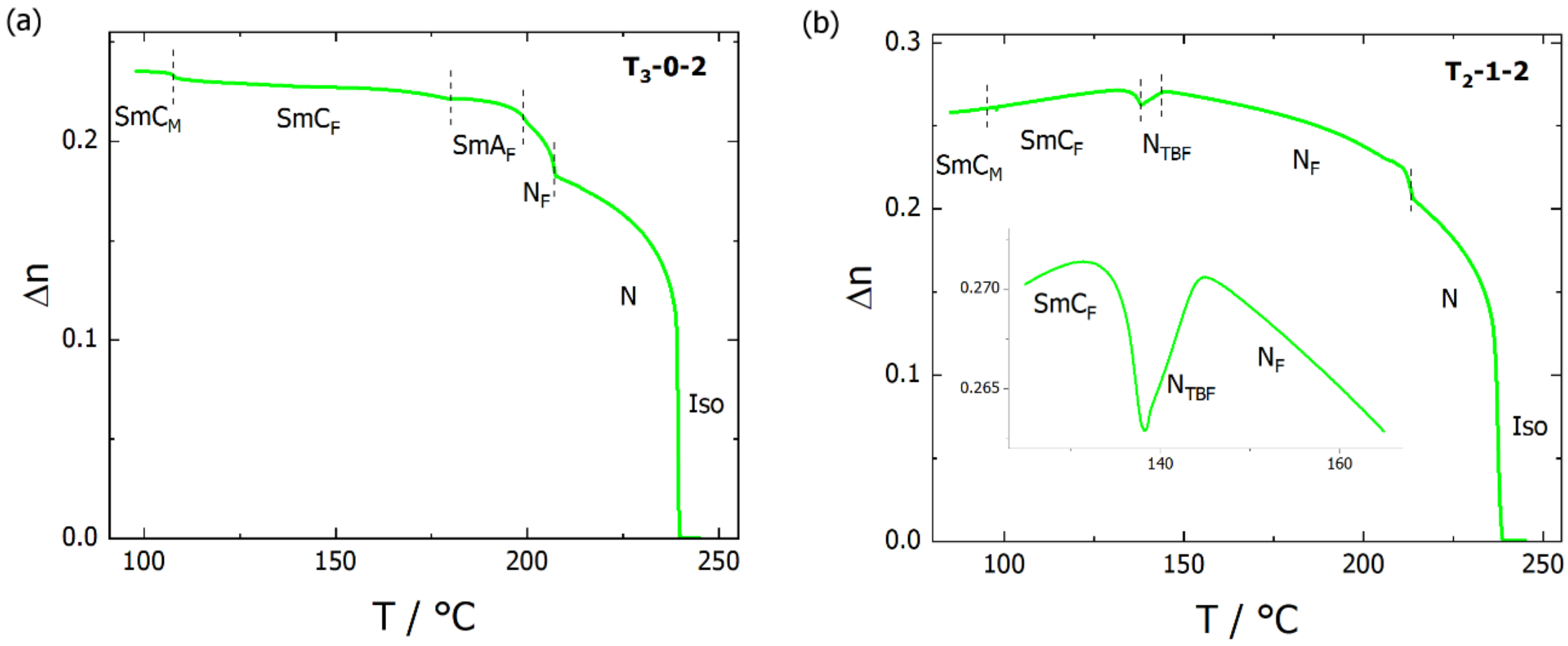

**Figure S1.** Optical birefringence vs. temperature measured with green light (λ=532 nm) for (a) **$T_3$-0-2** and (b) **$T_2$-1-2** compounds.

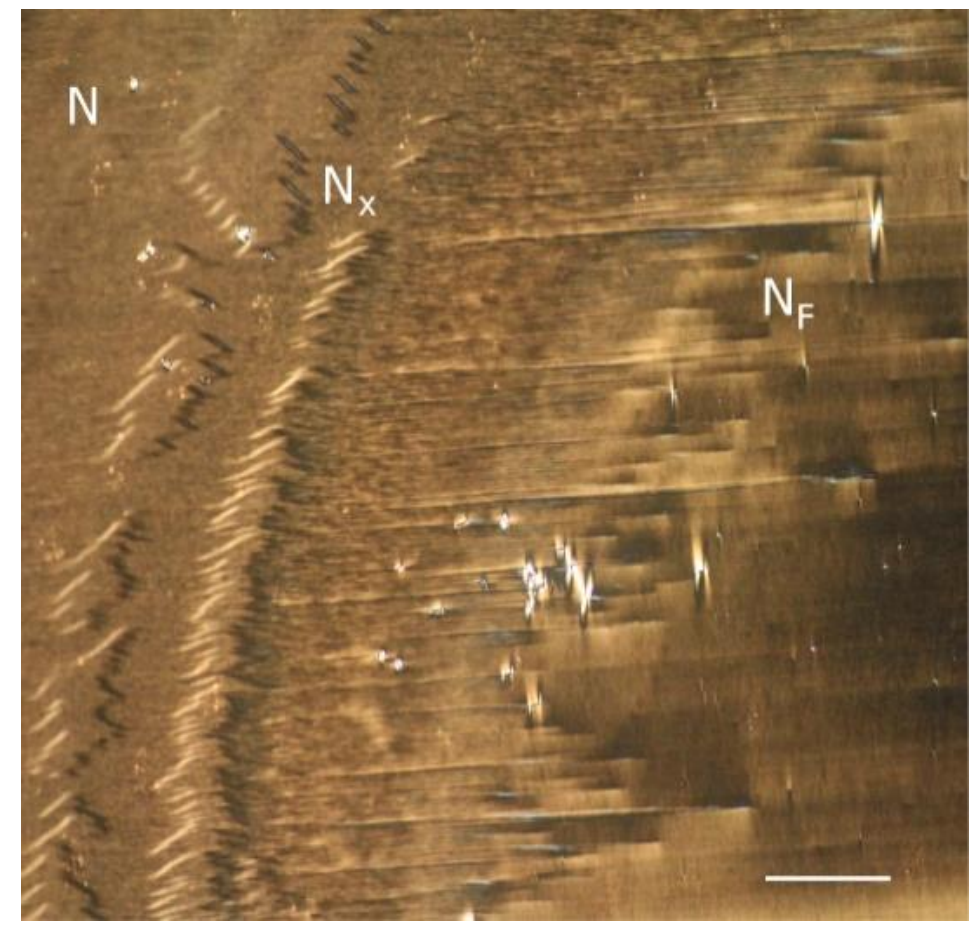

**Figure S2.** Optical texture observed between crossed polarizers for compound **$T_4$-0-2** in 1.8-μm-thick cell with planar anchoring condition at T=202 °C. Due to small temperature gradient (in horizontal direction) coexistence of three nematic phases is observed, with very narrow range of $N_x$ phase evidenced by characteristic chevron defects. Scale bar correspond to 100 μm.

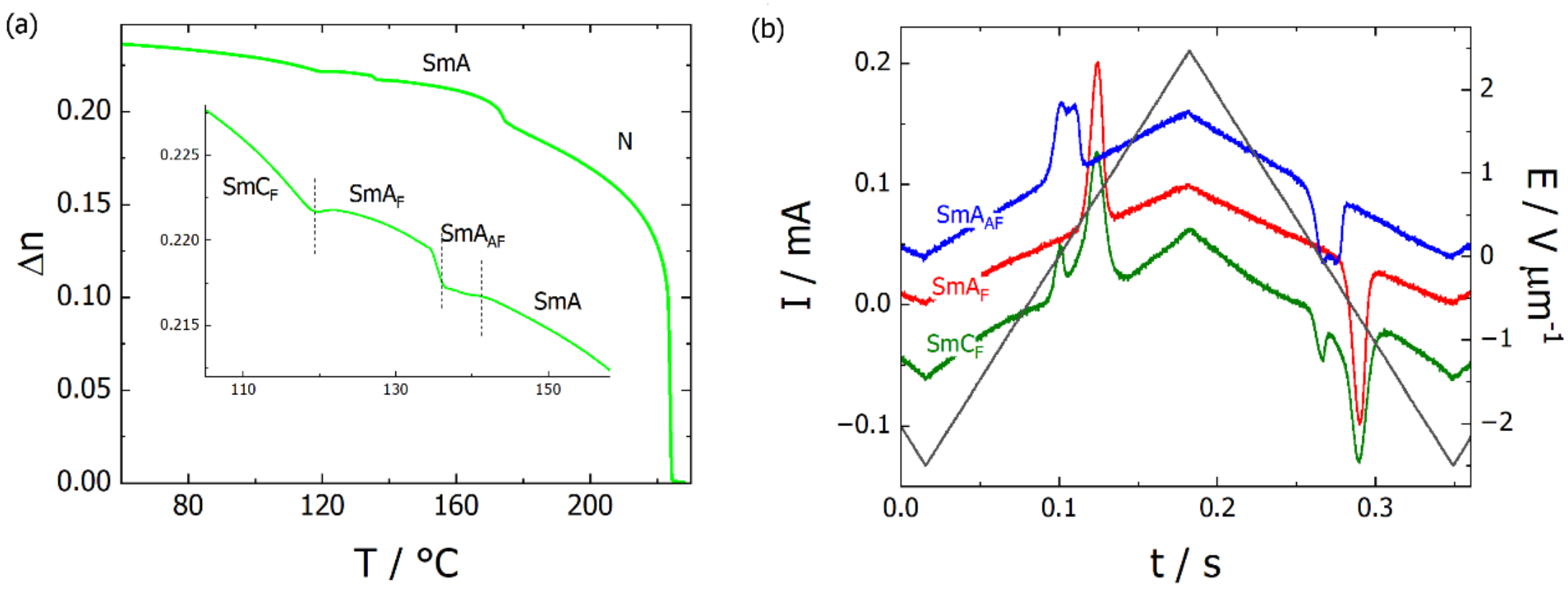

**Figure S3**. Compound **T$_5$-1-1**: (a) Optical birefringence vs. temperature measured with green light (λ=532 nm) indicating formation of four smectic phases below a nematic phase; (b) Polarisation switching current profiles recorded upon application of triangular-wave voltage in SmA$_{AF}$, SmA$_F$ and SmC$_F$ phases. Antiferroelectric character of SmA$_{AF}$ phase is evidenced by presence of two symmetric current peaks per half a cycle of applied voltage; note that in ferroelectric SmC$_F$ phases two peaks were also observed, however they are clearly asymmetric.

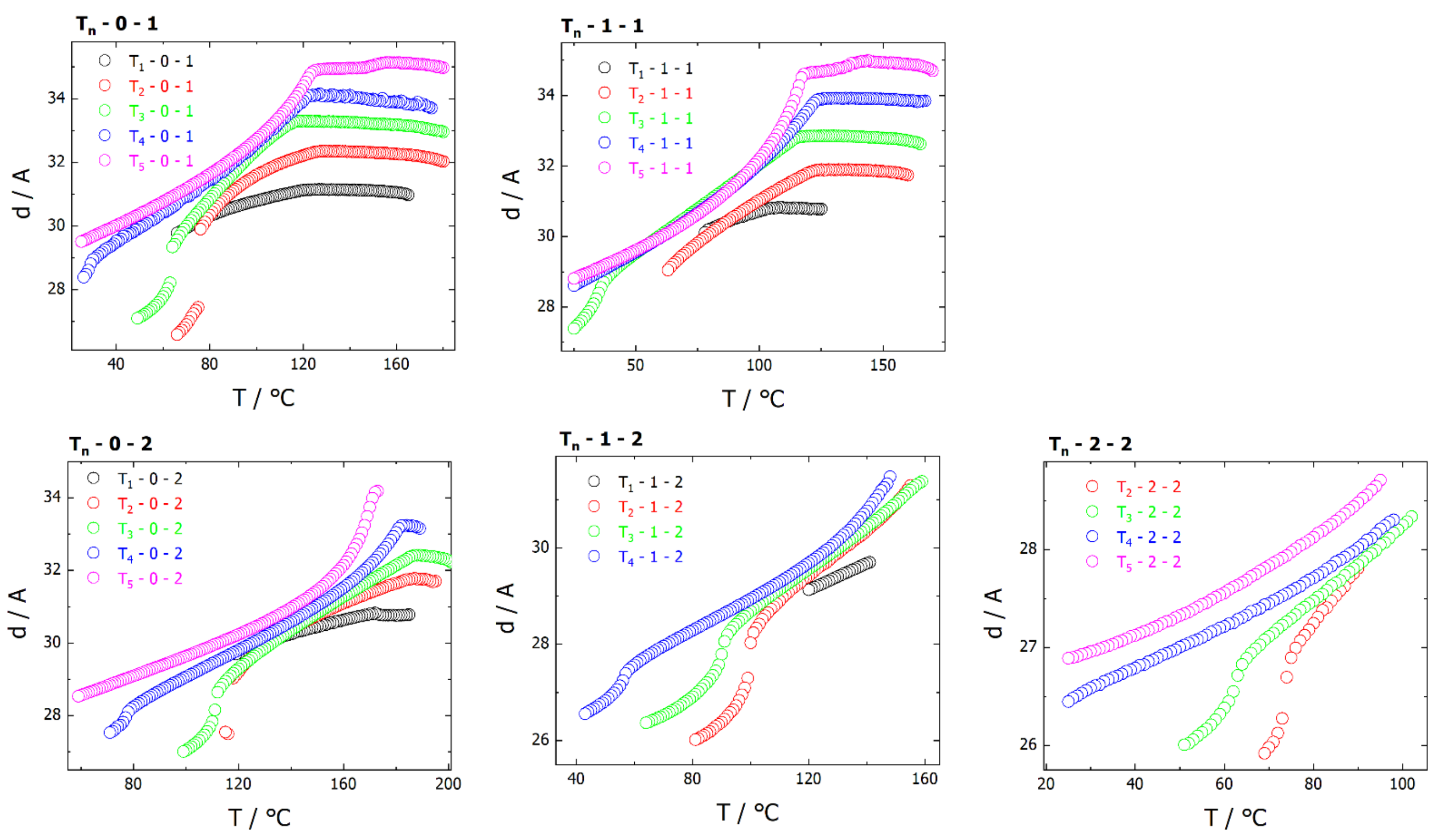

**Figure S4.** Smectic layer thickness, d, measured as a function of temperature for **T$_n$-X-Y** homologue series. In all series a transition to the lowest-temperature SmC$_M$ phase is manifested by a pronounced decrease of the layer thickness, the jump in d value at the SmC$_F$-SmC$_M$ phase transition decreases with elongation of terminal chain length.

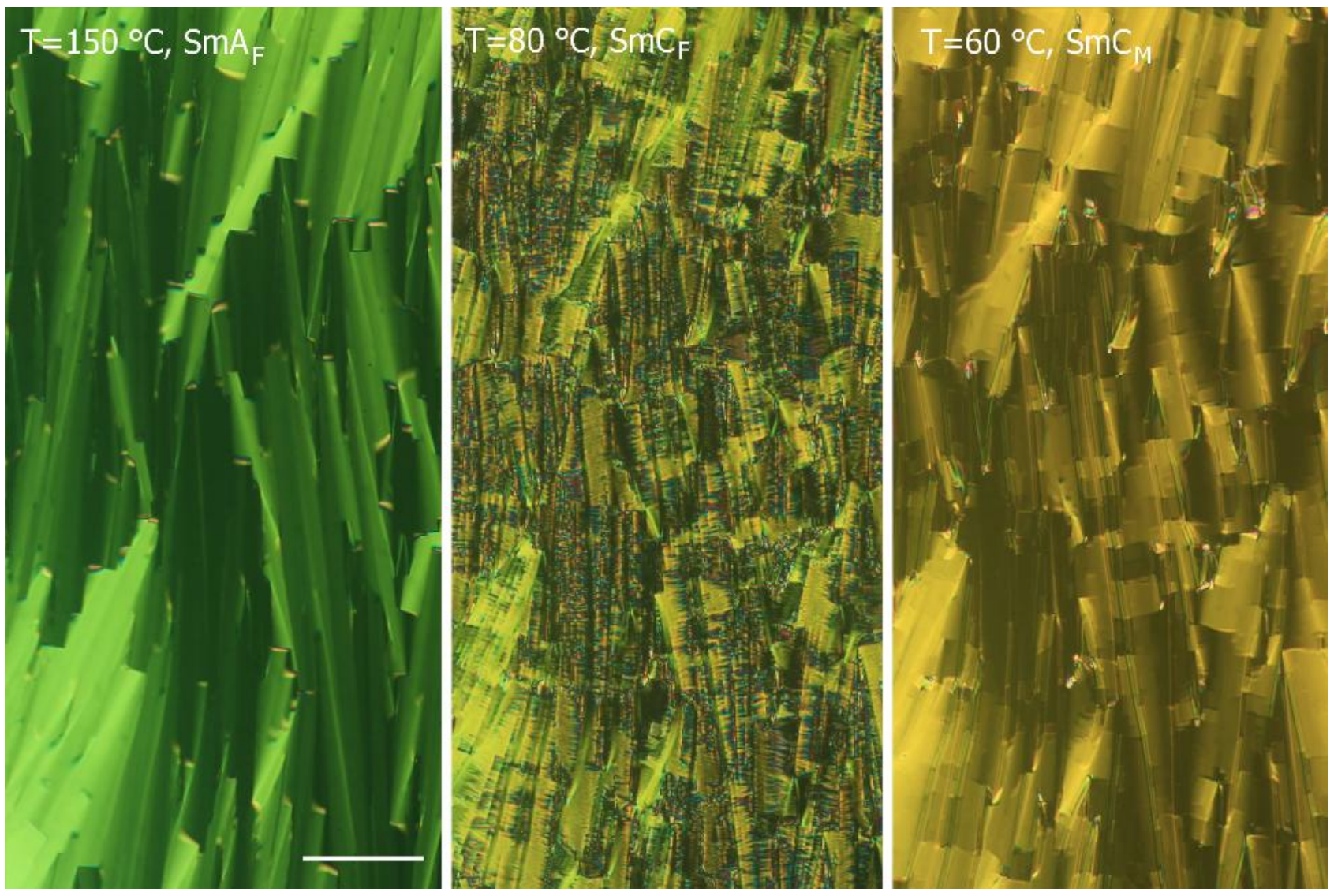

**Figure S5**. Optical textures observed between crossed polarizers for compound **T$_3$-0-1** in 3-μm-thick cell with planar anchoring condition in SmA$_F$, SmC$_F$ and SmC$_M$ phases. Scale bar correspond to 100 μm.

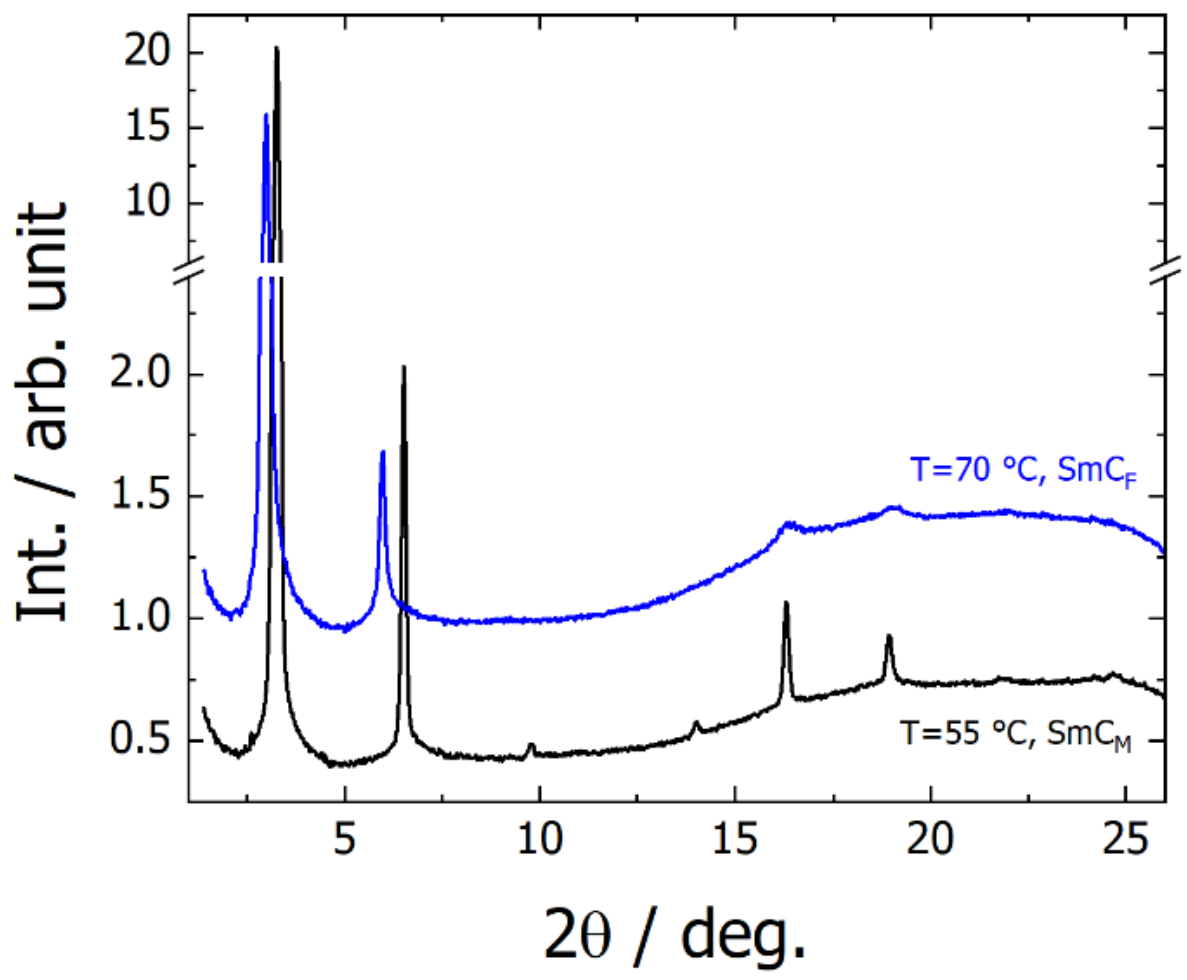

**Figure S6.** X-ray diffrectograms recorded in SmC$_F$ and SmC$_M$ phases of compound **T$_3$-0-1**, showing that already in SmC$_F$ phase there are fluctuations related to the modulated structure of SmC$_M$ phase.

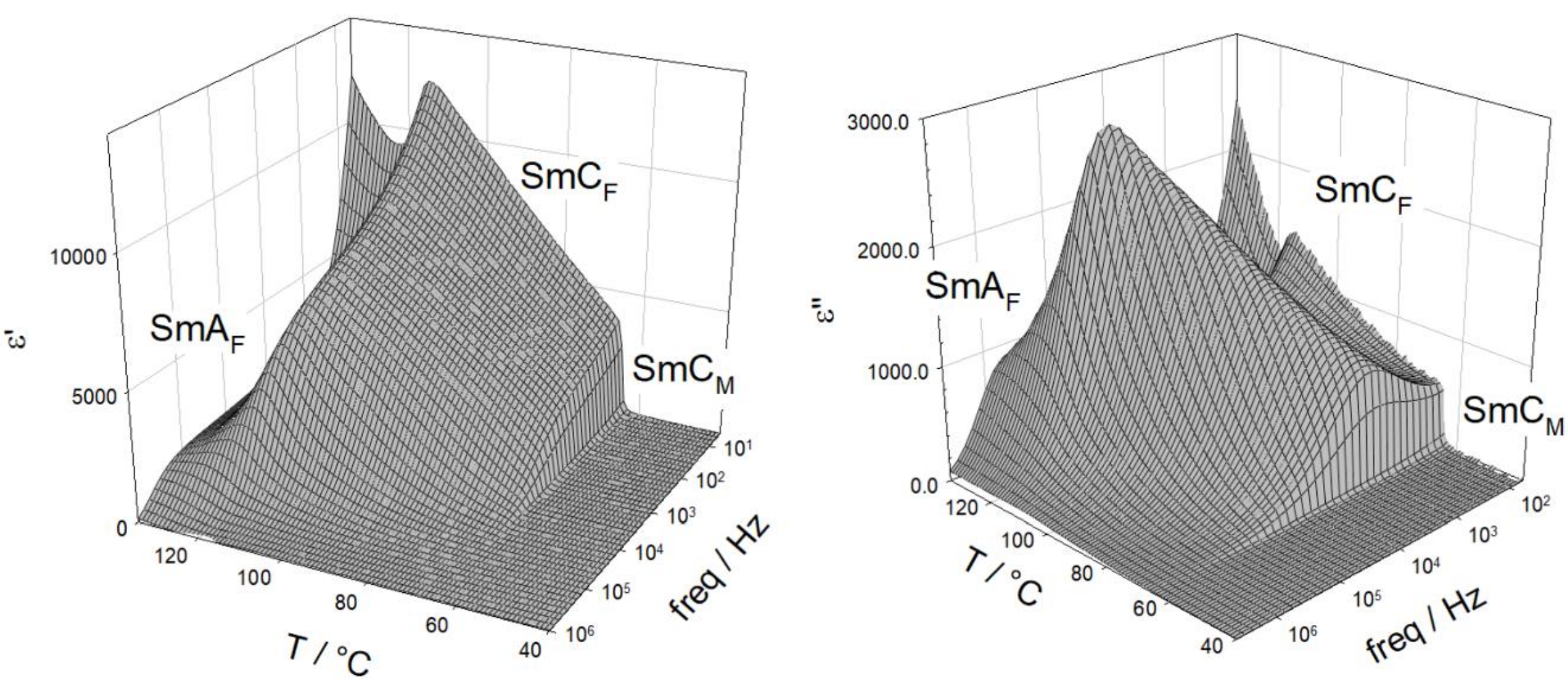

**Figure S7.** Real, ε' and imaginary, ε'' parts of apparent dielectric permittivity measured as a function of temperature and frequency for compound **T$_3$-0-1** in 10-μm-thick cell with gold electrodes and no surfactant layers. Note that in SmC$_M$ phase relaxation modes are suppressed.

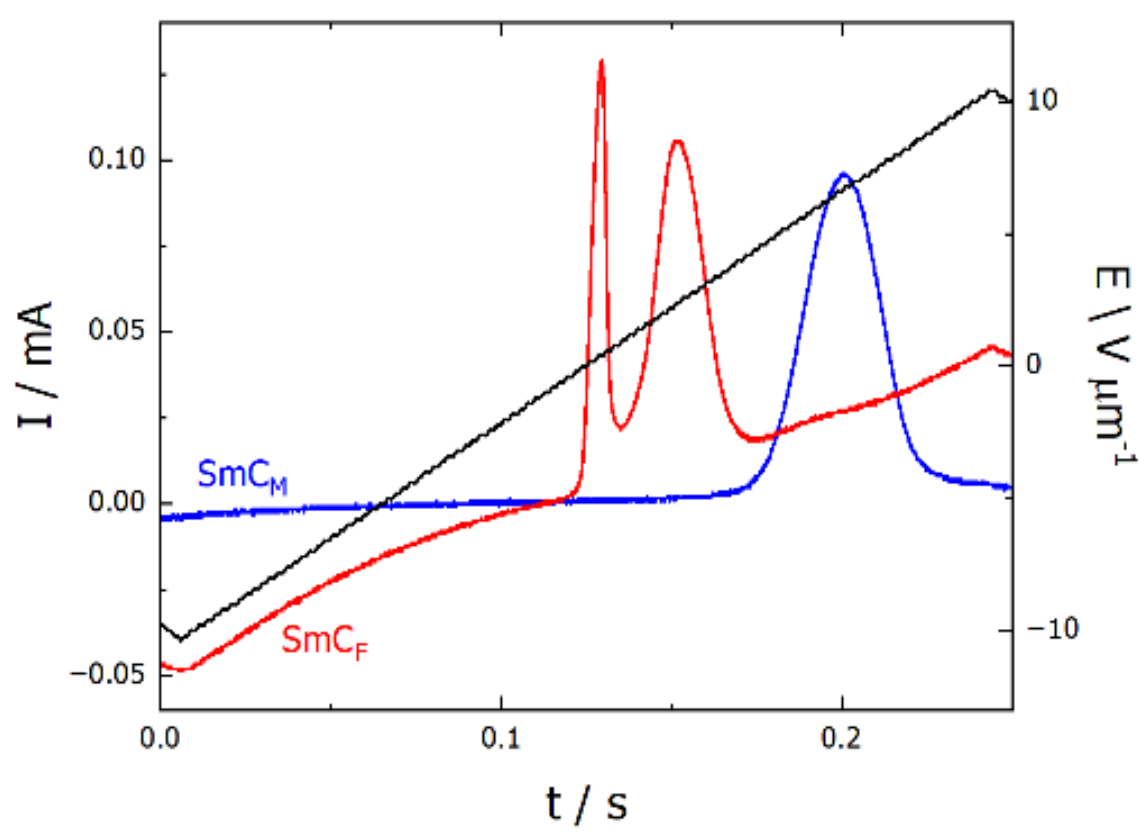

**Figure S8.** Polarization switching current recorded in SmC$_F$ and SmC$_M$ phases of compound **T$_3$-1-2**, upon application of triangular-wave voltage.

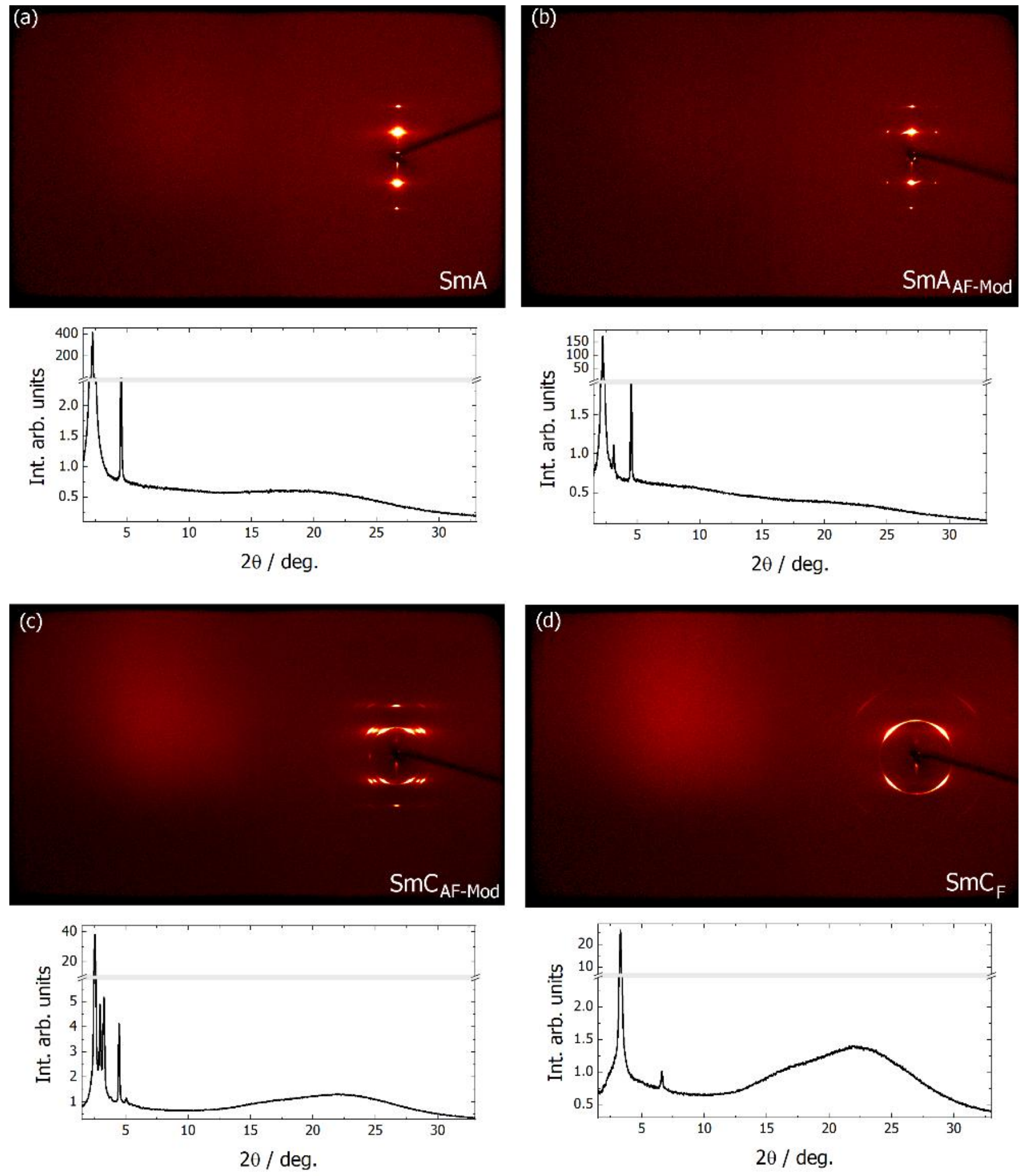

**Figure S9.** 2D XRD patterns recorded in SmA, $SmA_{AF-Mod}$, $SmC_{AF-Mod}$ and $SmC_F$ phases (a-d, respectively) of compound **$T_9$-2-2**, and corresponding diffractograms obtained by integration of 2D patterns over azimuthal angle. Note that in all phases no sharp diffraction signals were detected in the high-angle range, proving the presence of only short-range positional correlations between neighbouring molecules along the short molecular axes.

## 2. Comparison of paramorphotic $SmC_F$ textures

The polarised optical microscopy textures of the tilted ferroelectric smectic phases formed in these materials were highly dependent on the texture of the preceding phase. As a result, they vary highly between different materials and are particularly strongly influenced by phase sequence and sample preparation. For reference, examples of the different textures formed in the materials reported here are given in Table S1 below, along with a comparison to the POM texture observed in the higher temperature phase, where applicable. While it should be noted that similar textures have also been observed in materials forming the heliconical $SmC_P^H$ phase, no evidence for heliconical structure (such as selective reflection, or clear optical diffraction) was observed in the materials studied here, and therefore they have been assigned as $SmC_F$.

***Table S1***. POM textures observed for the $SmC_F$ phases of studied materials.

| sample | $SmC_F$ texture | Preceding texture |
|---|---|---|
| $T_2$-0-1 | $SmC_F$ between untreated glass. 200 μm scale bar | $SmA_F$ between untreated glass. 200 μm scale bar |
| $T_3$-0-1 | $SmC_F$ in 1.6 μm HG cell. 200 μm scale bar | $SmA_F$ in 1.6 μm HG cell. 200 μm scale bar |
| $T_3$-0-1 | $SmC_F$ in 3 μm cell with planar anchoring. 200 μm scale bar | $SmC_F$ in 3 μm cell with planar anchoring. 200 μm scale bar |
| $T_4$-0-1 | $SmC_F$ in 1.6 μm HG cell. 200 μm scale bar | $SmA_F$ in 1.6 μm HG cell. 200 μm scale bar |

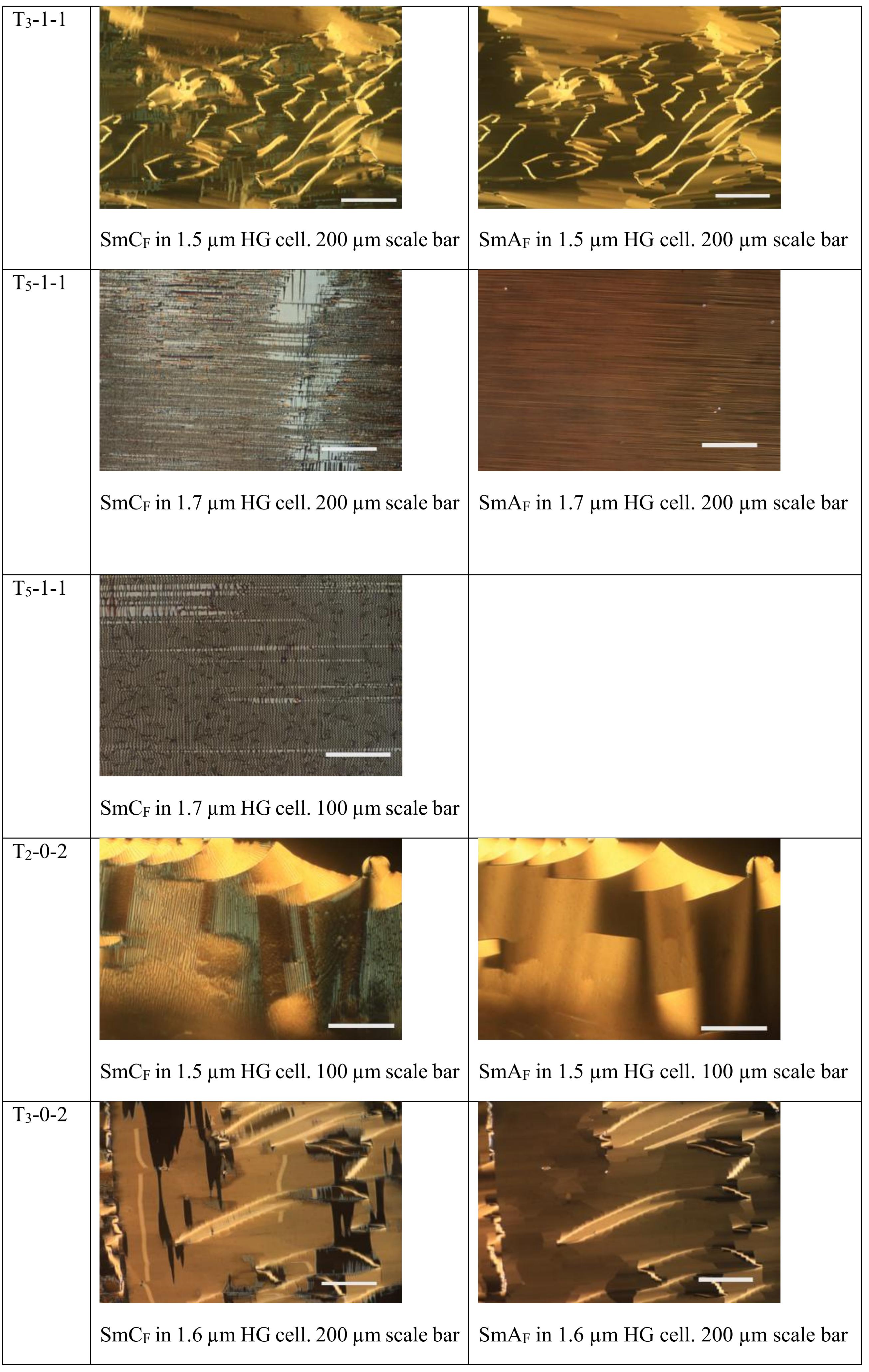

| | | |
|---|---|---|
| $T_3$-1-1 | $SmC_F$ in 1.5 μm HG cell. 200 μm scale bar | $SmA_F$ in 1.5 μm HG cell. 200 μm scale bar |
| $T_5$-1-1 | $SmC_F$ in 1.7 μm HG cell. 200 μm scale bar | $SmA_F$ in 1.7 μm HG cell. 200 μm scale bar |
| $T_5$-1-1 | $SmC_F$ in 1.7 μm HG cell. 100 μm scale bar | |
| $T_2$-0-2 | $SmC_F$ in 1.5 μm HG cell. 100 μm scale bar | $SmA_F$ in 1.5 μm HG cell. 100 μm scale bar |
| $T_3$-0-2 | $SmC_F$ in 1.6 μm HG cell. 200 μm scale bar | $SmA_F$ in 1.6 μm HG cell. 200 μm scale bar |

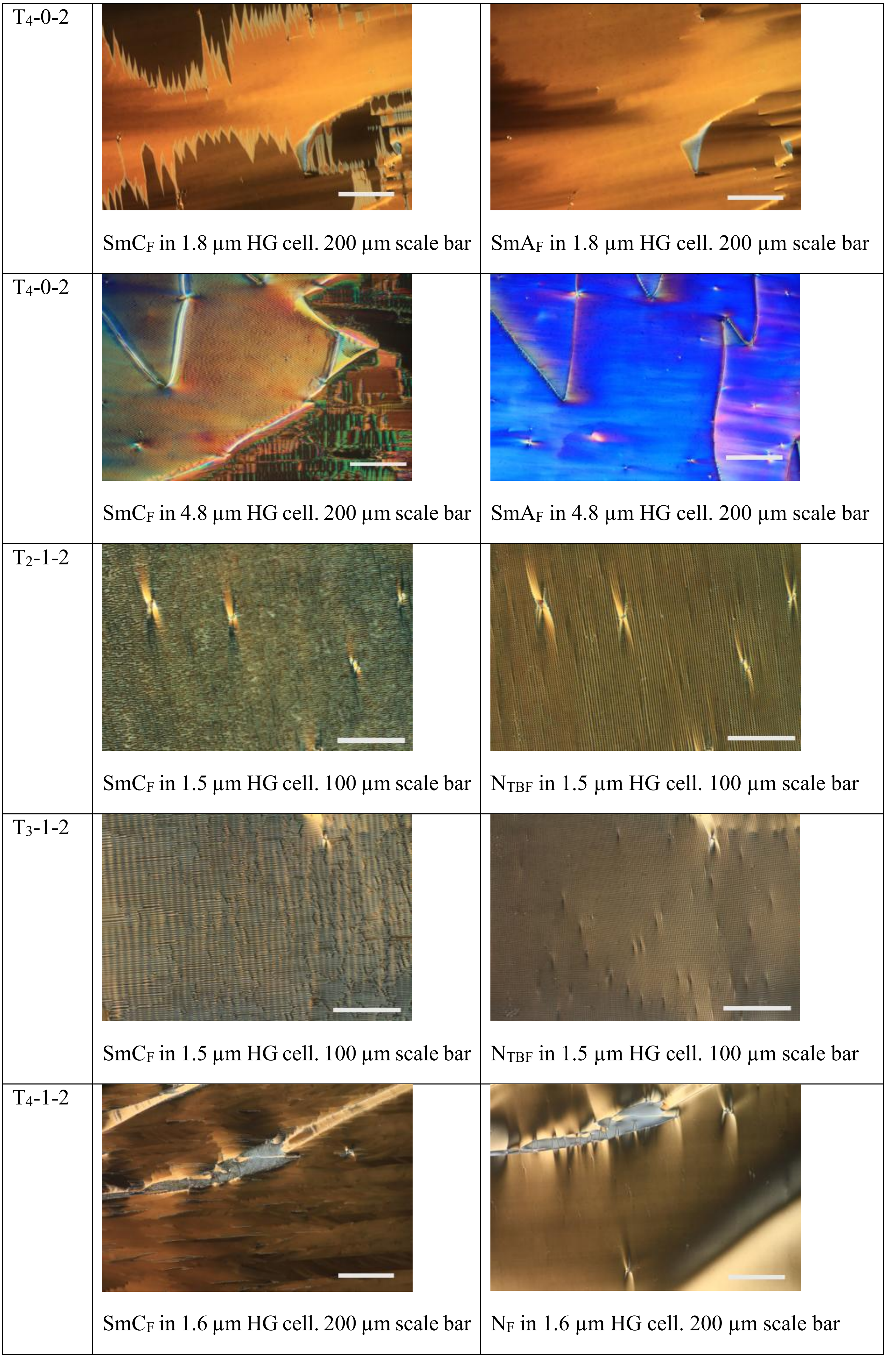

| | | |
|---|---|---|
| $T_4$-0-2 | $SmC_F$ in 1.8 µm HG cell. 200 µm scale bar | $SmA_F$ in 1.8 µm HG cell. 200 µm scale bar |
| $T_4$-0-2 | $SmC_F$ in 4.8 µm HG cell. 200 µm scale bar | $SmA_F$ in 4.8 µm HG cell. 200 µm scale bar |
| $T_2$-1-2 | $SmC_F$ in 1.5 µm HG cell. 100 µm scale bar | $N_{TBF}$ in 1.5 µm HG cell. 100 µm scale bar |
| $T_3$-1-2 | $SmC_F$ in 1.5 µm HG cell. 100 µm scale bar | $N_{TBF}$ in 1.5 µm HG cell. 100 µm scale bar |
| $T_4$-1-2 | $SmC_F$ in 1.6 µm HG cell. 200 µm scale bar | $N_F$ in 1.6 µm HG cell. 200 µm scale bar |
| $T_2$-2-2 | SmC$_F$ in 1.4 µm HG cell. 200 µm scale bar. (At transition to N$_F$ phase) | N$_F$ in 1.4 µm HG cell. 200 µm scale bar |
| $T_3$-2-2 | SmC$_F$ in 1.6 µm HG cell. 200 µm scale bar. | N$_F$ in 1.6 µm HG cell. 200 µm scale bar |
| $T_4$-2-2 | SmC$_F$ in 1.5 µm HG cell. 200 µm scale bar. | N$_F$ in 1.5 µm HG cell. 200 µm scale bar |
| $T_5$-2-2 | SmC$_F$ in 1.6 µm HG cell. 200 µm scale bar. | N$_F$ in 1.6 µm HG cell. 200 µm scale bar |
| $T_7$-2-2 | SmC$_F$ in 1.5 μm HG cell. 200 μm scale bar. | N$_F$ in 1.5 μm HG cell. 200 μm scale bar |
| $T_8$-2-2 | SmC$_F$ in 5 μm HT cell. 100 μm scale bar. | |
| $T_9$-2-2 | SmC$_F$ in 3 μm HT cell. 200 μm scale bar. | |
| $T_9$-2-2 | SmC$_F$ in 5 μm HG cell. 200 μm scale bar. | |

## 3. Experimental Methods

Differential scanning calorimetry has been utilized for the determination of phase transition temperatures and the associated enthalpy changes. Measurements were performed with a TA DSC Q200 instrument, under a nitrogen atmosphere with a heating/cooling rate of 10 K $min^{-1}$, unless otherwise specified.

Polarised-light optical microscopy allowed for observations of characteristic optical textures of liquid crystalline phases. Observations were carried out with a Zeiss AxioImager.A2m microscope equipped with a Linkam heating stage.

Measurements of optical birefringence were performed with a setup based on a photoelastic modulator (PEM-90, Hinds) working at a modulation frequency $f$ = 50 kHz. LED illuminator Photonic F5100 was used as a light source, equipped with narrow bandpass filters. The transmitted light intensity was monitored with a photodiode (FLC Electronics PIN-20) and the signal was deconvoluted with a lock-in amplifier (EG&G 7265) into 1$f$ and 2$f$ components to yield a retardation induced by the sample. Knowing the sample thickness, the retardation was recalculated into optical birefringence. Samples were prepared in 1.6-µm-thick cells with planar anchoring. The alignment quality was checked prior to measurement by inspection under the polarised-light optical microscope.

X-ray diffraction measurements of samples in liquid crystalline phases were carried out using a Bruker D8 GADDS system, equipped with micro-focus-type X-ray source with Cu anode and dedicated optics and VANTEC2000 area detector. Small angle diffraction experiments were performed with a Bruker Nanostar system (IµS microfocus source with copper target, MRI heating stage, Vantec 2000 area detector).

Spontaneous electric polarization was determined by integration of the current peaks recorded during polarization switching upon applying a triangular-wave voltage. 3- to 10-µm-thick cells with ITO or gold electrodes and no polymer aligning layers were used, and the switching current was determined by recording the voltage drop on a resistor connected in series with the sample, using Siglent SDS2104X oscilloscope.

The SHG response was investigated using a microscopic setup based on a solid-state laser EKSPLA NL202. Laser pulses (9 ns) at a 10 Hz repetition rate and max. 2 mJ pulse energy at $\lambda$=1064 nm were applied. The pulse energy was adjusted for each sample to avoid its decomposition. The infra-red beam was incident onto a LC homogenous cell of thickness 5 µm. An IR pass filter was placed at the entrance to the sample and a green pass filter at the exit of the sample.

The complex dielectric permittivity, $\varepsilon^*$, was measured using a Solartron 1260 impedance analyser, in the 1 Hz −10 MHz frequency range, and a probe voltage of 50 mV. The material was placed in a 5- or 10-µm-thick glass cell with gold or ITO electrodes and no polymer aligning layers. Lack of a surfactant layer resulted in a random configuration of the director in the LC phases.

## 4. Synthetic Procedures and Structural Characterisation

Unless otherwise stated, all materials were obtained from commercial sources and used without further purification. Flash column chromatography was carried out using silica grade 60 Å 40-63 micron, and monitored using thin layer chromatography (TLC) using aluminium-backed plates with a coating of Merck Kieselgel 60 F254 silica and an appropriate solvent system. Spots were visualised using UV light (254 nm). $^{1}H$, $^{19}F$, and $^{13}C$ NMR spectra were recorded on a 400 MHz Agilent NMR spectrometer using $CDCl_3$ as solvent and using residual non-deuterated trace solvents as reference. Chemical shifts ($\delta$) are given in ppm relative to TMS ($\delta$

= 0.00 ppm). Coupling constants (J) are given in Hz. Mass spectroscopy was conducted on a Micromass LCT instrument.

The synthetic route to the new materials reported here is given in Scheme S1. Intermediates 1a-f, 2a-f, 3.3a-f, 4.3a-f, 5.1-3,[1,2] and 6.1-2[3] have been reported previously.

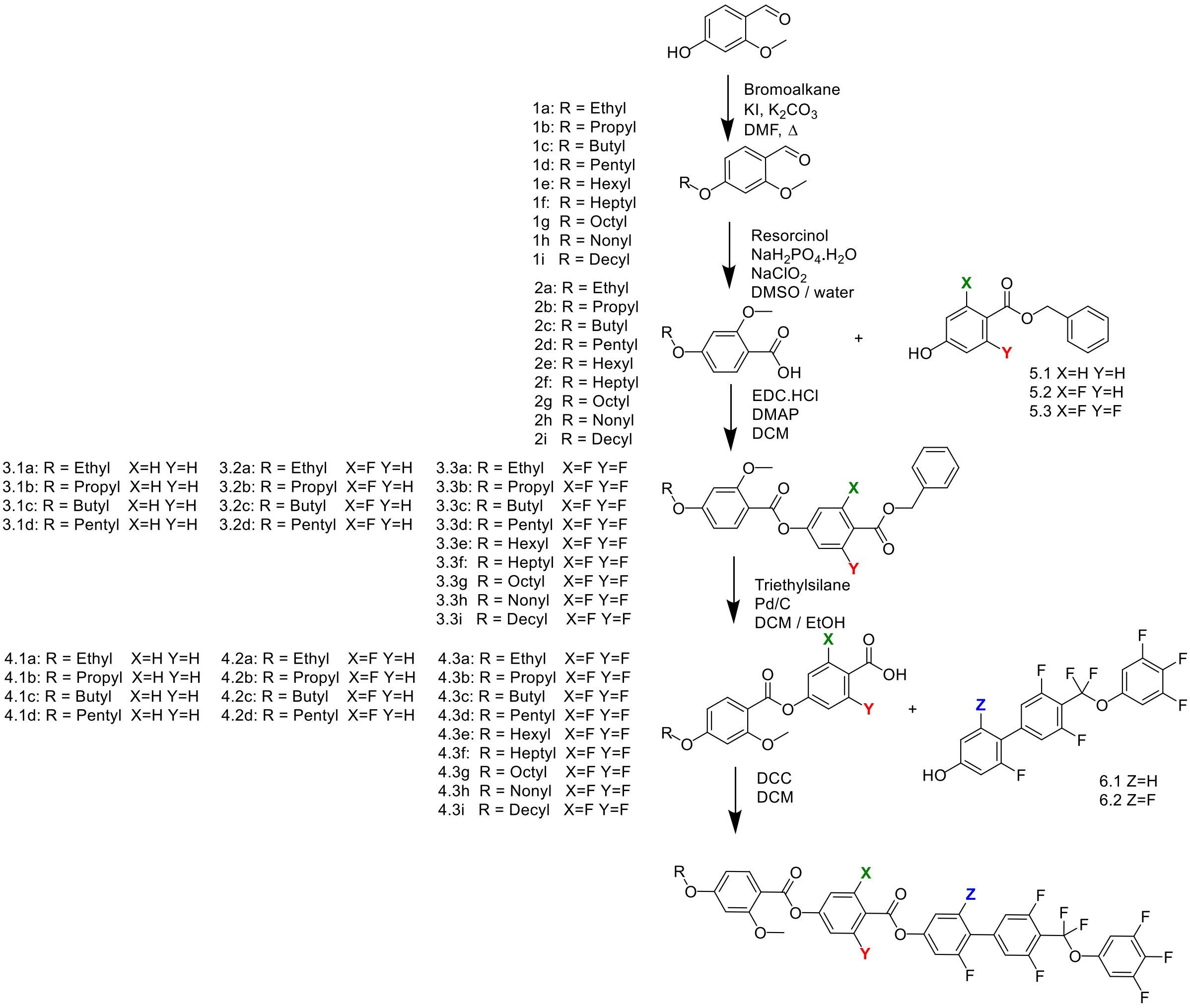

**Scheme S1**. The synthetic route to the new materials reported here.

## General Methods:

### A Esterification method 1

The appropriate benzoic acid and EDC.HCl were dissolved in DCM and stirred for 10 minutes. The corresponding phenol and 4-dimethylaminopyridine (DMAP) were added and the reaction was left stirring at room temperature overnight. The reaction was washed 3x with water and the solvent removed *in vacuo*. The crude product was recrystallised from ethanol.

### B Benzyl deprotection

Under an argon atmosphere triethylsilane was added dropwise to a stirred solution of 3-x and 5 % Pd/C in a 1:1 mix of ethanol and DCM. The reaction was stirred for 15 minutes after addition was complete, then filtered through celite and the solvent removed *in vacuo*. The crude product was washed with hexane to yield the product as a white powder.

**C Esterification method 2**

The appropriate benzoic acid and DCC were dissolved in DCM in an ice bath and stirred for 30 minutes. The corresponding phenol was added, and the reaction was left stirring overnight. The reaction was filtered and the solvent removed *in vacuo*. The crude product was recrystallised from ethanol.

**1g** 4-octyloxy-2-methoxybenzaldehyde

Potassium carbonate (3.815 g, 27 mmol), potassium iodide (115 mg, 0.7 mmol), and 4-hydroxy-2-methoxybenzaldehyde (2.066 g, 13.5 mmol) were added to dry DMF (30 ml) under an argon atmosphere and stirred for 10 minutes. 1-bromooctane (2.6 ml, 15 mmol) was added and the reaction mixture heated to 70 °C overnight. The mixture was cooled, added to 500 ml of water, and stirred until an orange precipitate formed. This was collected by vacuum filtration and the crude product was carried forward without further purification. Yield 3.4 g.

$^{1}$H NMR (400 MHz, $CDCl_3$) δ = 10.31 (s, 1H), 7.82 (d, *J*=8.7, 1H), 6.56 (dd, *J*=8.7, 2.2, 1H), 6.47 (d, *J*=2.2, 1H), 4.05 (t, *J*=6.5, 2H), 3.92 (s, 3H), 1.83 (p, *J*=6.8, 2H), 1.55 – 1.43 (m, 2H), 1.42 – 1.26 (m, 8H), 0.95 – 0.88 (m, 3H).

**1h** 4-nonyloxy-2-methoxybenzaldehyde

Potassium carbonate (3.790 g, 27 mmol), potassium iodide (180 mg, 1.1 mmol), and 4-hydroxy-2-methoxybenzaldehyde (2.066 g, 13.5 mmol) were added to dry DMF (30 ml) under an argon atmosphere and stirred for 10 minutes. 1-bromononane (2.9 ml, 15 mmol) was added and the reaction mixture heated to 85 °C for 16 hrs. The mixture was cooled, added to 500 ml of water, and stirred until an orange precipitate formed. This was collected by vacuum filtration, and the crude product was carried forward without further purification. Yield 4.84 g.

$^{1}$H NMR (400 MHz, $CDCl_3$) δ = 10.27 (s, 1H), 7.79 (d, *J*=8.6, 1H), 6.53 (dd, *J*=8.6, 2.2, 1H), 6.44 (d, *J*=2.2, 1H), 4.02 (t, *J*=6.6, 2H), 3.89 (s, 3H), 1.85 – 1.74 (m, 2H), 1.48 – 1.41 (m, 2H), 1.28 (s, 10H), 0.92 – 0.84 (m, 3H).

**1i** 4-decyloxy-2-methoxybenzaldehyde

Potassium carbonate (3.781 g, 27 mmol), potassium iodide (188 mg, 1.1 mmol), and 4-hydroxy-2-methoxybenzaldehyde (2.069 g, 13.5 mmol) were added to dry DMF (30 ml) under an argon atmosphere and stirred for 10 minutes. 1-bromodecane (3ml, 14 mmol) was added and the reaction mixture heated to 85 °C for 14 hrs. The mixture was cooled, added to 500 ml of water, and stirred until an orange precipitate formed. This was collected by vacuum filtration, and the crude product was carried forward without further purification. Yield 3.08 g.

$^{1}$H NMR (400 MHz, $CDCl_3$) δ = 10.27 (s, 1H), 7.79 (d, *J*=8.7, 1H), 6.53 (dd, *J*=8.7, 2.2, 1H), 6.44 (d, *J*=2.2, 1H), 4.02 (t, *J*=6.5, 2H), 3.89 (s, 3H), 1.85 – 1.74 (m, 2H), 1.63 – 1.56 (m, 2H), 1.46 (p, *J*=6.8, 2H), 1.36 – 1.25 (m, 12H), 0.92 – 0.84 (m, 3H).

**2g** 4-octyloxy-2-methoxybenzoic acid

The aldehyde **1g** (3.352 g, 13 mmol) and resorcinol (2.78 g 25 mmol) were dissolved in DMSO (60 ml). Sodium chlorite (4.753 g, 53 mmol) and sodium dihydrogen phosphate (6.57 g, 48 mmol) were dissolved in water (40 ml), and this was added carefully to the organics and the reaction left stirring overnight. The reaction mixture was diluted with 100 ml water, acidified, and the resulting precipitate collected by vacuum filtration. Yield 2.8 g.

$^{1}$H NMR (400 MHz, $CDCl_3$) $\delta$ = 8.15 (d, *J*=8.8, 1H), 6.66 (dd, *J*=8.8, 2.3, 1H), 6.55 (d, *J*=2.3, 1H), 4.14 – 4.01 (overlapping singlet and triplet, 5H), 1.89 – 1.77 (m, 2H), 1.49 (p, *J*=6.9, 2H), 1.40 – 1.26 (m, 8H), 0.95 – 0.88 (m, 3H).

**2h** 4-nonyloxy-2-methoxybenzoic acid

The aldehyde **1h** (2.047 g, 7.4 mmol) and resorcinol (2.64 g 24 mmol) were dissolved in DMSO (60 ml). Sodium chlorite (5.227 g, 58 mmol) and sodium dihydrogen phosphate (7.017 g, 51 mmol) were dissolved in water (40 ml), and this was added carefully to the organics and the reaction left stirring for 24 hours. The reaction mixture was diluted with 100 ml water, acidified, and the resulting precipitate collected by vacuum filtration. Yield 1.44 g.

$^{1}$H NMR (400 MHz, $CDCl_3$) $\delta$ = 8.12 (d, *J*=8.8, 1H), 6.63 (dd, *J*=8.8, 2.3, 1H), 6.53 (d, *J*=2.3, 1H), 4.04 (d, *J*=1.5, 3H), 4.01 (t, *J*=6.6, 2H), 1.87 – 1.74 (m, 2H), 1.52 – 1.40 (m, 2H), 1.39 – 1.24 (m, 10H), 0.92 – 0.84 (m, 3H).

**2i** 4-decyloxy-2-methoxybenzoic acid

The aldehyde **1i** (2.10 g, 7.2 mmol) and resorcinol (1.8 g 12 mmol) were dissolved in DMSO (30 ml) and acetonitrile (30 ml). Sodium chlorite (2.79 g, 30 mmol) and sodium dihydrogen phosphate (3.69 g, 27 mmol) were dissolved in water (30 ml), and this was added carefully to the organics and the reaction left stirring overnight. The acetonitrile was removed under vacuum, and then the reaction mixture was diluted with 400 ml water, acidified, and the resulting precipitate collected by vacuum filtration. Yield 1.97 g.

$^{1}$H NMR (400 MHz, $CDCl_3$) $\delta$ = 8.12 (d, *J*=8.8, 1H), 6.63 (dd, *J*=8.8, 2.2, 1H), 6.53 (d, *J*=2.2, 1H), 4.04 (s, 3H), 4.02 (t, *J*=6.7, 2H), 1.86 – 1.74 (m, 2H), 1.46 (p, *J*=6.8, 2H), 1.40 – 1.24 (m, 12H), 0.92 – 0.84 (m, 3H).

**3.1a** 4-((benzyloxy)carbonyl)phenyl-4’-ethoxy-2’-methoxybenzoate

Method A

| | | | |
|---|---|---|---|
| Benzoic acid 2a | 786 mg, | 4 mmol, | 1.1 eq. |
| Benzyl-4-hydroxybenzoate | 824 mg, | 3.6 mmol, | 1 eq. |
| EDC.HCl | 1.41 g, | 7.2 mmol, | 2 eq. |
| DMAP | 48 mg, | 0.4 mmol, | 0.11 eq. |
| DCM | 20 ml | | |

20 hrs

Yield 698 mg.

$^{1}$H NMR (400 MHz, $CDCl_3$) $\delta$ = 8.12 (d, *J*=8.5, 2H), 8.06 (d, *J*=8.6, 1H), 7.49 – 7.32 (m, 5H), 7.30 – 7.24 (overlapping, 2H), 6.58 – 6.50 (overlapping, 2H), 5.37 (s, 2H), 4.12 (q, *J*=7.0, 2H), 3.92 (s, 3H), 1.46 (t, *J*=7.0, 3H).

**3.1b** 4-((benzyloxy)carbonyl)phenyl-4’-propoxy-2’-methoxybenzoate

Method A
Benzoic acid 2b 420 mg, 2 mmol, 1.1 eq.
Benzyl-4-hydroxybenzoate 414 mg, 1.8 mmol, 1 eq.
EDC.HCl 706 mg, 3.6 mmol, 2 eq.
DMAP 22 mg, 0.18 mmol, 0.1 eq.
DCM 10 ml
20 hrs
Yield 302 mg.
$^{1}$H NMR (400 MHz, $CDCl_3$) δ = 8.13 (d, *J*=8.6, 2H), 8.06 (d, *J*=8.6, 1H), 7.49 – 7.32 (m, 4H), 7.28 (d, *J*=8.6, 2H), 6.54 (overlapping, 2H), 5.37 (s, 2H), 4.01 (t, *J*=6.6, 2H), 3.92 (s, 3H), 1.86 (p, *J*=6.9, 2H)1.07 (t, *J*=7.4, 3H).

**3.1c** 4-((benzyloxy)carbonyl)phenyl-4’-butoxy-2’-methoxybenzoate

Method A
Benzoic acid 2c 892 mg, 4 mmol, 1.1 eq.
Benzyl-4-hydroxybenzoate 829 mg, 3.6 mmol, 1 eq.
EDC.HCl 1.374 g, 7.2 mmol, 2 eq.
DMAP 44 mg, 0.36 mmol, 0.1 eq.
DCM 20 ml
20 hrs
Yield 314 mg.
$^{1}$H NMR (400 MHz, $CDCl_3$) δ = 8.13 (d, *J*=8.5, 2H), 8.06 (d, *J*=8.6, 1H), 7.49 – 7.24 (m, 7H), 6.58 – 6.50 (m, 2H), 5.37 (s, 2H), 4.05 (t, *J*=6.5, 2H), 3.92 (s, 3H), 1.81 (p, *J*=6.7, 2H), 1.51 (p, *J*=7.3, 2H), 1.00 (t, *J*=7.4, 3H).

**3.1d** 4-((benzyloxy)carbonyl)phenyl-4’-pentoxy-2’-methoxybenzoate

Method A
Benzoic acid 2d 1.011 g, 4.2 mmol, 1.1 eq.
Benzyl-4-hydroxybenzoate 870 mg, 3.8 mmol, 1 eq.
EDC.HCl 1.490 g, 7.6 mmol, 2 eq.
DMAP 46 mg, 0.38 mmol, 0.1 eq.
DCM 15 ml
16 hrs
Yield 175 mg.
$^{1}$H NMR (400 MHz, $CDCl_3$) δ = 8.12 (d, *J*=8.6, 2H), 8.06 (d, *J*=8.6, 1H), 7.49 – 7.30 (m, 5H), 7.27 (d, *J*=8.6, 2H), 6.58 – 6.50 (m, 2H), 5.37 (s, 2H), 4.04 (t, *J*=6.6, 2H), 3.92 (s, 3H), 1.82 (p, *J*=6.8, 2H), 1.53 – 1.35 (m, 4H), 0.95 (t, *J*=7.0, 3H).

**3.2a** 4-((benzyloxy)carbonyl)-3-fluorophenyl-4’-ethoxy-2’-methoxybenzoate

Method A
Benzoic acid 2a 387 mg, 1.8 mmol, 1 eq.
Phenol 5.2 441 mg, 1.8 mmol, 1 eq.
EDC.HCl 694 mg, 3.5 mmol, 1.9 eq.
DMAP 22 mg, 0.18 mmol, 0.1 eq.
DCM 10 ml
20 hrs
Yield 215 mg.

$^1$H NMR (400 MHz, $CDCl_3$) δ = 8.07 – 7.98 (overlapping, 2H), 7.46 (d, *J*=7.1, 2H), 7.43 – 7.34 (overlapping, 2H), 7.38 – 7.30 (overlapping, 1H), 7.08 (d, *J*=9.4, 2H), 6.58 – 6.50 (overlapping, 2H), 5.39 (s, 2H), 4.12 (q, *J*=7.0, 2H), 3.92 (s, 3H), 1.46 (t, *J*=7.0, 3H).

**3.2b** 4-((benzyloxy)carbonyl)-3-fluorophenyl-4'-propoxy-2'-methoxybenzoate

Method A

| | | | |
|---|---|---|---|
| Benzoic acid 2b | 422 mg, | 1.8 mmol, | 1 eq. |
| Phenol 5.2 | 442 mg, | 1.8 mmol, | 1 eq. |
| EDC.HCl | 723 mg, | 3.7 mmol, | 2 eq. |
| DMAP | 17 mg, | 0.14 mmol, | 0.08 eq. |
| DCM | 15 ml | | |

16 hrs
Yield 263 mg.
$^1$H NMR (400 MHz, $CDCl_3$) δ = 8.07 – 7.98 (overlapping, 2H), 7.53 – 7.30 (overlapping, 5H), 7.09 (d, *J*=9.7, 2H), 6.58 – 6.51 (overlapping, 2H), 5.39 (s, 2H), 4.01 (t, *J*=6.6, 2H), 3.92 (s, 3H), 1.85 (h, *J*=7.0, 2H), 1.07 (t, *J*=7.4, 3H).

**3.2c** 4-((benzyloxy)carbonyl)-3-fluorophenyl-4'-butoxy-2'-methoxybenzoate

Method A

| | | | |
|---|---|---|---|
| Benzoic acid 2c | 895 mg, | 4 mmol, | 1.1 eq. |
| Phenol 5.2 | 884 mg, | 3.6 mmol, | 1 eq. |
| EDC.HCl | 1.402 g, | 7.2 mmol, | 2 eq. |
| DMAP | 44 mg, | 0.36 mmol, | 0.1 eq. |
| DCM | 20 ml | | |

16 hrs
Yield 472 mg.
$^1$H NMR (400 MHz, $CDCl_3$) δ = 8.07 – 7.98 (overlapping, 2H), 7.50 – 7.30 (overlapping, 5H), 7.08 (d, *J*=9.3, 2H), 6.58 – 6.50 (overlapping, 2H), 5.39 (s, 2H), 4.05 (t, *J*=6.5, 2H), 3.92 (s, 3H), 1.86 – 1.75 (m, 2H), 1.59 – 1.45 (m, 2H), 1.00 (t, *J*=7.4, 3H).

**3.2d** 4-((benzyloxy)carbonyl)-3-fluorophenyl-4'-pentoxy-2'-methoxybenzoate

Method A

| | | | |
|---|---|---|---|
| Benzoic acid 2d | 1.008 g, | 4.2 mmol, | 1.1 eq. |
| Phenol 5.2 | 932 mg, | 3.8 mmol, | 1 eq. |
| EDC.HCl | 1.513 g, | 7.7 mmol, | 2 eq. |
| DMAP | 44 mg, | 0.36 mmol, | 0.1 eq. |
| DCM | 15 ml | | |

16 hrs
Yield 483 mg.
Carried forward without characterisation.

**3.3g** 4-((benzyloxy)carbonyl)-3.5-difluorophenyl-4'-octyloxy-2'-methoxybenzoate

Method A

| | | | |
|---|---|---|---|
| Benzoic acid 2g | 700 mg, | 2.5 mmol, | 1.1 eq. |
| Phenol 5.3 | 603 mg, | 2.3 mmol, | 1 eq. |
| EDC.HCl | 950 mg, | 4.8 mmol, | 2.1 eq. |
| DMAP | 30 mg, | 0.25 mmol, | 0.11 eq. |
| DCM | 15 ml | | |

16 hrs.
Yield 318 mg.
$^{1}$H NMR (400 MHz, $CDCl_3$) δ = 8.00 (d, *J*=8.7, 1H), 7.49 – 7.30 (m, 5H), 6.95 – 6.87 (m, 2H), 6.58 – 6.49 (m, 2H), 5.40 (s, 2H), 4.04 (t, *J*=6.5, 2H), 3.92 (s, 3H), 1.87 – 1.76 (m, 2H), 1.53 – 1.42 (m, 2H), 1.39 – 1.27 (m, 8H), 0.93 – 0.84 (m, 3H).

**3.3h** 4-((benzyloxy)carbonyl)-3.5-difluorophenyl-4'-nonyloxy-2'-methoxybenzoate

Method C

| | | | |
|---|---|---|---|
| Benzoic acid **2h** | 996 mg, | 3.4 mmol, | 1.3 eq. |
| Phenol **5.3** | 660 mg, | 2.5 mmol, | 1 eq. |
| DCC | 561 mg, | 2.7 mmol, | 1.1 eq. |
| DCM | 15 ml | | |

14 hrs.
Yield 536 mg.

$^{1}$H NMR (400 MHz, $CDCl_3$) δ = 8.00 (d, *J*=8.7, 1H), 7.48 – 7.30 (m, 4H), 6.91 (d, *J*=8.8, 1H), 6.57 – 6.49 (m, 2H), 5.40 (s, 2H), 4.04 (t, *J*=6.5, 2H), 3.92 (s, 3H), 1.81 (dt, *J*=14.1, 6.7, 2H), 1.53 – 1.41 (m, *J*=8.0, 2H), 1.40 – 1.27 (m, 10H), 0.92 – 0.85 (m, 3H).

**3.3i** 4-((benzyloxy)carbonyl)-3.5-difluorophenyl-4'-decyloxy-2'-methoxybenzoate

Method C

| | | | |
|---|---|---|---|
| Benzoic acid **2i** | 730 mg, | 2.4 mmol, | 1.3 eq. |
| Phenol **5.3** | 479 mg, | 1.8 mmol, | 1 eq. |
| DCC | 378 mg, | 1.8 mmol, | 1 eq. |
| DCM | 15 ml | | |

14 hrs.
Yield 240 mg.

$^{1}$H NMR (400 MHz, $CDCl_3$) δ = 8.00 (d, *J*=8.7, 1H), 7.48 – 7.39 (m, 2H), 7.41 – 7.30 (m, 3H), 6.91 (d, *J*=9.0, 2H), 6.57 – 6.49 (m, 2H), 5.40 (s, 2H), 4.04 (t, *J*=6.5, 2H), 3.92 (s, 3H), 1.85 – 1.76 (m, 2H), 1.49 – 1.45 (m, 2H), 1.42 – 1.26 (m, 12H), 0.88 (t, *J*=6.6, 3H).

**4.1a** 4-carboxylphenyl-4'-ethoxy-2'-methoxybenzoate
Method B

| | | | |
|---|---|---|---|
| Benzyl ester 3.1a | 648 mg, | 1.6 mmol, | 1 eq. |
| Triethylsilane | 0.8 ml, | 5.2 mmol, | 3.3 eq. |
| 5 % Pd/C | 155 mg, | | |
| Ethanol | 5 ml | | |
| DCM | 5 ml | | |

Yield 395 mg.
$^{1}$H NMR (400 MHz, $CDCl_3$) δ = 8.16 (d, *J*=8.6, 2H), 8.07 (d, *J*=8.5, 1H), 7.32 (d, *J*=8.6, 2H), 6.59 – 6.51 (m, 2H), 4.13 (q, *J*=7.0, 2H), 3.92 (s, 3H), 1.47 (t, *J*=7.0, 3H).

**4.1b** 4-carboxylphenyl-4'-propoxy-2'-methoxybenzoate
Method B

| | | | |
|---|---|---|---|
| Benzyl ester 3.1b | 274 mg, | 0.6 mmol, | 1 eq. |

Triethylsilane 0.2 ml, 1.3 mmol, 2 eq.
5 % Pd/C 83 mg,
Ethanol 3 ml
DCM 3 ml
Yield 184 mg.
$^1$H NMR (400 MHz, $CDCl_3$) δ = 8.17 (d, *J*=8.0, 2H), 8.07 (d, *J*=8.6, 1H), 7.32 (d, *J*=8.6, 2H), 7.26 (s, 2H), 6.59 – 6.52 (m, 2H), 4.01 (t, *J*=6.6, 2H), 3.93 (s, 3H), 1.91 – 1.81 (m, 2H), 1.07 (t, *J*=7.4, 3H).

**4.1c** 4-carboxylphenyl-4'-butoxy-2'-methoxybenzoate
Method B
Benzyl ester 3.1c 280 mg, 0.6 mmol, 1 eq.
Triethylsilane 0.4 ml, 2.6 mmol, 4 eq.
5 % Pd/C 70 mg,
Ethanol 3 ml
DCM 3 ml
Yield 172 mg.
$^1$H NMR (400 MHz, $CDCl_3$) δ = 8.17 (d, *J*=8.5, 2H), 8.07 (d, *J*=8.7, 1H), 7.32 (d, *J*=8.5, 2H), 6.59 – 6.51 (overlapping, 2H), 4.05 (t, *J*=6.5, 2H), 3.93 (s, 3H), 1.81 (p, *J*=6.8, 2H), 1.59 – 1.45 (m, 2H), 1.00 (t, *J*=7.4, 3H).

**4.1d** 4-carboxylphenyl-4'-pentoxy-2'-methoxybenzoate
Method B
Benzyl ester 3.1d 170 mg, 0.4 mmol, 1 eq.
Triethylsilane 0.35 ml, 2.2 mmol, 5.5 eq.
5 % Pd/C 50 mg,
Ethanol 3 ml
DCM 3 ml
Yield 122 mg.
$^1$H NMR (400 MHz, $CDCl_3$) δ = 8.16 (d, *J*=8.2, 2H), 8.07 (d, *J*=8.7, 21H), 7.32 (d, *J*=8.3, 2H), 6.56 – 6.51 (m, 2H), 4.07 – 4.02 (m, 2H), 3.93 (s, 3H), 1.87 – 1.74 (m, 2H), 1.45 – 1.41 (m, 2H), 1.29 – 1.25 (m, 2H), 0.98 – 0.93 (m, 3H).

**4.2a** 4-carboxyl-3-fluorophenyl-4'-ethoxy-2'-methoxybenzoate
Method B
Benzyl ester 3.2a 180 mg, 0.4 mmol, 1 eq.
Triethylsilane 0.2 ml, 1.3 mmol, 3 eq.
5 % Pd/C 63 mg,
Ethanol 3 ml
DCM 3 ml
Yield 114 mg.
$^1$H NMR (400 MHz, $CDCl_3$) δ = 8.12 – 8.01 (m, 2H), 7.17 – 7.09 (m, 2H), 6.55 (d, *J*=10.8, 2H), 4.13 (q, *J*=7.0, 2H), 3.93 (s, 3H), 1.47 (t, *J*=7.0, 3H).

**4.2b** 4-carboxyl-3-fluorophenyl-4'-propoxy-2'-methoxybenzoate
Method B
Benzyl ester 3.2b 229 mg, 0.5 mmol, 1 eq.
Triethylsilane 0.25 ml, 1.6 mmol, 3 eq.
5 % Pd/C 85 mg,
Ethanol 3 ml
DCM 3 ml

Yield 152 mg.
$^1$H NMR (400 MHz, $CDCl_3$) δ = 8.13 – 8.01 (overlapping, 2H), 7.13 (d, *J*=11.6, 2H), 6.58 – 6.51 (overlapping, 2H), 4.01 (t, *J*=6.6, 2H), 3.93 (s, 3H), 1.91 – 1.81 (m, 2H), 1.07 (t, *J*=7.4, 3H).

**4.2c** 4-carboxyl-3-fluorophenyl-4'-butoxy-2'-methoxybenzoate
Method B

| | | | |
|---|---|---|---|
| Benzyl ester 3.2c | 160 mg, | 0.35 mmol, | 1 eq. |
| Triethylsilane | 0.2 ml, | 1.3 mmol, | 3.7 eq. |
| 5 % Pd/C | 40 mg, | | |
| Ethanol | 3 ml | | |
| DCM | 3 ml | | |

Yield 108 mg.
$^1$H NMR (400 MHz, $CDCl_3$) δ = 8.13 – 8.01 (m, 2H), 7.17 – 7.10 (m, 2H), 6.59 – 6.51 (m, 2H), 4.08 – 4.03 (m, 2H), 3.93 (s, 3H), 1.94 – 1.61 (m, 2H), 1.57 – 1.49 (m, 2H), 1.00 (t, *J*=7.8, 3H).

**4.2d** 4-carboxyl-3-fluorophenyl-4'-pentoxy-2'-methoxybenzoate
Method B

| | | | |
|---|---|---|---|
| Benzyl ester 3.2d | 463 mg, | 1 mmol, | 1 eq. |
| Triethylsilane | 0.55 ml, | 3.6 mmol, | 3.6 eq. |
| 5 % Pd/C | 117 mg, | | |
| Ethanol | 5 ml | | |
| DCM | 5 ml | | |

Yield 333 mg.
$^1$H NMR (400 MHz, $CDCl_3$) δ = 8.13 – 8.01 (m, 2H), 7.17 – 7.09 (m, 2H), 6.59 – 6.51 (m, 2H), 4.05 (t, *J*=6.5, 2H), 3.93 (s, 3H), 1.83 (p, *J*=6.8, 2H), 1.51 – 1.36 (m, 4H), 0.95 (t, *J*=6.9, 3H).

**4.3g** 4-carboxyl-3.5-difluorophenyl-4'-octyloxy-2'-methoxybenzoate

Method B

| | | | |
|---|---|---|---|
| Benzyl ester 3.3g | 318 mg, | 0.6 mmol, | 1 eq. |
| Triethylsilane | 0.3 ml, | 2 mmol, | 3.3 eq. |
| 5 % Pd/C | 60 mg, | | |
| Ethanol | 3 ml | | |
| DCM | 3 ml | | |

Yield 237 mg.
$^1$H NMR (400 MHz, $CDCl_3$) δ = 8.01 (d, *J*=8.7, 1H), 6.96 (d, *J*=9.2, 2H), 6.59 – 6.50 (m, 2H), 4.04 (t, *J*=6.5, 2H), 3.93 (s, 3H), 1.82 (p, *J*=6.8, 2H), 1.52 – 1.42 (m, 2H), 1.34– 1.28 (m, 8H), 0.93 – 0.85 (m, 3H).

**4.3h** 4-carboxyl-3.5-difluorophenyl-4'-nonyloxy-2'-methoxybenzoate

Method B

| | | | |
|---|---|---|---|
| Benzyl ester **3.3h** | 240 mg, | 0.4 mmol, | 1 eq. |
| Triethylsilane | 0.25 ml, | 1.6 mmol, | 4 eq. |
| 5 % Pd/C | 50 mg, | | |
| Ethanol | 3 ml | | |
| DCM | 3 ml | | |

Yield 138 mg.
$^1$H NMR (400 MHz, $CDCl_3$) δ = 8.01 (d, *J*=8.7, 1H), 6.96 (d, *J*=9.9, 2H), 6.58 – 6.50 (m, 2H), 4.04 (t, *J*=6.5, 2H), 3.93 (s, 3H), 1.86 – 1.78 (m, 2H), 1.55 – 1.44 (m, 2H), 1.29 (s, 10H), 0.93 – 0.82 (m, 3H).

**4.3i** 4-carboxyl-3.5-difluorophenyl-4'-decyloxy-2'-methoxybenzoate

Method B

| | | | |
|---|---|---|---|
| Benzyl ester **3.3i** | 240 mg, | 0.4 mmol, | 1 eq. |
| Triethylsilane | 0.25 ml, | 1.6 mmol, | 4 eq. |
| 5 % Pd/C | 50 mg, | | |
| Ethanol | 5 ml | | |
| DCM | 5 ml | | |

Yield 116 mg.
$^1$H NMR (400 MHz, $CDCl_3$) δ = 8.01 (d, *J*=8.7, 1H), 6.95 (d, *J*=9.2, 2H), 6.58 – 6.49 (m, 2H), 4.04 (t, *J*=6.5, 2H), 3.93 (s, 3H), 1.86 – 1.78 (m, 2H), 1.39 – 1.26 (m, 14 H), 0.92 – 0.85 (m, 3H).

| | Method | EDC.HCl | DMAP | DCC | Acid | Phenol | Isolated Yield |
|---|---|---|---|---|---|---|---|
| **$T_2$-0-1** | A | 240 mg, 1.1 mmol | 10 mg, 0.08 mmol | - | 204 mg, 0.65 mmol | 244 mg, 0.6 mmol | 90 mg |
| **$T_3$-0-1** | A | 92 mg, 0.5 mmol | 6 mg, 0.05 mmol | - | 86 mg, 0.26 mmol | 98 mg, 0.2 mmol | 45 mg |
| **$T_4$-0-1** | A | 119 mg, 0.6 mmol | 4 mg, 0.03 mmol | - | 82 mg, 0.24 mmol | 96 mg, 0.2 mmol | 40 mg |
| **$T_5$-0-1** | A | 44 mg, 0.2 mmol | 3 mg, 0.025 mmol | - | 69 mg, 0.2 mmol | 63 mg, 0.15 mmol | 37 mg |
| **$T_2$-1-1** | A | 50 mg, 0.3 mmol | 2 mg, 0.016 mmol | - | 50 mg, 0.15 mmol | 57 mg, 0.14 mmol | 38 mg |
| **$T_3$-1-1** | A | 91 mg, 0.5 mmol | 4 mg, 0.03 mmol | - | 71 mg, 0.2 mmol | 78 mg, 0.19 mmol | 57 mg |
| **$T_4$-1-1** | A | 106 mg, 0.5 mmol | 2 mg, 0.016 mmol | - | 103 mg, 0.3 mmol | 105 mg, 0.25 mmol | 80 mg |
| **$T_5$-1-1** | B | 58 mg, 0.3 mmol | 3 mg, 0.025 mmol | - | 83 mg, 0.22 mmol | 84 mg, 0.2 mmol | 87 mg |
| **$T_2$-0-2** | C | - | - | 165 mg, 0.8 mmol | 163 mg, 0.5 mmol | 166 mg, 0.4 mmol | 66 mg |
| **$T_3$-0-2** | C | - | - | 42 mg, 0.2 mmol | 92 mg, 0.27 mmol | 94 mg 0.2 mmol | 45 mg |
| **$T_4$-0-2** | C | - | - | 38 mg, 0.18 mmol | 85 mg, 0.25 mmol | 83 mg 0.2 mmol | 36 mg |
| **$T_5$-0-2** | C | - | - | 34 mg, 0.17 mmol | 69 mg, 0.19 mmol | 54 mg 0.14 mmol | 17 mg |
| **$T_2$-1-2** | C | - | - | 31 mg, 0.15 mmol | 60 mg, 0.18 mmol | 60 mg 0.14 mmol | 17 mg |
| **$T_3$-1-2** | C | - | - | 42 mg, 0.2 mmol | 77 mg, 0.22 mmol | 73 mg 0.16 mmol | 47 mg |
| **$T_4$-1-2** | C | - | - | 37 mg, 0.18 mmol | 81 mg, 0.22 mmol | 72 mg, 0.16 mmol | 38 mg |
| **$T_5$-1-2** | C | - | - | 61 mg, 0.3 mmol | 151 mg, 0.4 mmol | 133 mg, 0.3 mmol | 49 mg |
| **$T_2$-2-2** | C | - | - | 28 mg, 0.14 mmol | 55 mg, 0.16 mmol | 52 mg, 0.12 mmol | 36 mg |
| **$T_3$-2-2** | C | - | - | 25 mg, 0.12 mmol | 49 mg, 0.13 mmol | 49 mg, 0.11 mmol | 28 mg |
| **$T_4$-2-2** | C | - | - | 48 mg, 0.24 mmol | 95 mg, 0.26 mmol | 86 mg, 0.2 mmol | 35 mg |
| **$T_5$-2-2** | C | - | - | 59 mg, 0.29 mmol | 150 mg, 0.39 mmol | 124 mg, 0.28 mmol | 56 mg |
| **$T_6$-2-2** | C | - | - | 16 mg, 0.01 mmol | 43 mg, 0.11 mmol | 35 mg, 0.08 mmol | 12 mg |
| **$T_7$-2-2** | C | - | - | 28 mg, 0.14 mmol | 74 mg, 0.18 mmol | 59 mg, 0.14 mmol | 22 mg |
| **$T_8$-2-2** | C | - | - | 83 mg, 0.4 mmol | 233 mg, 0.54 mmol | 172 mg, 0.4 mmol | 92 mg[1] |
| **$T_9$-2-2** | C | - | - | 39 mg, 0.19 mmol | 105 mg, 0.24 mmol | 68 mg, 0.16 mg | 45 mg[2] |
| **$T_{10}$-2-2** | C | - | - | 43 mg, 0.21 mmol | 116 mg, 0.25 mmol | 84 mg, 0.2 mmol | 72 mg[2] |

[1] Purified by flash column chromatography with DCM as eluent

[2] Purified by flash column chromatography with 2:1 DCM/ Hexane as eluent.

**$T_2$-0-1**

HRMS (ESI) m/z Calculated for $C_{36}H_{22}O_7F_8$:
$[M+H]^+$ theoretical mass: 719.1311, found 719.13112, difference 0.091 ppm.

$^1$H NMR (400 MHz, $CDCl_3$) δ = 8.26 (d, *J*=8.6, 2H), 8.09 (d, *J*=8.6, 1H), 7.50 (t, *J*=8.6, 1H), 7.39 (d, *J*=8.6, 2H), 7.27 – 7.14 (m, 4H), 7.04 – 6.96 (m, 2H), 6.60 – 6.53 (m, 2H), 4.14 (q, *J*=6.9, 2H), 3.94 (s, 3H), 1.47 (t, *J*=7.0, 3H).

$^{19}$F NMR (376 MHz, $CDCl_3$) δ = -61.81 (t, *J*=26.4, 2F), -110.43 (td, *J*=26.4, 10.3, 2F), -113.81 – -113.91 (m, 1F), -132.37 – -132.51 (m, 2F), -163.12 (tt, *J*=20.8, 5.8, 1F).

$^{13}$C NMR (101 MHz, $CDCl_3$) δ 164.79, 163.99, 162.83, 162.53, 161.26, 161.20, 160.80, 158.73, 158.68, 157.98, 155.88, 152.43, 152.30, 149.65, 144.60, 140.77, 134.63, 131.82, 130.57, 125.76, 122.46, 118.51, 113.22, 113.03, 111.08, 110.82, 110.16, 107.59, 107.19, 105.35, 99.43, 63.98, 56.03, 14.66.

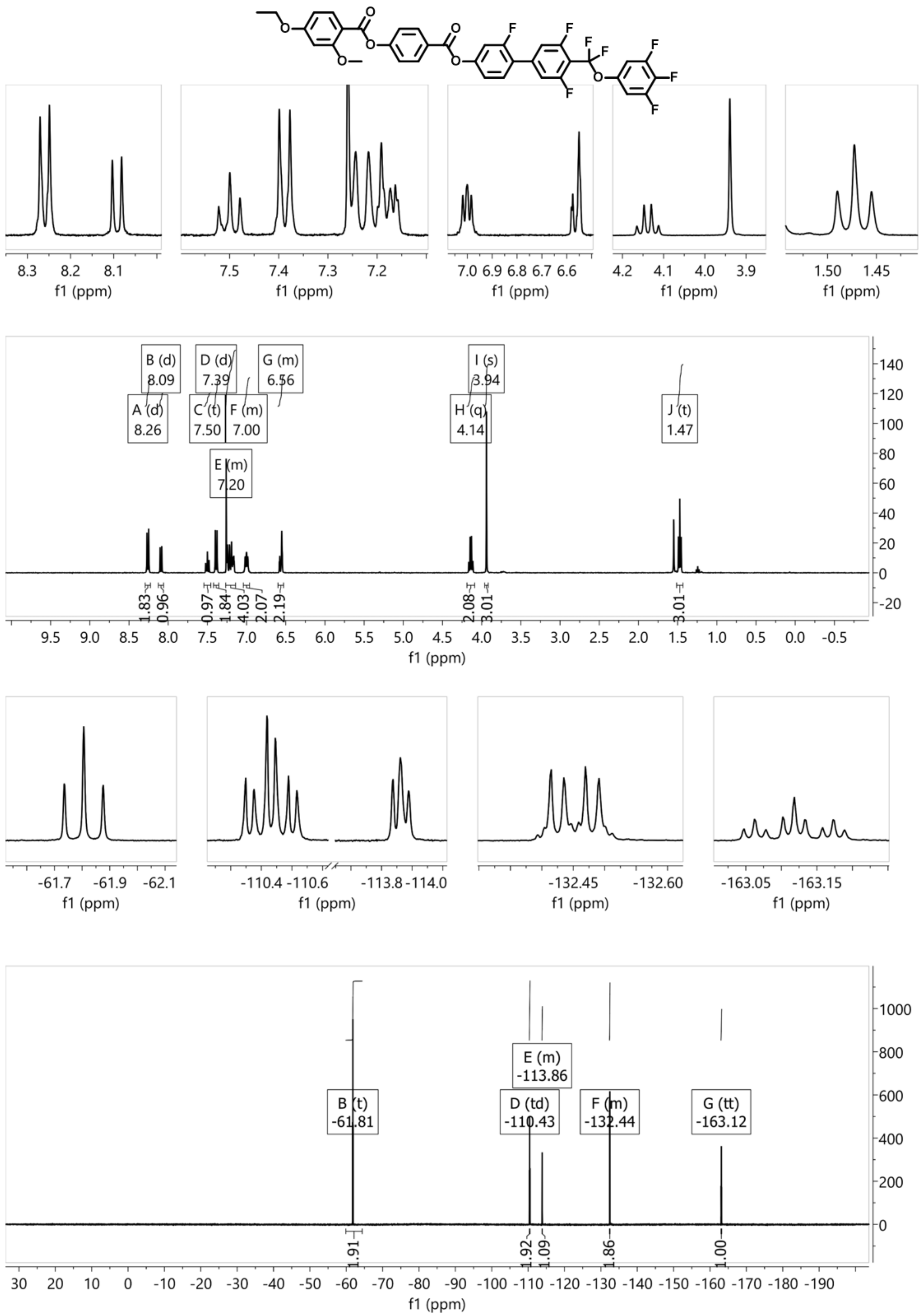

A (d)
8.26
B (d)
8.09
C (t)
7.50
D (d)
7.39
E (m)
7.20
F (m)
7.00
G (m)
6.56
H (q)
4.14
I (s)
3.94
J (t)
1.47
f1 (ppm)
B (t)
-61.81
D (td)
-110.43
E (m)
-113.86
F (m)
-132.44
G (tt)
-163.12

*Figure S 10. $^{1}H$ and $^{19}F$ NMR spectra of $T_2$-0-1.*

**$T_3$-0-1**

HRMS (ESI) m/z Calculated for $C_{37}H_{24}O_7F_8$:
$[M+H]^+$ theoretical mass: 733.1467, found 733.14669, difference -0.020 ppm.

$^{1}H$ NMR (400 MHz, $CDCl_3$) δ = 8.26 (d, *J*=8.7, 1H), 8.09 (d, *J*=8.5, 1H), 7.50 (t, *J*=8.6, 1H), 7.39 (d, *J*=8.7, 2H), 7.28 – 7.14 (m, 4H), 7.04 – 6.96 (m, 2H), 6.60 – 6.53 (m, 2H), 4.02 (t, *J*=6.5, 2H), 3.94 (s, 3H), 1.91 – 1.81 (m, 2H), 1.08 (t, *J*=7.4, 3H).

$^{19}F$ NMR (376 MHz, $CDCl_3$) δ = -61.81 (t, *J*=26.3, 2F), -110.43 (td, *J*=26.3, 10.4, 2F), -113.86 (t, *J*=9.7, 1F), -132.45 (dd, *J*=20.8, 8.0, 2F), -163.12 (tt, *J*=20.8, 5.8, 1F).

$^{13}C$ NMR (101 MHz, $CDCl_3$) δ 165.00, 163.99, 162.83, 162.55, 161.25, 161.21, 160.80, 158.67, 158.61, 158.29, 155.89, 152.42, 152.31, 152.27, 152.16, 149.82, 149.71, 147.02, 140.94, 140.78, 137.40, 137.14, 134.61, 131.82, 130.57, 125.75, 122.47, 118.48, 113.29, 113.22, 113.20, 113.04, 112.98, 112.94, 111.08, 110.82, 110.08, 107.59, 107.55, 107.53, 107.46, 107.41, 105.42, 99.40, 69.91, 56.03, 22.46, 10.48.

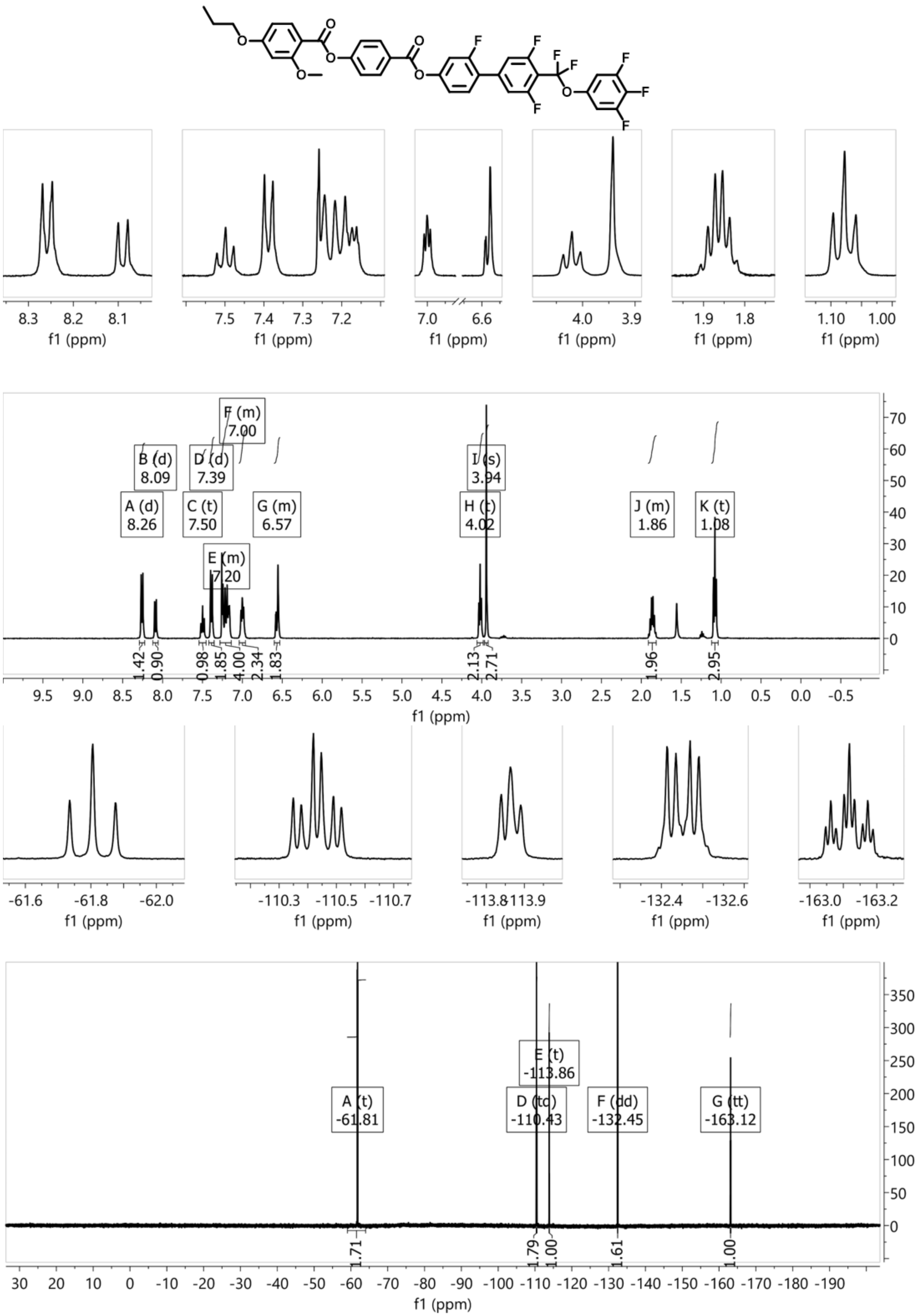

A (d) 8.26
B (d) 8.09
C (t) 7.50
D (d) 7.39
E (m) 7.20
F (m) 7.00
G (m) 6.57
H (t) 4.02
I (s) 3.94
J (m) 1.86
K (t) 1.08
f1 (ppm)
A (t) -61.81
D (tc) -110.43
E (t) -113.86
F (dd) -132.45
G (tt) -163.12

*Figure S 11. $^{1}H$ and $^{19}F$ NMR spectra of $T_3$-0-1.*

**$T_4$-0-1**

HRMS (ESI) m/z Calculated for $C_{38}H_{26}O_7F_8$:
$[M+H]^+$ theoretical mass: 747.1624, found 747.16228, difference -0.100 ppm

$^{1}H$ NMR (400 MHz, $CDCl_3$) δ = 8.26 (d, *J*=8.6, 2H), 8.09 (d, *J*=8.6, 1H), 7.50 (t, *J*=8.6, 1H), 7.39 (d, *J*=8.6, 2H), 7.27 – 7.14 (m, 4H), 7.04 – 6.96 (m, 2H), 6.60 – 6.52 (m, 2H), 4.06 (t, *J*=6.5, 2H), 3.94 (s, 3H), 1.88 – 1.76 (m, 2H), 1.56 – 1.46 (m, 2H), 1.01 (t, *J*=7.4, 3H).

$^{19}F$ NMR (376 MHz, $CDCl_3$) δ = -61.81 (t, *J*=26.4, 2F), -110.43 (td, *J*=26.4, 10.2, 2F), -113.82 – -113.91 (m, 1F), -132.45 (dd, *J*=20.8, 7.9, 2F), -163.12 (tt, *J*=20.9, 5.9, 1F).

$^{13}C$ NMR (101 MHz, $CDCl_3$) δ 165.01, 163.99, 162.84, 162.54, 160.85, 158.29, 155.88, 151.86, 149.73, 134.61, 131.82, 130.53, 125.76, 122.47, 118.48, 113.30, 113.27, 113.22, 113.05, 113.00, 112.94, 111.08, 110.83, 109.92, 107.65, 107.55, 107.42, 107.37, 105.42, 99.40, 68.14, 56.03, 31.13, 19.19, 13.82.

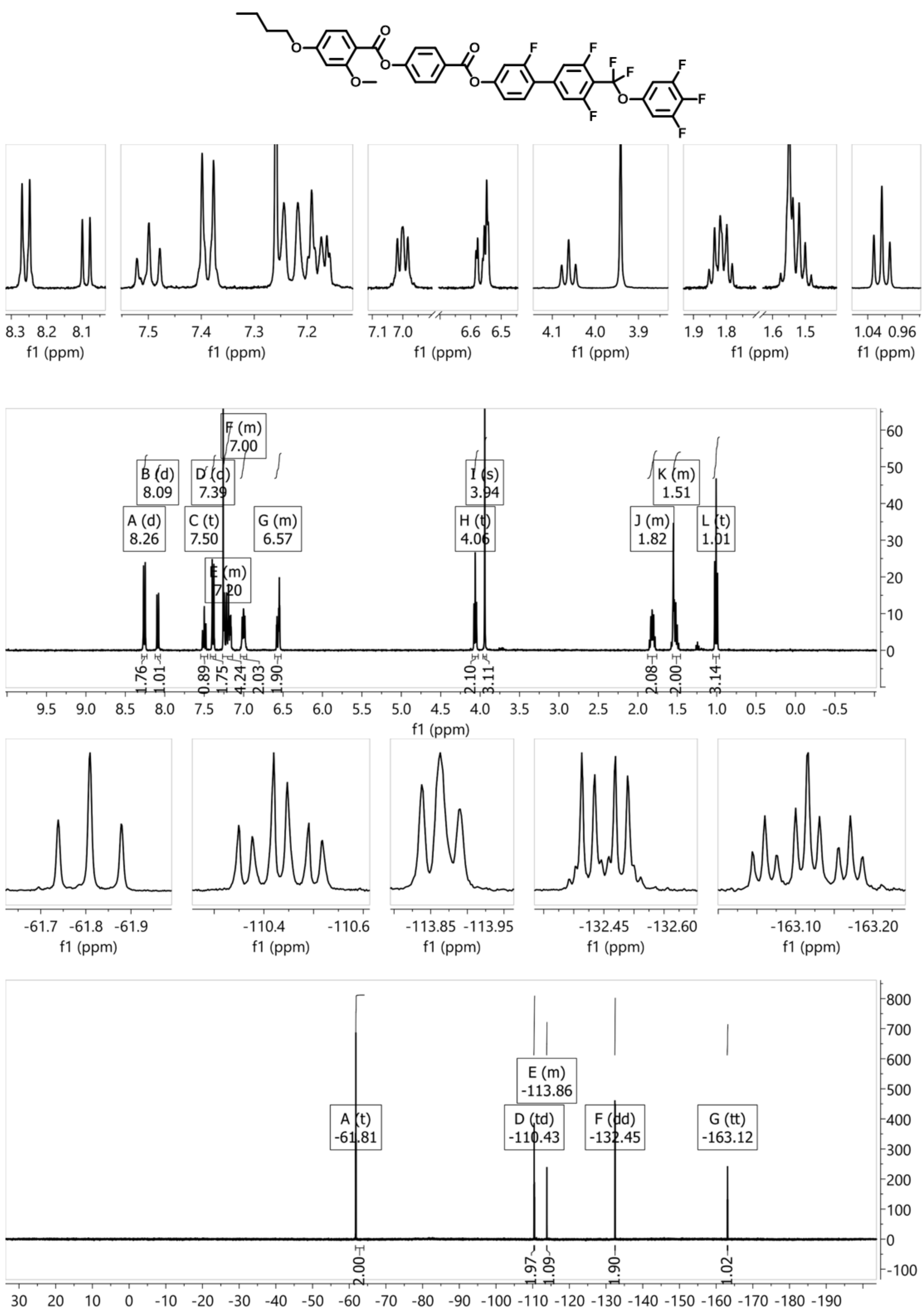

A (d) 8.26
B (d) 8.09
C (t) 7.50
D (d) 7.39
E (m) 7.20
F (m) 7.00
G (m) 6.57
H (t) 4.06
I (s) 3.94
J (m) 1.82
K (m) 1.51
L (t) 1.01
f1 (ppm)
A (t) -61.81
D (td) -110.43
E (m) -113.86
F (dd) -132.45
G (tt) -163.12
f1 (ppm)

*Figure S 12. $^{1}H$ and $^{19}F$ NMR spectra of $T_4$-0-1.*

**$T_5$-0-1**

HRMS (ESI) m/z Calculated for $C_{39}H_{28}O_7F_8$:
$[M+H]^+$ theoretical mass: 761.1780, found 761.17794, difference -0.085 ppm

$^{1}H$ NMR (400 MHz, $CDCl_3$) δ = 8.26 (d, *J*=8.3, 2H), 8.09 (d, *J*=8.4, 1H), 7.50 (t, *J*=8.6, 1H), 7.39 (d, *J*=8.3, 2H), 7.27 – 7.15 (m, 4H), 7.04 – 6.96 (m, 2H), 6.60 – 6.52 (m, 2H), 4.05 (t, *J*=6.4, 2H), 3.94 (s, 3H), 1.88 – 1.80 (m, 2H), 1.54 – 1.36 (m, 4H), 0.96 (t, *J*=6.9, 3H).

$^{19}F$ NMR (376 MHz, $CDCl_3$) δ = -61.81 (t, *J*=26.3, 2F), -110.43 (td, *J*=26.3, 10.7, 2F), -113.86 (t, *J*=9.8, 1F), -132.36 – -132.53 (m, 2F), -163.12 (tt, *J*=20.8, 5.9, 1F).

$^{13}C$ NMR (101 MHz, $CDCl_3$) δ 165.01, 163.99, 162.83, 162.54, 161.31, 161.27, 161.19, 161.11, 160.80, 158.72, 158.68, 158.62, 158.29, 155.89, 152.43, 152.42, 152.33, 152.31, 152.28, 152.26, 152.18, 149.86, 149.73, 144.73, 140.98, 140.80, 140.73, 137.44, 137.18, 134.61, 131.82, 130.57, 130.53, 125.75, 122.47, 118.52, 113.27, 113.22, 113.21, 113.03, 112.98, 112.96, 111.08, 110.83, 110.07, 107.59, 107.53, 107.42, 105.41, 99.40, 68.44, 56.03, 28.80, 28.12, 22.43, 14.01.

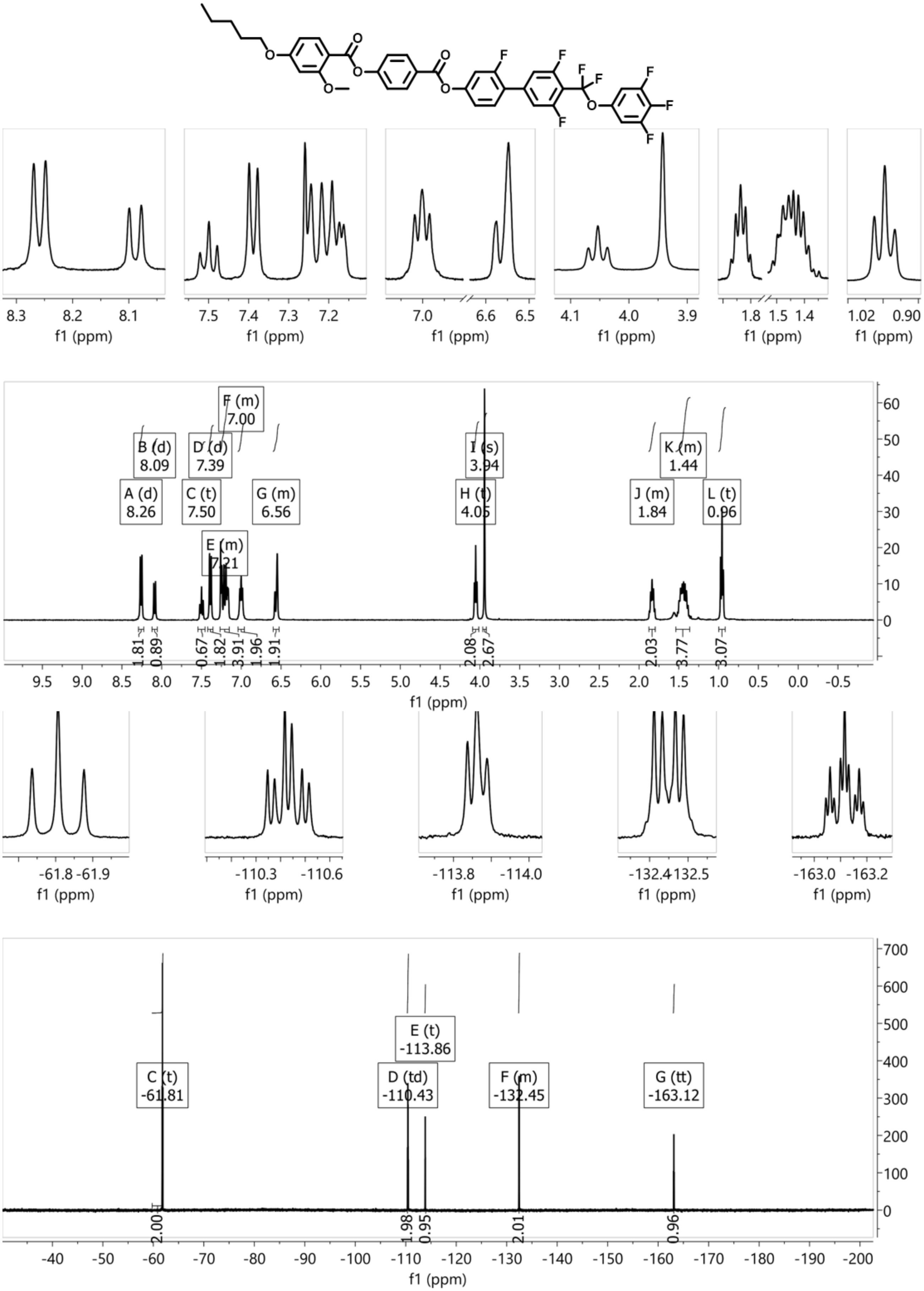

O
O
O
O
O
F
F
F
F
F
O
F
F
F
F
8.3
8.2
8.1
f1 (ppm)
7.5
7.4
7.3
7.2
f1 (ppm)
7.0
6.6
6.5
f1 (ppm)
4.1
4.0
3.9
f1 (ppm)
1.8
1.5
1.4
f1 (ppm)
1.02
0.90
f1 (ppm)
F (m)
7.00
B (d)
8.09
D (d)
7.39
I (s)
3.94
K (m)
1.44
A (d)
8.26
C (t)
7.50
G (m)
6.56
H (t)
4.05
J (m)
1.84
L (t)
0.96
E (m)
7.21
1.81
0.89
0.67
1.82
3.91
1.96
1.91
2.08
2.67
2.03
3.77
3.07
9.5
9.0
8.5
8.0
7.5
7.0
6.5
6.0
5.5
5.0
4.5
4.0
3.5
3.0
2.5
2.0
1.5
1.0
0.5
0.0
-0.5
f1 (ppm)
60
50
40
30
20
10
0
-61.8
-61.9
f1 (ppm)
-110.3
-110.6
f1 (ppm)
-113.8
-114.0
f1 (ppm)
-132.4
-132.5
f1 (ppm)
-163.0
-163.2
f1 (ppm)
E (t)
-113.86
C (t)
-61.81
D (td)
-110.43
F (m)
-132.45
G (tt)
-163.12
2.00
1.98
0.95
2.01
0.96
-40
-50
-60
-70
-80
-90
-100
-110
-120
-130
-140
-150
-160
-170
-180
-190
-200
f1 (ppm)
700
600
500
400
300
200
100
0

*Figure S 13. $^{1}H$ and $^{19}F$ NMR spectra of $T_5$-0-1.*

**$T_2$-1-1**

HRMS (ESI) m/z Calculated for $C_{36}H_{21}O_7F_9$:
$[M+H]^+$ theoretical mass: 737.1216, found 737.12152, difference -0.153 ppm

$^{1}H$ NMR (400 MHz, $CDCl_3$) δ = 8.16 (t, *J*=8.3, 1H), 8.06 (d, *J*=8.6, 1H), 7.50 (t, *J*=8.5, 1H), 7.26 – 7.15 (m, 4H), 7.04 – 6.96 (m, 2H), 6.60 – 6.52 (m, 2H), 4.19 – 4.11 (m, 2H), 3.94 (s, 3H), 1.47 (t, *J*=7.0, 3H).

$^{19}F$ NMR (376 MHz, $CDCl_3$) δ = -61.82 (t, *J*=26.3, 2F), -104.42 (t, *J*=9.4, 1F), -110.42 (td, *J*=26.3, 10.8, 2F), -113.81 (d, *J*=9.0, 1F), -132.45 (dd, *J*=20.7, 8.0, 2F), -163.11 (t, *J*=20.7, 1F).

$^{13}C$ NMR (101 MHz, $CDCl_3$) δ 165.01, 164.23, 162.68, 162.24, 161.58, 161.23, 160.81, 158.69, 158.59, 158.27, 156.80, 156.70, 152.37, 152.31, 152.10, 151.90, 146.94, 144.51, 140.72, 134.69, 133.31, 130.53, 123.46, 123.28, 118.54, 118.45, 118.26, 114.28, 114.16, 113.22, 113.17, 113.05, 112.99, 111.82, 111.60, 111.13, 110.80, 109.72, 107.62, 107.59, 107.50, 107.38, 105.43, 99.41, 64.02, 56.03, 14.65.

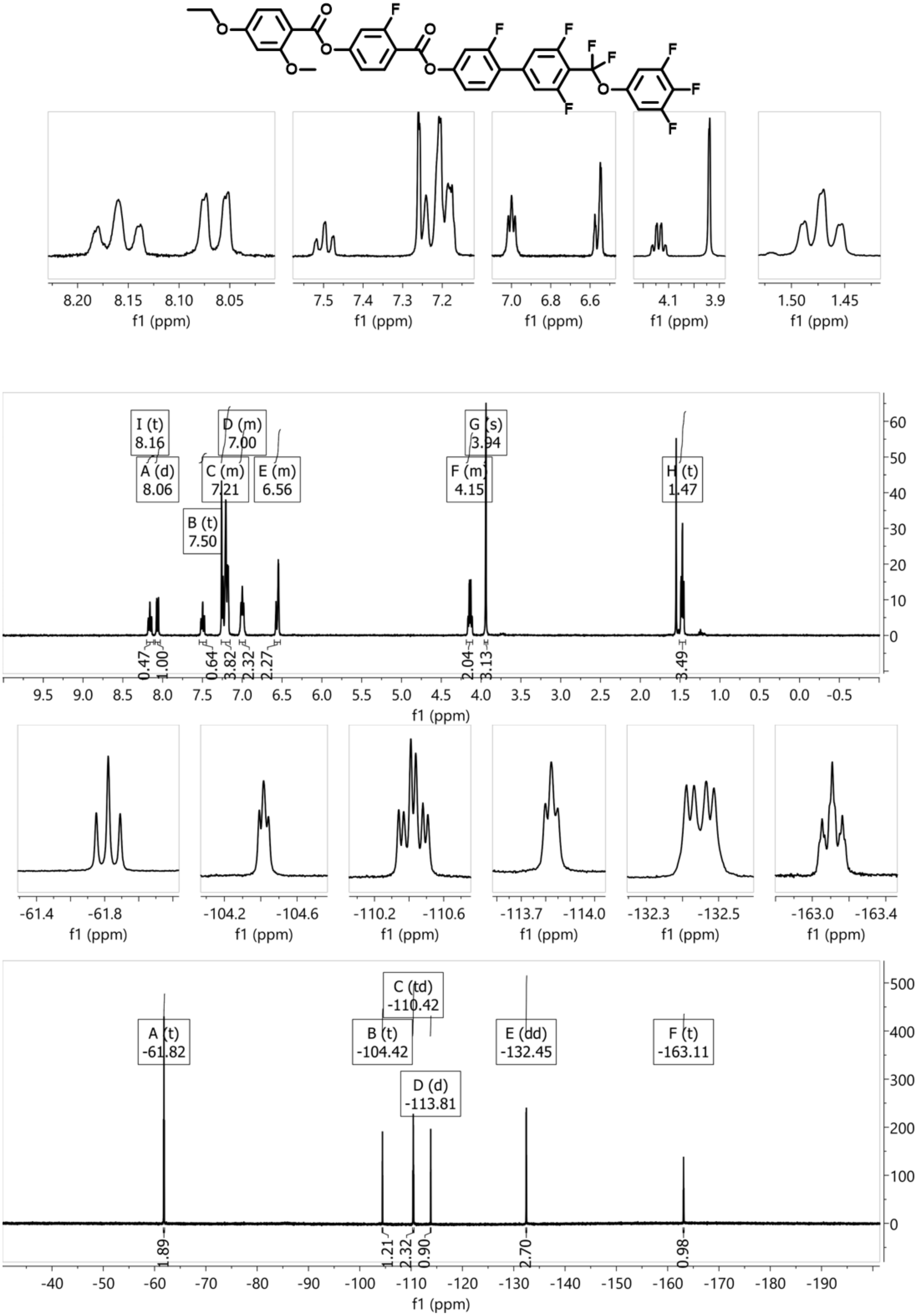

*Figure S 14. $^1H$ and $^{19}F$ NMR spectra of $T_2$-1-1.*

**$T_3$-1-1**

HRMS (ESI) m/z Calculated for $C_{37}H_{23}O_7F_9$:
$[M+H]^+$ theoretical mass: 751.1373, found 751.13731, difference 0.036 ppm

$^1H$ NMR (400 MHz, $CDCl_3$) δ = 8.16 (t, *J*=8.4, 1H), 8.06 (d, *J*=8.6, 1H), 7.50 (t, *J*=8.5, 1H), 7.26 – 7.16 (m, 7H), 7.04 – 6.96 (m, 1H), 6.60 – 6.53 (m, 2H), 4.02 (t, *J*=6.6, 2H), 3.94 (s, 3H), 1.91 – 1.81 (m, 2H), 1.08 (t, *J*=7.4, 3H).

$^{19}F$ NMR (376 MHz, $CDCl_3$) δ = -61.82 (t, *J*=26.3, 2F), -104.35 – -104.49 (m, 1F), -110.42 (td, *J*=26.3, 10.8, 2F), -113.82 (t, *J*=9.7, 1F), -132.45 (dd, *J*=20.7, 8.2, 2F), -163.11 (t, *J*=20.7, 1F).

$^{13}C$ NMR (101 MHz, $CDCl_3$) δ 165.23, 164.53, 164.21, 162.70, 162.25, 161.79, 161.59, 161.32, 161.31, 161.25, 161.21, 160.84, 158.33, 158.32, 156.90, 156.79, 156.78, 156.68, 152.26, 152.08, 152.03, 151.99, 140.86, 140.83, 140.81, 140.70, 134.67, 133.31, 130.60, 124.36, 124.29, 123.50, 123.35, 118.49, 118.26, 114.27, 114.11, 113.24, 111.76, 111.05, 110.82, 109.56, 107.61, 107.37, 105.51, 99.38, 69.95, 56.03, 22.45, 10.48.

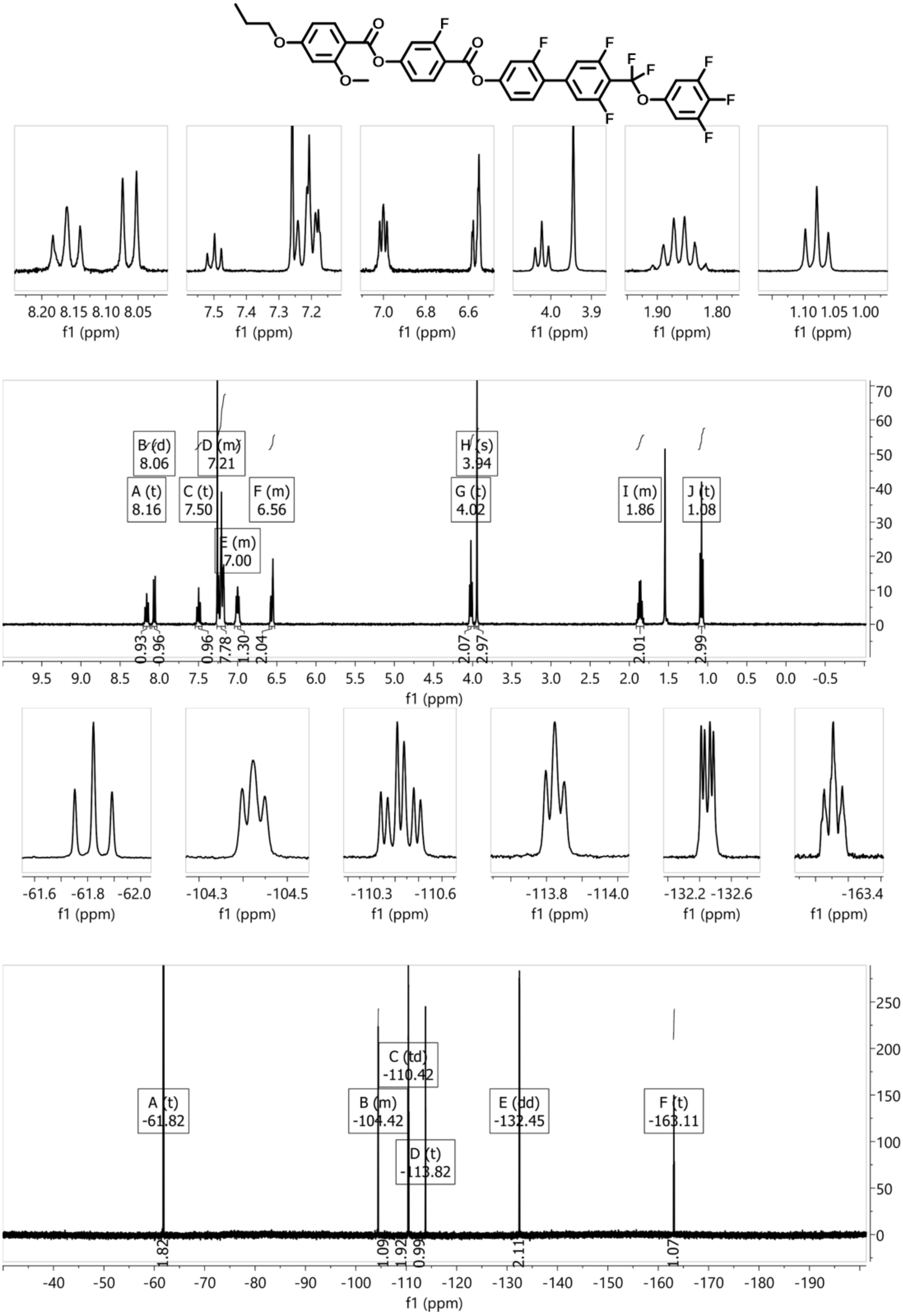

A (t) 8.16
B (d) 8.06
C (t) 7.50
D (m) 7.21
E (m) 7.00
F (m) 6.56
G (t) 4.02
H (s) 3.94
I (m) 1.86
J (t) 1.08
f1 (ppm)
A (t) -61.82
B (m) -104.42
C (td) -110.42
D (t) -113.82
E (dd) -132.45
F (t) -163.11
f1 (ppm)

*Figure S 15. $^{1}H$ and $^{19}F$ NMR spectra of $T_3$-1-1.*

**$T_4$-1-1**

HRMS (ESI) m/z Calculated for $C_{38}H_{25}O_7F_9$:
$[M+H]^+$ theoretical mass: 765.1529, found 765.15299, difference 0.074 ppm

$^{1}H$ NMR (400 MHz, $CDCl_3$) δ = 8.16 (t, *J*=8.5, 1H), 8.06 (d, *J*=8.6, 1H), 7.50 (t, *J*=8.5, 1H), 7.26 – 7.15 (m, 6H), 7.04 – 6.96 (m, 2H), 6.60 – 6.52 (m, 2H), 4.06 (t, *J*=6.5, 2H), 3.94 (s, 3H), 1.82 (p, *J*=6.6, 2H), 1.58 – 1.46 (m, 2H), 1.01 (t, *J*=7.4, 3H).

$^{19}F$ NMR (376 MHz, $CDCl_3$) δ = -61.81 (t, *J*=26.4, 2F), -104.42 (dd, *J*=11.3, 8.0, 1F), -110.42 (td, *J*=26.4, 10.5, 2F), -113.82 (t, *J*=9.7, 1F), -132.38 – -132.51 (m, 2F), -163.12 (tt, *J*=21.0, 6.0, 1F).

$^{13}C$ NMR (101 MHz, $CDCl_3$) δ 165.23, 164.21, 162.69, 162.24, 161.68, 161.59, 161.24, 161.19, 160.79, 158.72, 158.62, 158.26, 156.79, 152.35, 152.00, 151.96, 149.78, 140.79, 134.66, 133.30, 130.57, 130.53, 118.51, 118.44, 118.27, 118.23, 114.26, 114.19, 114.15, 113.30, 113.23, 113.21, 113.02, 112.99, 112.96, 111.72, 111.47, 111.05, 110.80, 109.56, 108.96, 107.59, 107.56, 107.43, 107.35, 105.50, 99.39, 68.17, 56.03, 31.11, 19.19, 13.82.

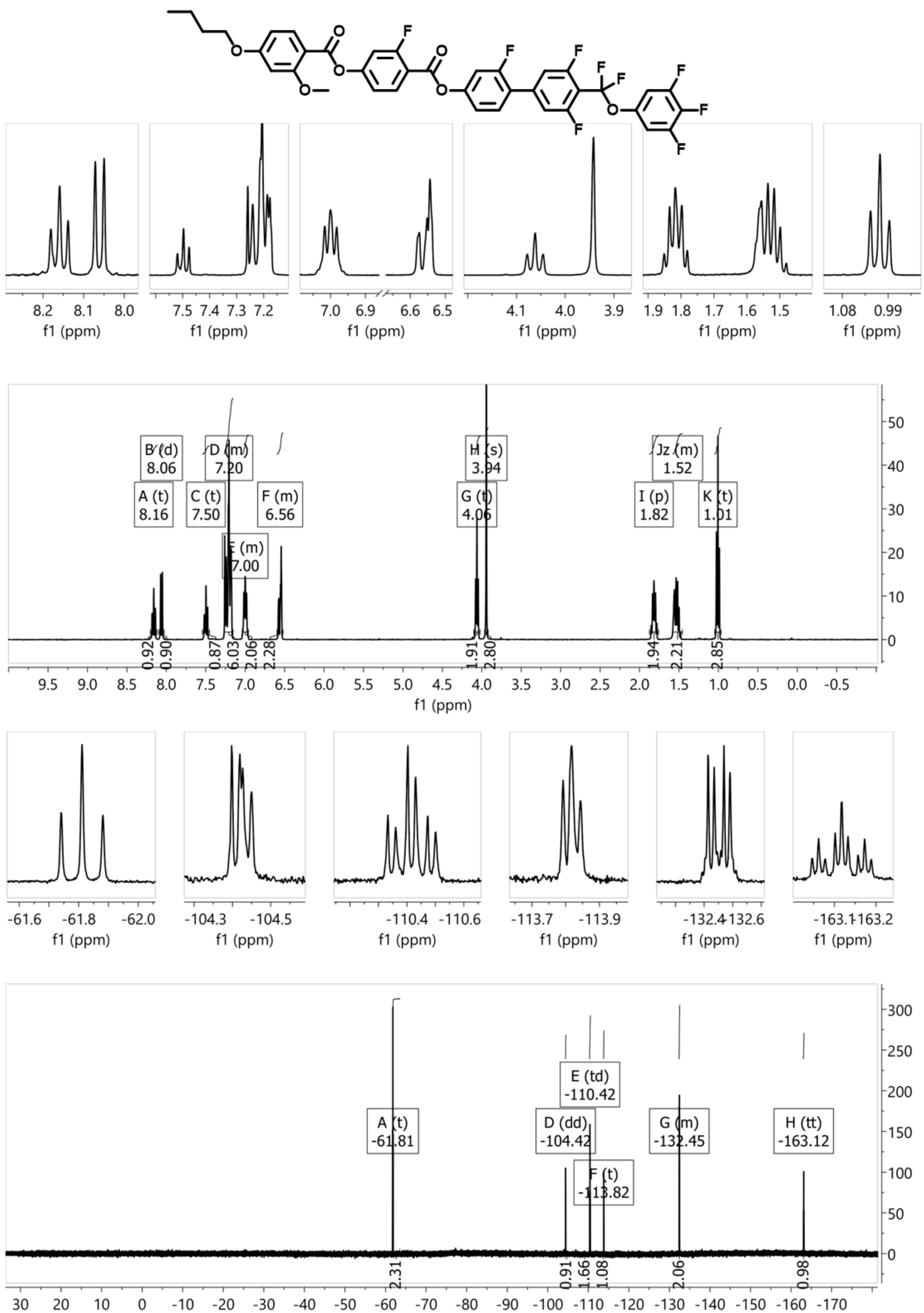
F
O
O
O
F
O
O
F
F
F
F
O
F
F
F
8.2 8.1 8.0
7.5 7.4 7.3 7.2
7.0 6.9 6.6 6.5
4.1 4.0 3.9
1.9 1.8 1.7 1.6 1.5
1.08 0.99
f1 (ppm)
B (d) 8.06
D (m) 7.20
H (s) 3.94
Jz (m) 1.52
A (t) 8.16
C (t) 7.50
F (m) 6.56
G (t) 4.06
I (p) 1.82
K (t) 1.01
E (m) 7.00
0.92 0.90 0.87 6.03 2.06 2.28 1.91 2.80 1.94 2.21 2.85
9.5 9.0 8.5 8.0 7.5 7.0 6.5 6.0 5.5 5.0 4.5 4.0 3.5 3.0 2.5 2.0 1.5 1.0 0.5 0.0 -0.5
-61.6 -61.8 -62.0
-104.3 -104.5
-110.4 -110.6
-113.7 -113.9
-132.4 -132.6
-163.1 -163.2
E (td) -110.42
A (t) -61.81
D (dd) -104.42
G (m) -132.45
H (tt) -163.12
F (t) -113.82
2.31 0.91 1.66 1.08 2.06 0.98
30 20 10 0 -10 -20 -30 -40 -50 -60 -70 -80 -90 -100 -110 -120 -130 -140 -150 -160 -170

*Figure S 16. $^{1}H$ and $^{19}F$ NMR spectra of $T_4$-1-1.*

**$T_5$-1-1**

HRMS (ESI) m/z Calculated for $C_{39}H_{27}O_7F_9$:
$[M+H]^+$ theoretical mass: 779.1686, found 779.16851, difference -0.094 ppm

$^{1}H$ NMR (400 MHz, $CDCl_3$) δ = 8.16 (t, *J*=8.5, 1H), 8.06 (d, *J*=8.6, 1H), 7.50 (t, *J*=8.6, 1H), 7.26 – 7.16 (m, 6H), 7.04 – 6.96 (m, 2H), 6.60 – 6.52 (m, 2H), 4.05 (t, *J*=6.6, 2H), 3.94 (s, 3H), 1.84 (p, *J*=6.7, 2H), 1.54 – 1.36 (m, 4H), 0.96 (t, *J*=7.0, 3H).

$^{19}F$ NMR (376 MHz, $CDCl_3$) δ = -61.81 (t, *J*=26.3, 2F), -104.42 (dd, *J*=11.4, 8.2, 1F), -110.42 (td, *J*=26.3, 11.0, 2F), -113.82 (t, *J*=9.8, 1F), -132.45 (dd, *J*=21.0, 8.2, 2F), -163.12 (tt, *J*=21.0, 6.0, 1F).

$^{13}C$ NMR (101 MHz, $CDCl_3$) δ 165.23, 164.21, 162.69, 162.24, 161.67, 161.62, 161.59, 161.24, 161.17, 160.77, 158.67, 158.63, 158.26, 156.80, 156.68, 152.32, 152.25, 152.03, 151.92, 149.90, 149.81, 149.76, 149.71, 144.64, 144.60, 140.87, 140.75, 139.76, 137.36, 137.27, 137.08, 134.66, 133.31, 130.57, 130.53, 123.39, 123.29, 118.48, 118.44, 118.27, 118.24, 114.25, 114.16, 113.30, 113.23, 113.20, 113.02, 112.99, 112.96, 111.72, 111.47, 111.06, 110.80, 109.55, 107.59, 107.54, 107.44, 107.35, 105.50, 99.39, 68.48, 56.03, 28.78, 28.12, 22.43, 14.00.

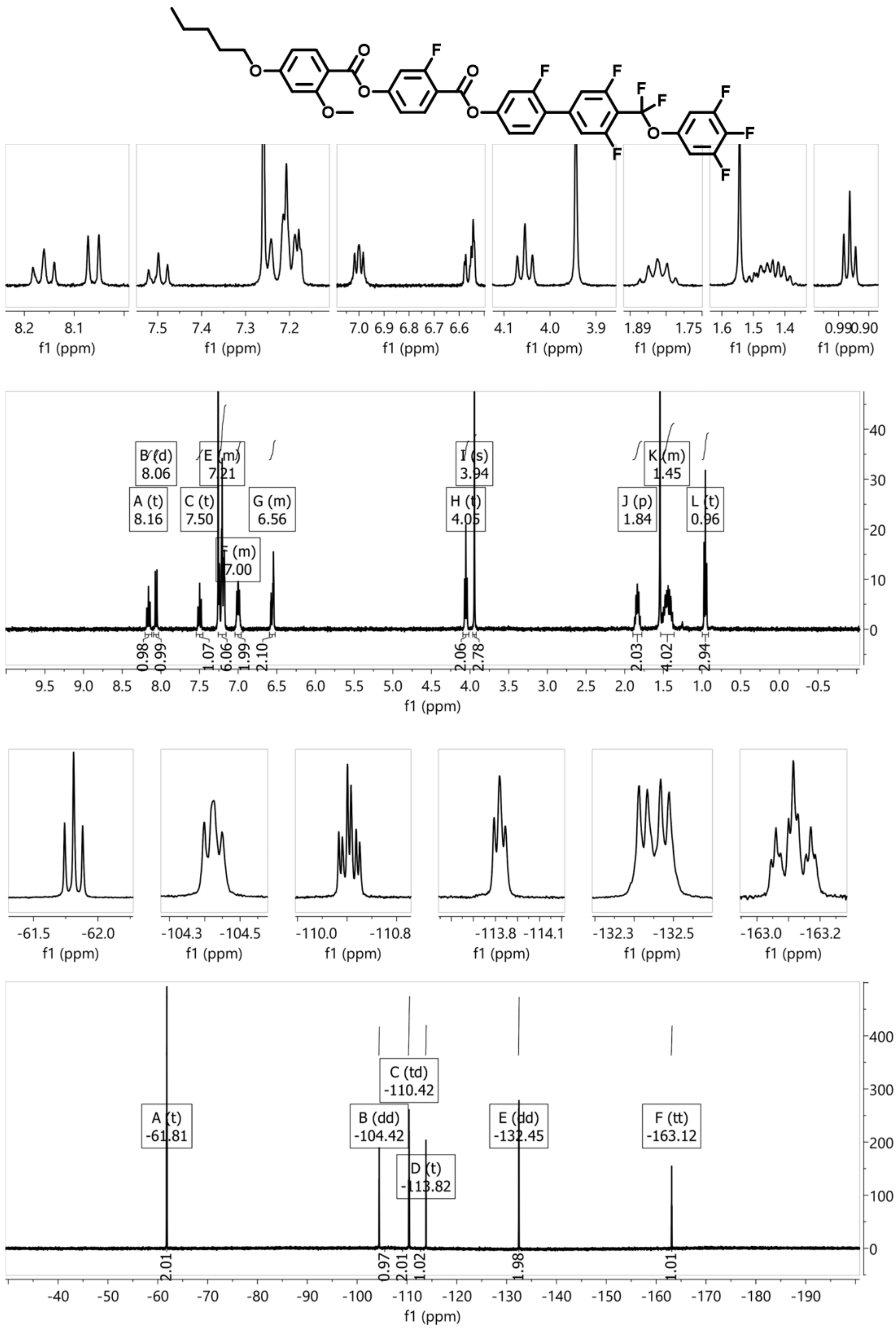

F
O
O
F
F
F
F
F
O
F
F
F
F
f1 (ppm)
A (t)
8.16
B (d)
8.06
C (t)
7.50
E (m)
7.21
F (m)
7.00
G (m)
6.56
H (t)
4.05
I (s)
3.94
J (p)
1.84
K (m)
1.45
L (t)
0.96
A (t)
-61.81
B (dd)
-104.42
C (td)
-110.42
D (t)
-113.82
E (dd)
-132.45
F (tt)
-163.12

*Figure S 17. $^{1}H$ and $^{19}F$ NMR spectra of $T_5$-1-1.*

**$T_2$-0-2**
HRMS (ESI) m/z Calculated for $C_{36}H_{21}O_7F_9$:
$[M+H]^+$ theoretical mass: 737.1216, found 737.12150, difference -0.180 ppm.

$^{1}H$ NMR (400 MHz, $CDCl_3$) δ = 8.24 (d, *J*=8.7, 2H), 8.09 (d, *J*=8.6, 1H), 7.39 (d, *J*=8.7, 2H), 7.17 (d, *J*=10.3, 2H), 7.06 – 6.97 (m, 4H), 6.60 – 6.52 (m, 2H), 4.14 (q, *J*=6.9, 2H), 3.94 (s, 3H), 1.47 (t, *J*=7.0, 3H).

$^{19}F$ NMR (376 MHz, $CDCl_3$) δ = -61.98 (t, *J*=26.5, 2F), -110.63 (td, *J*=26.5, 10.9, 2F), -112.06 (d, *J*=8.8, 2F), -132.44 (dd, *J*=21.0, 8.6, 2F), -163.07 (t, *J*=21.0, 1F).

$^{13}C$ NMR (101 MHz, $CDCl_3$) δ 164.82, 163.60, 162.78, 162.55, 161.05, 161.00, 160.91, 158.49, 158.40, 156.04, 152.17, 152.07, 149.87, 149.78, 144.69, 144.52, 137.27, 134.65, 134.47, 132.25, 131.88, 125.35, 122.54, 114.89, 114.63, 112.92, 110.07, 107.64, 107.61, 107.47, 107.40, 106.97, 106.94, 106.89, 106.77, 106.72, 106.68, 105.35, 99.42, 63.99, 56.03, 14.66.

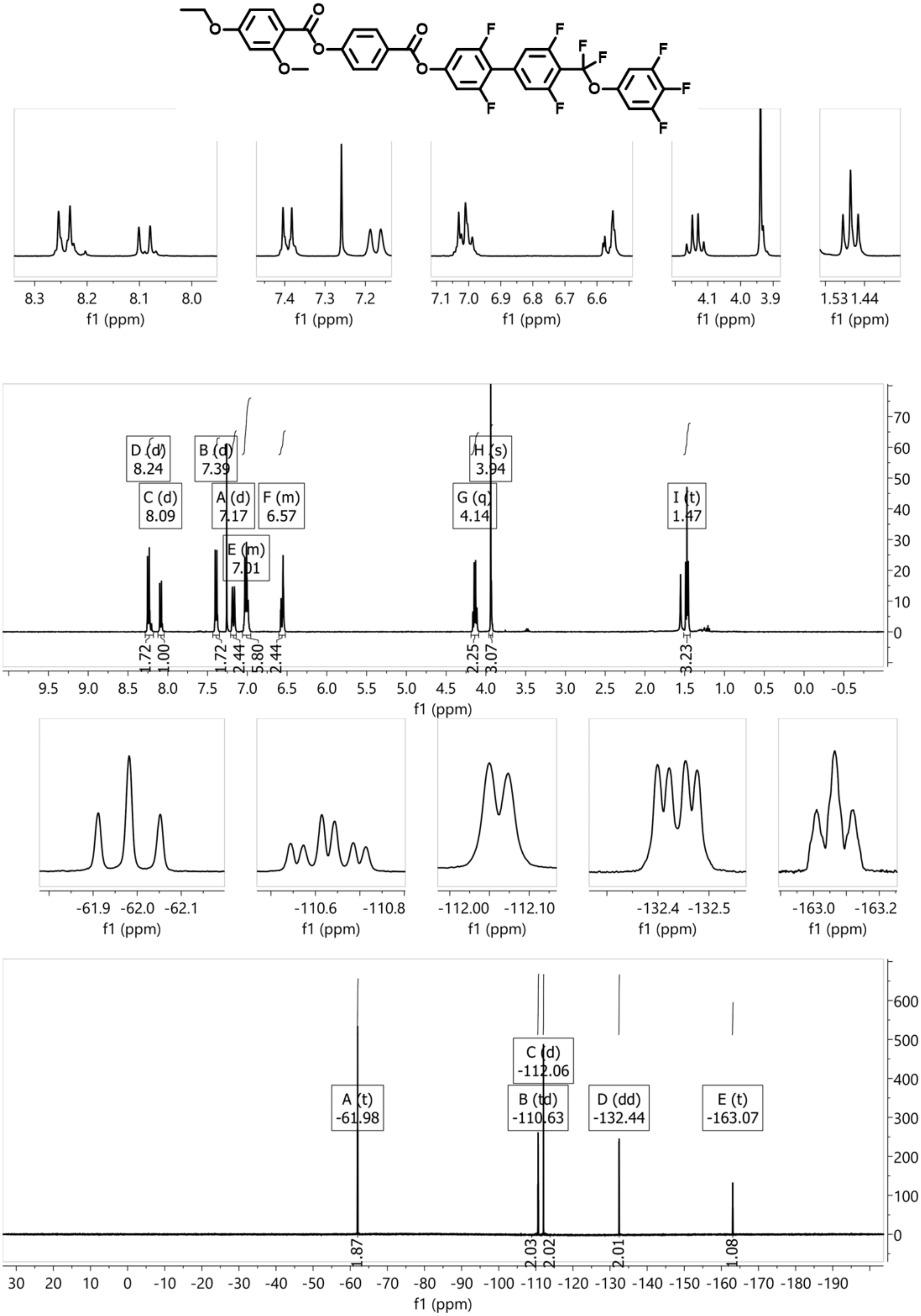
D (d) 8.24
C (d) 8.09
B (d) 7.39
A (d) 7.17
F (m) 6.57
E (m) 7.01
H (s) 3.94
G (q) 4.14
I (t) 1.47
f1 (ppm)
A (t) -61.98
B (td) -110.63
C (d) -112.06
D (dd) -132.44
E (t) -163.07

*Figure S 18. $^{1}H$ and $^{19}F$ NMR spectra of $T_2$-0-2.*

**$T_3$-0-2**

HRMS (ESI) m/z Calculated for $C_{37}H_{23}O_7F_9$:
$[M+H]^+$ theoretical mass: 751.1373, found 751.13685, difference -0.576 ppm.

$^{1}H$ NMR (400 MHz, $CDCl_3$) δ = 8.24 (d, *J*=7.9, 2H), 8.09 (d, *J*=8.9, 1H), 7.39 (d, *J*=7.9, 2H), 7.18 (d, *J*=10.5, 2H), 7.05 – 6.97 (m, 4H), 6.60 – 6.53 (m, 2H), 4.02 (t, *J*=6.6, 2H), 3.94 (s, 3H), 1.91 – 1.79 (m, 2H), 1.10 – 1.04 (m, 3H).

$^{19}F$ NMR (376 MHz, $CDCl_3$) δ = -61.99 (t, *J*=26.6, 2F), -110.63 (td, *J*=26.6, 10.7, 2F), -112.06 (d, *J*=8.6, 2F), -132.44 (dd, *J*=20.6, 8.2, 2F), -163.06 (t, *J*=20.6, 1F).

$^{13}C$ NMR (101 MHz, $CDCl_3$) δ 165.03, 163.61, 163.02, 162.57, 158.56, 158.43, 156.06, 152.01, 149.80, 134.63, 131.88, 125.35, 122.55, 114.88, 114.62, 114.31, 114.08, 112.78, 112.11, 110.00, 107.66, 107.44, 107.08, 106.78, 105.42, 99.39, 69.92, 56.04, 22.46, 10.49.

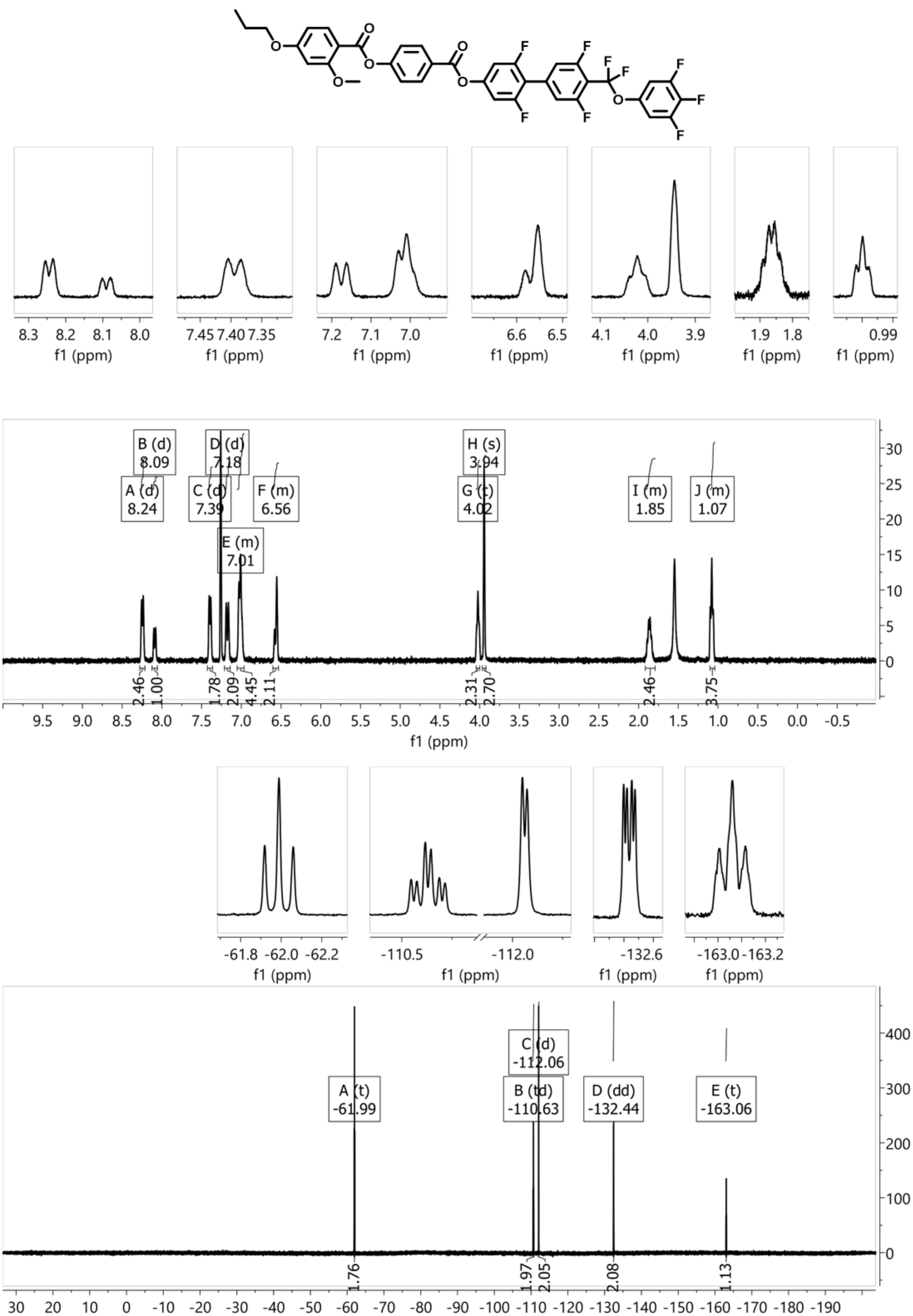

A (d) 8.24
B (d) 8.09
C (d) 7.39
D (d) 7.18
E (m) 7.01
F (m) 6.56
G (t) 4.02
H (s) 3.94
I (m) 1.85
J (m) 1.07
f1 (ppm)
A (t) -61.99
B (td) -110.63
C (d) -112.06
D (dd) -132.44
E (t) -163.06
f1 (ppm)

*Figure S 19. $^{1}H$ and $^{19}F$ NMR spectra of $T_3$-0-2.*

**$T_4$-0-2**

HRMS (ESI) m/z Calculated for $C_{38}H_{25}O_7F_9$:
$[M+H]^+$ theoretical mass: 765.1529, found 765.15276, difference -0.226ppm.

$^{1}H$ NMR (400 MHz, $CDCl_3$) δ = 8.24 (d, *J*=8.8, 2H), 8.09 (d, *J*=8.6, 1H), 7.39 (d, *J*=8.8, 2H), 7.18 (d, *J*=10.2, 2H), 7.05 – 6.97 (m, 4H), 6.61 – 6.52 (m, 2H), 4.06 (t, *J*=6.5, 2H), 3.94 (s, 3H), 1.88 – 1.76 (m, 2H), 1.60 – 1.46 (m, 2H), 1.01 (t, *J*=7.4, 3H).

$^{19}F$ NMR (376 MHz, $CDCl_3$) δ = -61.99 (t, *J*=26.7, 2F), -110.63 (td, *J*=26.7, 10.8, 2F), -112.06 (d, *J*=8.7, 2F), -132.44 (dd, *J*=20.7, 8.2, 2F), -163.06 (t, *J*=20.7, 1F).

$^{13}C$ NMR (101 MHz, $CDCl_3$) δ 165.05, 163.61, 162.79, 162.56, 161.13, 161.00, 160.95, 158.54, 158.41, 156.05, 152.31, 152.25, 152.12, 152.00, 149.84, 144.76, 144.58, 138.96, 134.62, 134.42, 131.88, 125.34, 122.55, 114.88, 114.62, 109.98, 107.64, 107.40, 106.98, 106.68, 105.42, 99.39, 68.14, 56.03, 31.13, 19.19, 13.83.

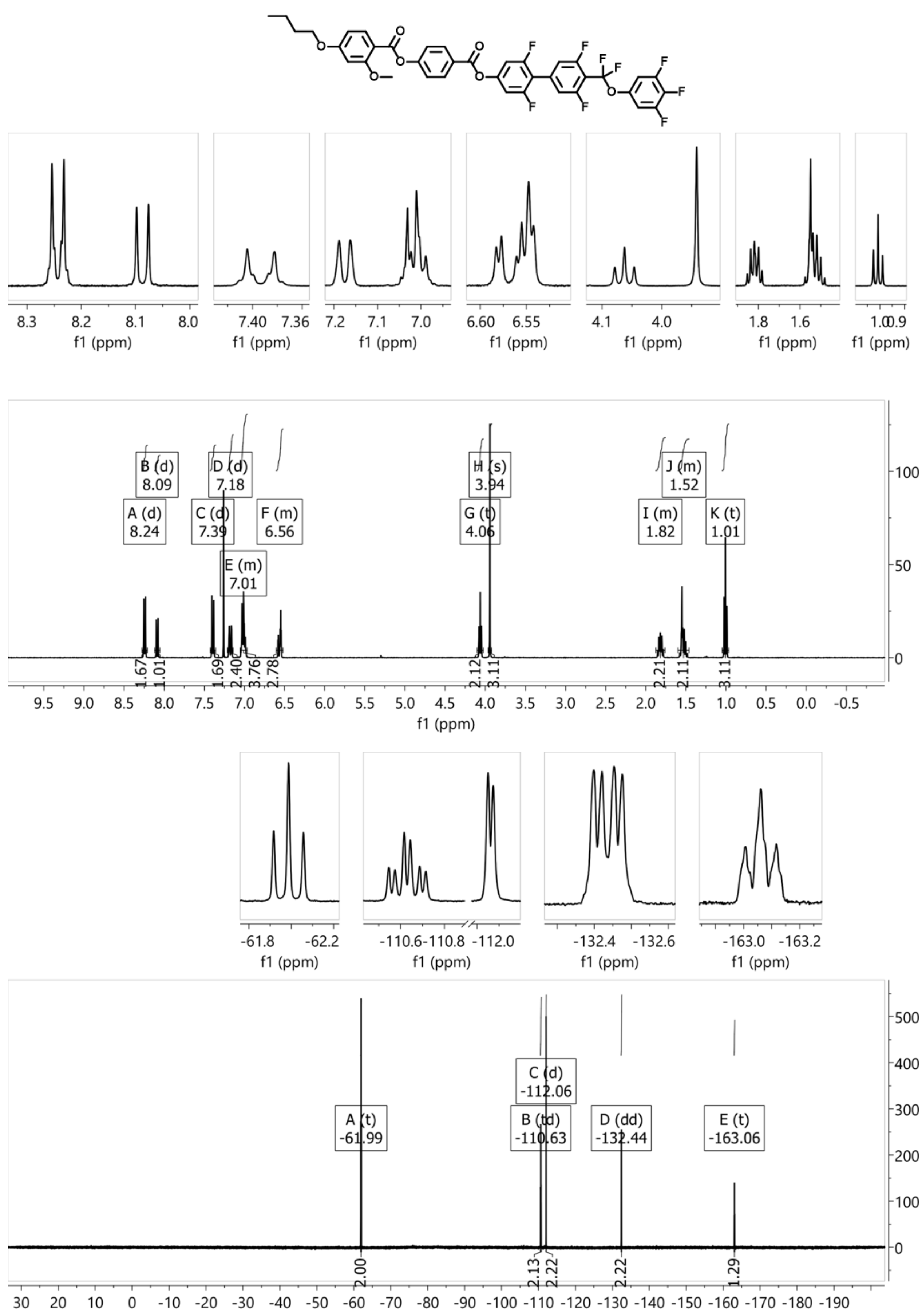

A (d)
8.24
B (d)
8.09
C (d)
7.39
D (d)
7.18
E (m)
7.01
F (m)
6.56
G (t)
4.06
H (s)
3.94
I (m)
1.82
J (m)
1.52
K (t)
1.01
f1 (ppm)
A (t)
-61.99
B (td)
-110.63
C (d)
-112.06
D (dd)
-132.44
E (t)
-163.06
f1 (ppm)

*Figure S 20. $^{1}H$ and $^{19}F$ NMR spectra of $T_4$-0-2.*

**$T_5$-0-2**
HRMS (ESI) m/z Calculated for $C_{39}H_{27}O_7F_9$:
$[M+H]^+$ theoretical mass: 779.1686, found 779.16858, difference -0.004 ppm.

$^{1}H$ NMR (400 MHz, $CDCl_3$) δ = 8.24 (d, *J*=8.6, 2H), 8.09 (d, *J*=8.6, 1H), 7.39 (d, *J*=8.6, 2H), 7.26 (s, 2H), 7.18 (d, *J*=10.6, 2H), 7.05 – 6.97 (m, 4H), 6.60 – 6.52 (m, 2H), 4.05 (t, *J*=6.6, 2H), 3.94 (s, 3H), 1.84 (p, *J*=6.8, 2H), 1.52 – 1.36, (m, 4H), 0.96 (t, *J*=7.0, 3H).

$^{19}F$ NMR (376 MHz, $CDCl_3$) δ = -61.97 (t, *J*=26.5, 2F), -110.62 (td, *J*=26.5, 10.2, 2F), -112.06 (d, *J*=8.5, 2F), -132.44 (dd, *J*=20.8, 7.9, 2F), -163.07 (tt, *J*=20.8, 5.9, 1F).

$^{13}C$ NMR (101 MHz, $cdcl_3$) δ 165.04, 165.00, 163.60, 163.56, 163.54, 162.78, 162.74, 162.56, 162.53, 161.66, 161.00, 159.63, 158.49, 158.41, 156.26, 156.06, 134.63, 132.15, 131.88, 125.81, 125.35, 122.59, 122.55, 114.88, 114.78, 114.65, 114.61, 110.08, 110.02, 107.63, 107.39, 106.98, 106.96, 106.71, 106.67, 105.44, 105.40, 99.40, 68.43, 56.02, 28.79, 28.12, 22.43, 14.01.

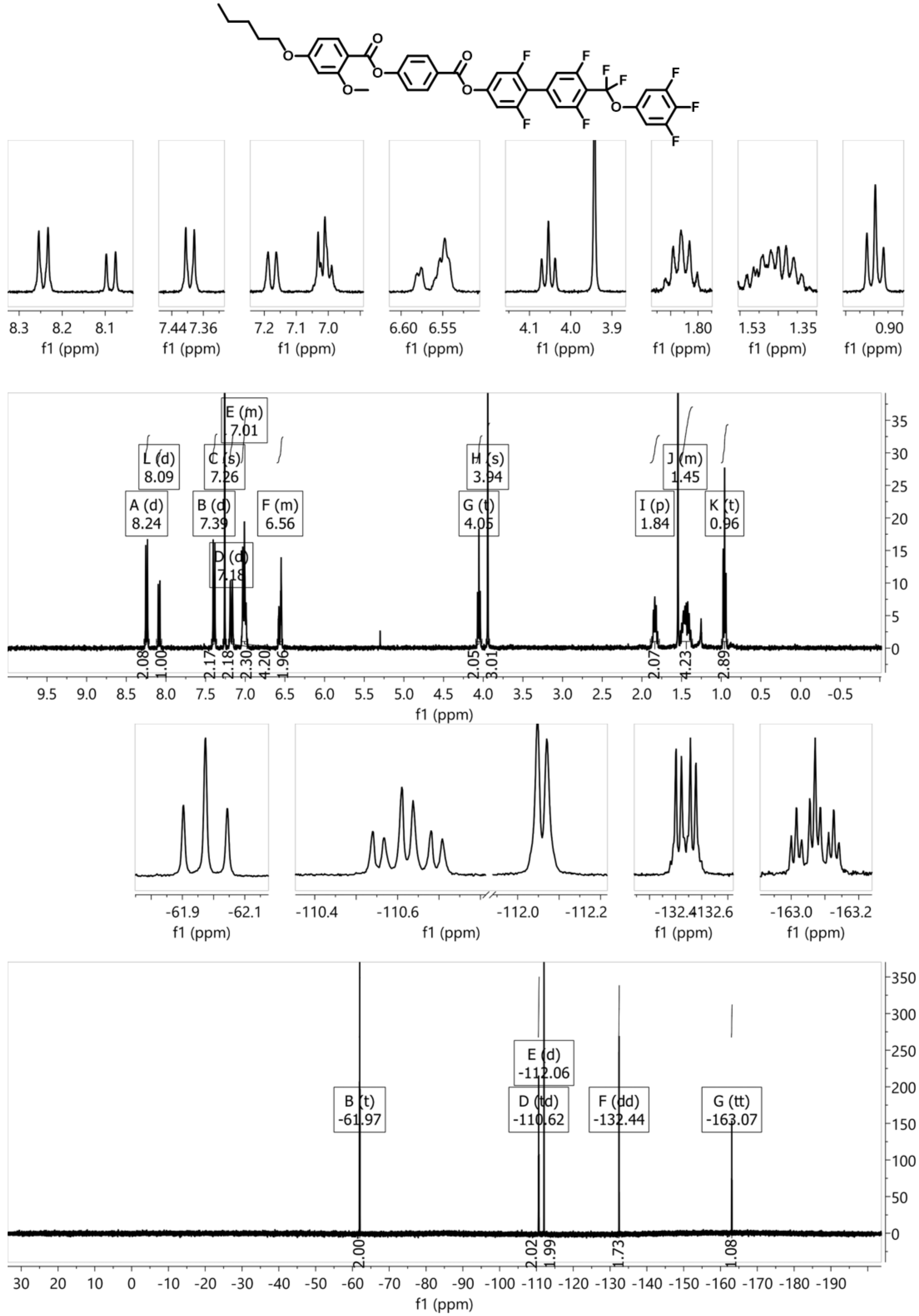

O
O
O
O
O
F
F
F
F
F
F
O
F
F
F
8.3
8.2
8.1
7.44
7.36
7.2
7.1
7.0
6.60
6.55
4.1
4.0
3.9
1.80
1.53
1.35
0.90
f1 (ppm)
E (m)
7.01
L (d)
8.09
C (s)
7.26
H (s)
3.94
J (m)
1.45
A (d)
8.24
B (d)
7.39
F (m)
6.56
G (t)
4.05
I (p)
1.84
K (t)
0.96
D (d)
7.18
2.08
1.00
2.17
2.18
2.30
4.20
1.96
2.05
3.01
2.07
4.23
2.89
-61.9
-62.1
-110.4
-110.6
-112.0
-112.2
-132.4
132.6
-163.0
-163.2
E (d)
-112.06
B (t)
-61.97
D (td)
-110.62
F (dd)
-132.44
G (tt)
-163.07
2.00
2.02
1.99
1.73
1.08

*Figure S 21. $^{1}H$ and $^{19}F$ NMR spectra of $T_5$-0-2.*

**$T_2$-1-2**
HRMS (ESI) m/z Calculated for $C_{36}H_{20}O_7F_{10}$:
$[M+H]^+$ theoretical mass: 755.1122, found 755.11234, difference 0.171 ppm.

$^{1}H$ NMR (400 MHz, $CDCl_3$) δ = 8.15 (t, *J*=8.6, 1H), 8.06 (d, *J*=8.6, 1H), 7.23 – 7.14 (m, 4H), 7.02 (overlapping multiplets, 4H), 6.60 – 6.52 (m, 2H), 4.14 (q, *J*=7.0, 2H), 3.94 (s, 3H), 1.47 (t, *J*=7.0, 3H).

$^{19}F$ NMR (376 MHz, $CDCl_3$) δ = -61.99 (t, *J*=26.7, 2F), -104.13 (d, *J*=9.4, 1F), -110.61 (td, *J*=26.7, 10.8, 2F), -111.99 (d, *J*=8.7, 2F), -132.44 (dd, *J*=20.6, 8.5, 2F), -163.06 (t, *J*=20.9, 1F).

$^{13}C$ NMR (101 MHz, $CDCl_3$) δ 165.05, 164.24, 162.70, 162.19, 161.62, 161.29, 161.08, 160.94, 158.58, 158.36, 157.00, 152.33, 152.19, 151.77, 151.49, 144.39, 139.80, 137.22, 134.70, 134.35, 133.34, 118.35, 114.87, 114.62, 113.89, 113.83, 111.78, 111.53, 109.55, 107.64, 106.96, 106.65, 105.44, 99.40, 64.03, 56.03, 14.65.

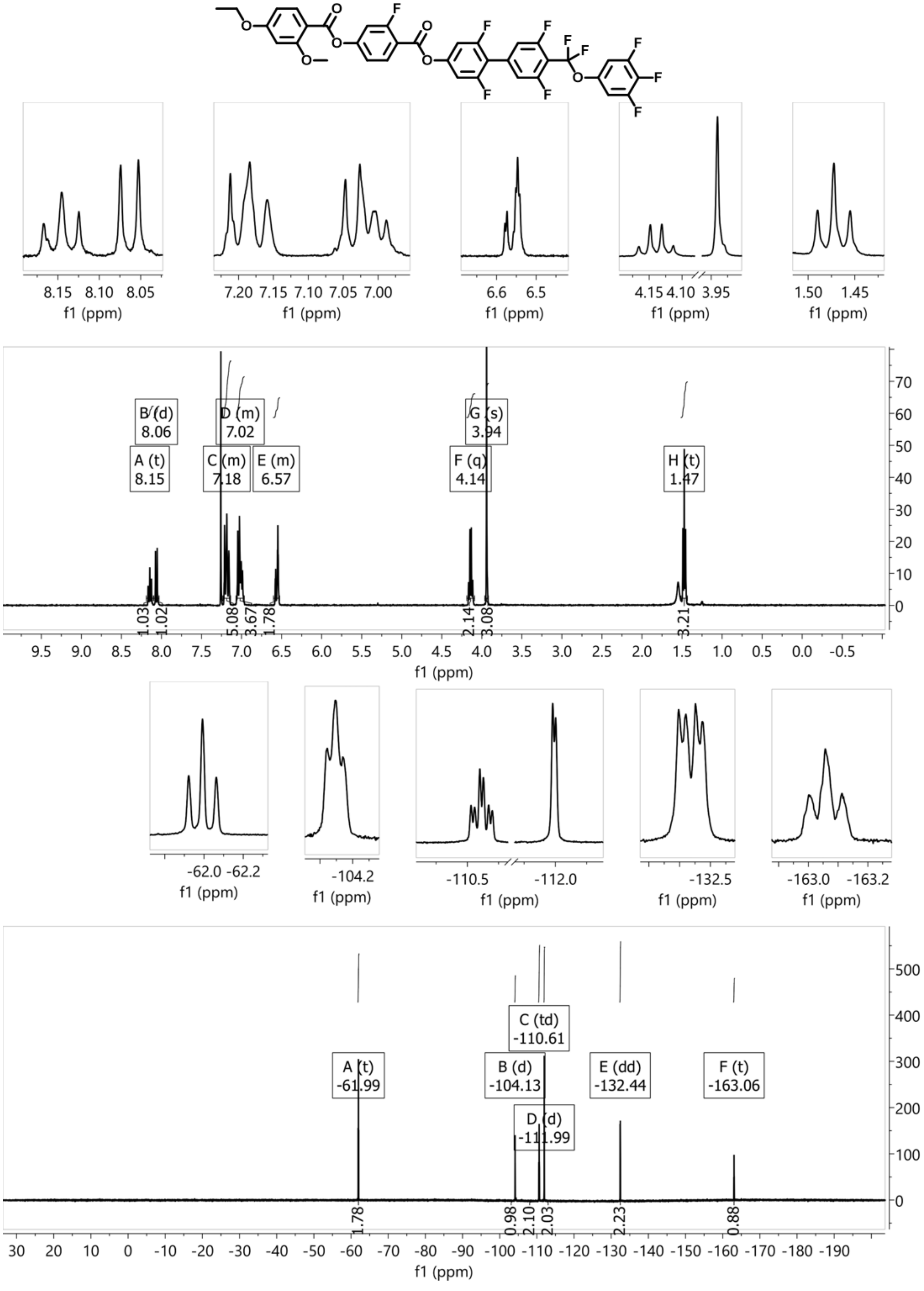

8.15 8.10 8.05
f1 (ppm)
7.20 7.15 7.10 7.05 7.00
f1 (ppm)
6.6 6.5
f1 (ppm)
4.15 4.10 3.95
f1 (ppm)
1.50 1.45
f1 (ppm)
B (d)
8.06
D (m)
7.02
G (s)
3.94
A (t)
8.15
C (m)
7.18
E (m)
6.57
F (q)
4.14
H (t)
1.47
1.03
1.02
5.08
3.67
1.78
2.14
3.08
3.21
f1 (ppm)
-62.0 -62.2
f1 (ppm)
-104.2
f1 (ppm)
-110.5 -112.0
f1 (ppm)
-132.5
f1 (ppm)
-163.0 -163.2
f1 (ppm)
C (td)
-110.61
A (t)
-61.99
B (d)
-104.13
E (dd)
-132.44
F (t)
-163.06
D (d)
-111.99
1.78
0.98
2.10
2.03
2.23
0.88
f1 (ppm)

*Figure S 22. $^{1}H$ and $^{19}F$ NMR spectra of $T_2$-1-2.*

**$T_3$-1-2**

HRMS (ESI) m/z Calculated for $C_{37}H_{22}O_7F_{10}$:
$[M+H]^+$ theoretical mass: 769.1279, found 769.12812, difference 0.337 ppm.

$^{1}H$ NMR (400 MHz, $CDCl_3$) δ = 8.15 (t, *J*=8.4, 1H), 8.06 (d, *J*=8.6, 1H), 7.19 (overlapping multiplets, 4H), 7.08 – 6.97 (m, 4H), 6.60 – 6.53 (m, 2H), 4.02 (t, *J*=6.6, 2H), 3.94 (s, 3H), 1.85 (p, *J*=7.0, 2H), 1.08 (t, *J*=7.4, 3H).

$^{19}F$ NMR (376 MHz, $CDCl_3$) δ = -61.99 (t, *J*=26.4, 2F), -104.16 (t, *J*=9.3, 1F), -110.62 (td, *J*=26.4, 10.6, 2F), -111.99 (d, *J*=8.6, 2F), -132.44 (dd, *J*=20.8, 8.2, 2F), -163.06 (t, *J*=20.8, 1F).

$^{13}C$ NMR (101 MHz, $CDCl_3$) δ 165.26, 164.29, 163.71, 162.72, 162.21, 161.70, 161.35, 160.98, 158.54, 158.47, 157.11, 156.89, 151.65, 144.61, 134.68, 134.35, 133.33, 118.32, 114.88, 114.63, 113.86, 113.82, 111.78, 111.53, 109.47, 107.64, 107.40, 106.96, 106.65, 105.51, 99.37, 69.95, 56.03, 22.45, 10.48.

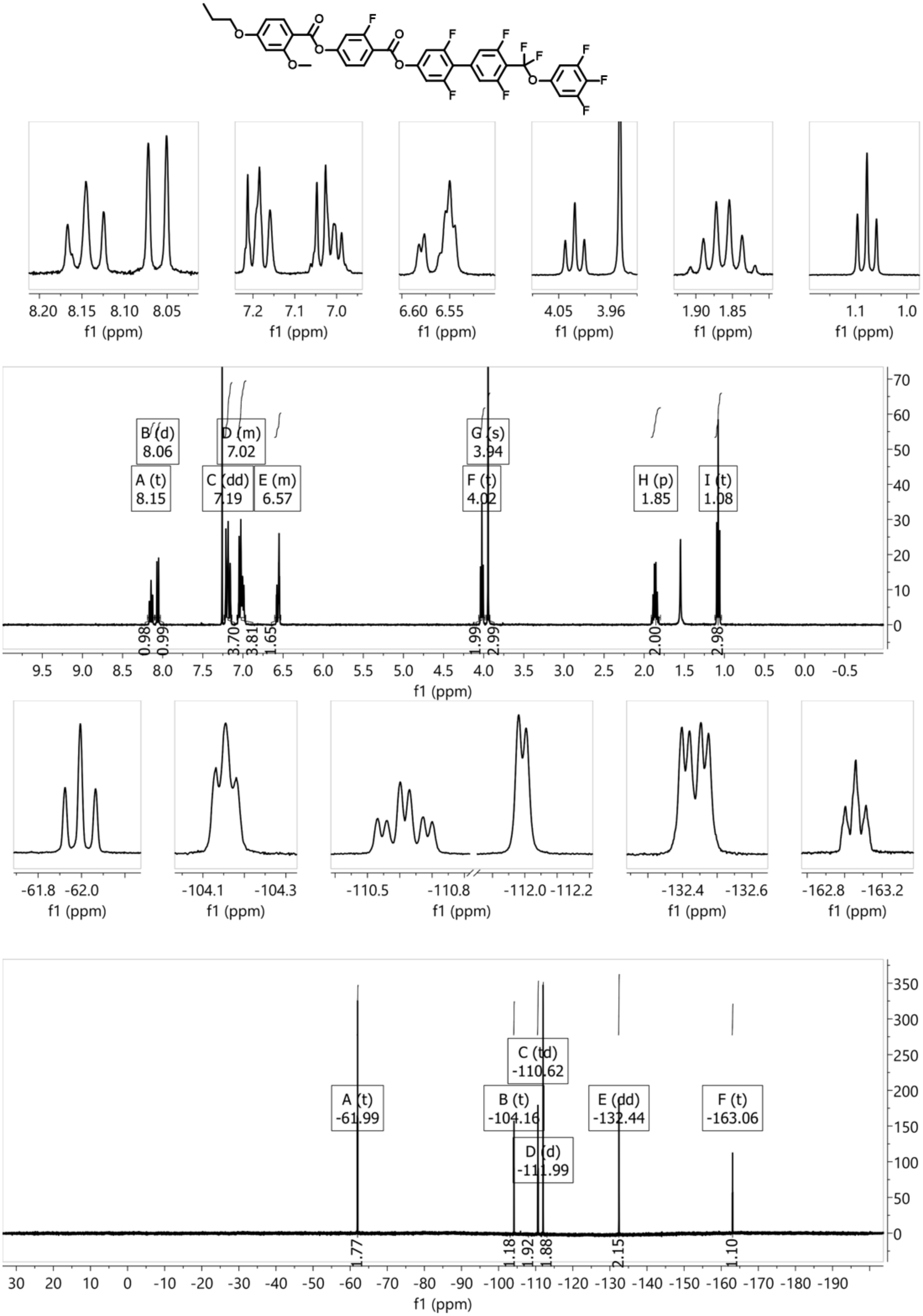

8.20 8.15 8.10 8.05
7.2 7.1 7.0
6.60 6.55
4.05 3.96
1.90 1.85
1.1 1.0
f1 (ppm)
B (d) 8.06
D (m) 7.02
G (s) 3.94
A (t) 8.15
C (dd) 7.19
E (m) 6.57
F (t) 4.02
H (p) 1.85
I (t) 1.08
0.98 0.99 3.70 3.81 1.65 1.99 2.99 2.00 2.98
-61.8 -62.0
-104.1 -104.3
-110.5 -110.8 -112.0 -112.2
-132.4 -132.6
-162.8 -163.2
C (td) -110.62
A (t) -61.99
B (t) -104.16
E (dd) -132.44
F (t) -163.06
D (d) -111.99
1.77 1.18 1.92 1.88 2.15 1.10

*Figure S 23. $^{1}H$ and $^{19}F$ NMR spectra of $T_3$-1-2.*

**$T_4$-1-2**

HRMS (ESI) m/z Calculated for $C_{38}H_{24}O_7F_{10}$:
$[M+H]^+$ theoretical mass: 783.1435, found 783.14369, difference 0.228 ppm.

$^{1}H$ NMR (400 MHz, $CDCl_3$) δ = 8.14 (t, *J*=8.5, 1H), 8.06 (d, *J*=8.6, 1H), 7.24 – 7.14 (m, 4H), 7.07 – 6.97 (m, 4H), 6.60 – 6.52 (m, 3H), 4.06 (t, *J*=6.4, 2H), 3.94 (s, 3H), 1.87 – 1.76 (m, 2H), 1.58 – 1.46 (m, 4H), 1.01 (t, *J*=7.4, 3H).

$^{19}F$ NMR (376 MHz, $CDCl_3$) δ = -61.99 (t, *J*=26.4, 2F), -104.15 (t, *J*=8.2, 1F), -110.62 (td, *J*=26.2, 10.8, 2F), -111.99 (d, *J*=8.2, 2F), -132.43 (dd, *J*=20.3, 8.1, 2F), -163.06 (t, *J*=20.8, 1F).

$^{13}C$ NMR (101 MHz, $CDCl_3$) δ 165.27, 162.71, 161.80, 161.70, 161.44, 161.08, 160.99, 158.50, 158.47, 157.00, 151.41, 134.68, 133.33, 118.32, 114.87, 114.62, 113.86, 111.78, 111.53, 109.46, 107.65, 107.40, 106.95, 106.65, 105.51, 99.37, 68.18, 56.04, 31.11, 19.19, 13.82.

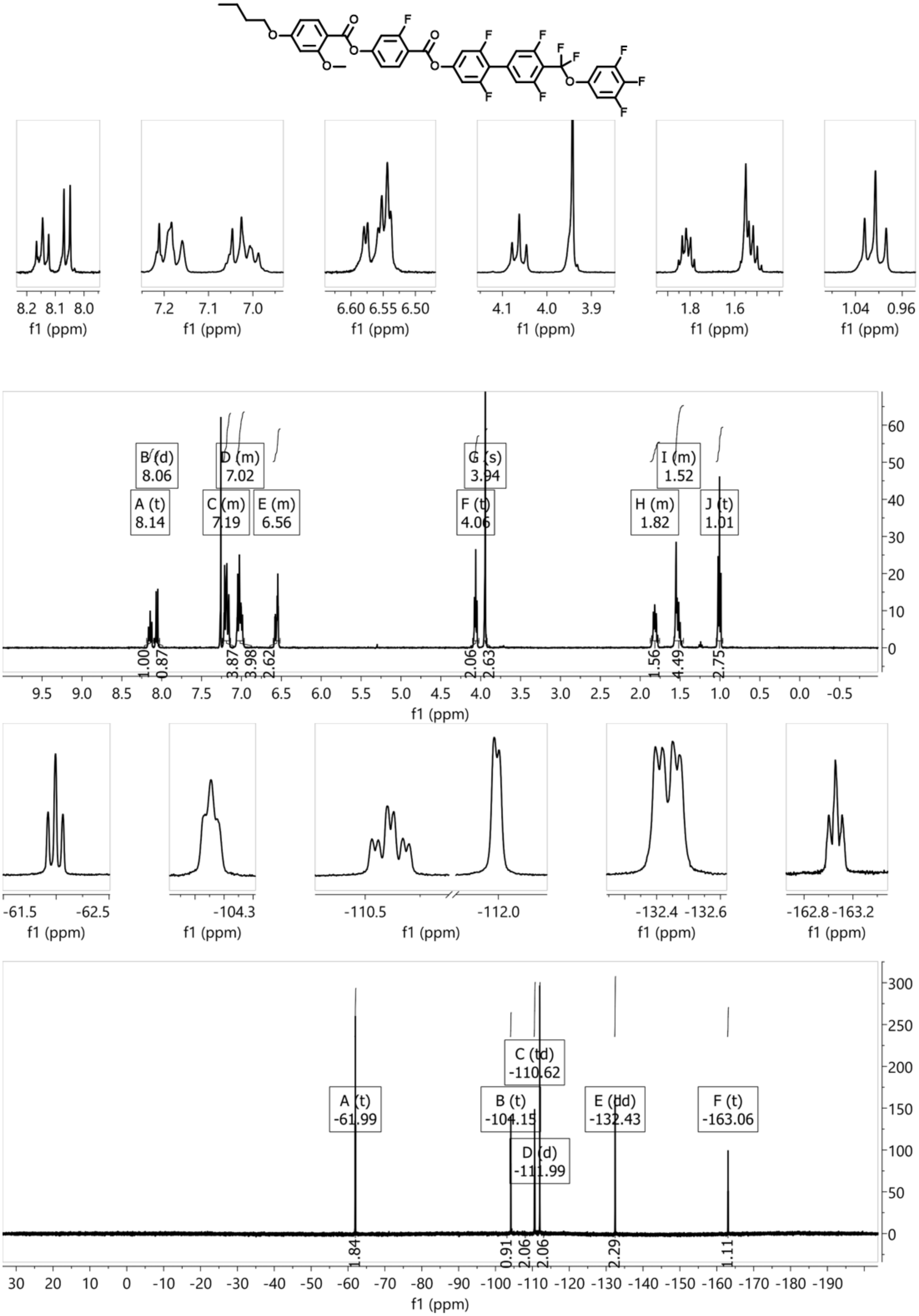

f1 (ppm)
B (d) 8.06
D (m) 7.02
A (t) 8.14
C (m) 7.19
E (m) 6.56
G (s) 3.94
F (t) 4.06
I (m) 1.52
H (m) 1.82
J (t) 1.01
1.00
0.87
3.87
3.98
2.62
2.06
2.63
1.56
4.49
2.75
C (td) -110.62
A (t) -61.99
B (t) -104.15
E (dd) -132.43
F (t) -163.06
D (d) -111.99
1.84
0.91
2.06
2.06
2.29
1.11

*Figure S 24. $^{1}H$ and $^{19}F$ NMR spectra of $T_4$-1-2.*

**$T_5$-1-2**

HRMS (ESI) m/z Calculated for $C_{39}H_{26}O_7F_{10}$:
$[M+H]^+$ theoretical mass: 797.1592, found 797.15935, difference 0.237 ppm.

$^{1}H$ NMR (400 MHz, $CDCl_3$) δ = 8.15 (t, *J*=8.4, 1H), 8.06 (d, *J*=8.6, 1H), 7.24 – 7.14 (m, 4H), 7.07 – 6.97 (m, 4H), 6.60 – 6.52 (m, 2H), 4.05 (t, *J*=6.6, 2H), 3.94 (s, 3H), 1.88 – 1.78 (m, 2H), 1.53 – 1.36 (m, 5H), 0.96 (t, *J*=7.1, 3H).

$^{19}F$ NMR (376 MHz, $CDCl_3$) δ = -61.98 (t, *J*=26.5, 2F), -104.10 – -104.22 (m, 1F), -110.61 (td, *J*=26.3, 10.5, 2F), -111.99 (d, *J*=8.6, 2F), -132.44 (dd, *J*=20.8, 8.0, 2F), -162.98 – -163.17 (m, 1F).

$^{13}C$ NMR (101 MHz, $CDCl_3$) δ 165.26, 164.24, 162.71, 162.20, 161.62, 161.34, 161.28, 161.24, 161.23, 160.97, 160.89, 158.47, 158.41, 157.06, 156.96, 151.90, 151.70, 151.64, 134.67, 133.32, 118.31, 114.86, 114.61, 113.93, 113.77, 111.77, 111.52, 109.49, 107.64, 107.39, 106.94, 106.64, 105.52, 99.39, , 68.49, 56.03, 28.78, 28.11, 22.43, 14.00.

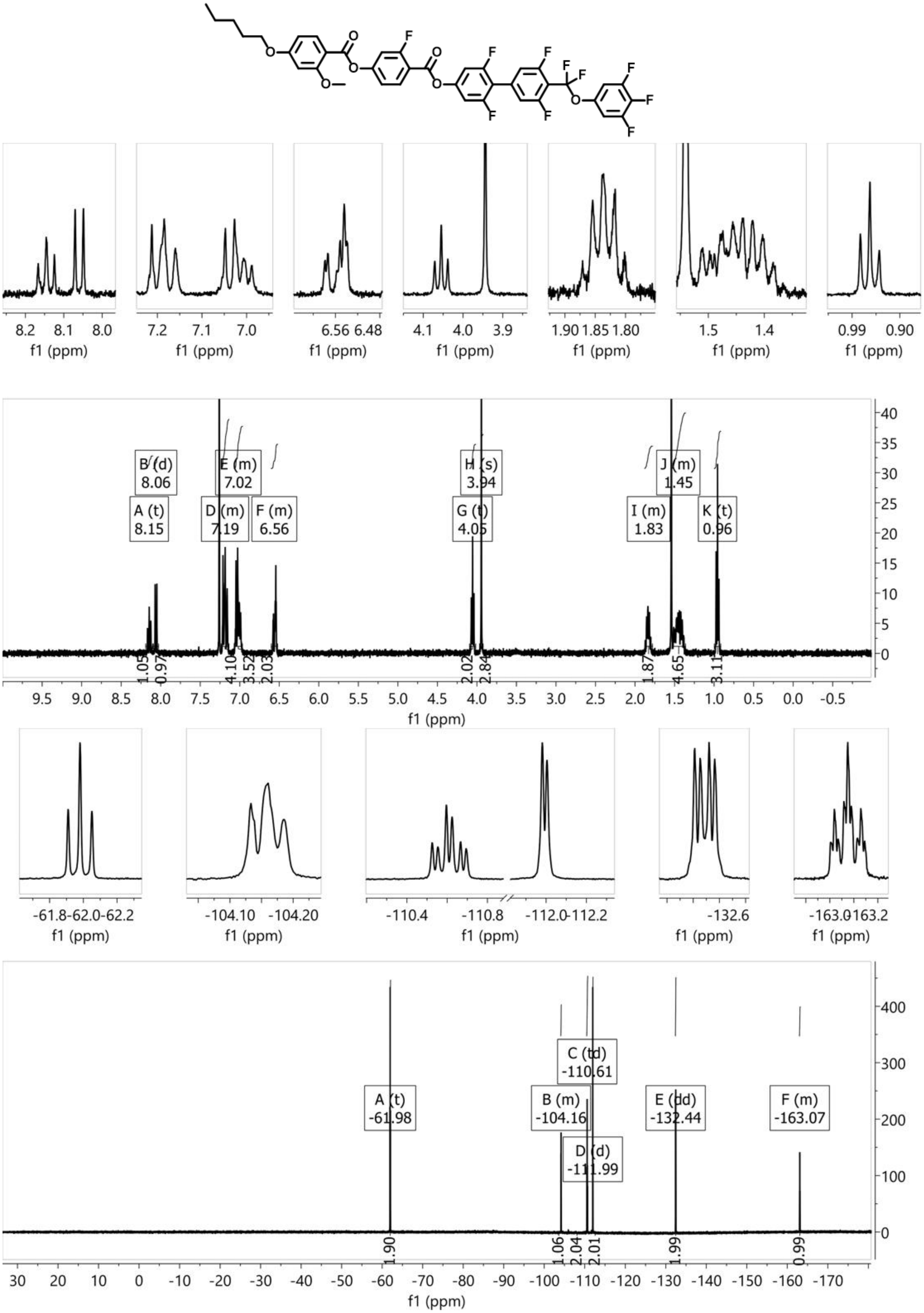
A (t)
8.15
B (d)
8.06
D (m)
7.19
E (m)
7.02
F (m)
6.56
G (t)
4.05
H (s)
3.94
I (m)
1.83
J (m)
1.45
K (t)
0.96
f1 (ppm)
A (t)
-61.98
B (m)
-104.16
C (td)
-110.61
D (d)
-111.99
E (dd)
-132.44
F (m)
-163.07
f1 (ppm)

*Figure S 25. $^{1}H$ and $^{19}F$ NMR spectra of $T_5$-1-2.*

**$T_2$-2-2**
HRMS (ESI) m/z Calculated for $C_{36}H_{19}O_7F_{11}$:
$[M+H]^+$ theoretical mass: 773.1028, found 773.10291, difference 0.156 ppm.

$^{1}H$ NMR (400 MHz, $CDCl_3$) δ = 8.03 (d, *J*=8.7, 1H), 7.17 (d, *J*=10.4, 2H), 7.09 – 6.97 (m, 6H), 6.60 – 6.51 (m, 2H), 4.14 (q, *J*=7.0, 2H), 3.94 (s, 3H), 1.47 (t, *J*=7.0, 3H).

$^{19}F$ NMR (376 MHz, $CDCl_3$) δ = -61.99 (t, *J*=26.5, 2F), -106.61 (d, *J*=9.5, 2F), -110.55 (td, *J*=26.5, 10.3, 2F), -111.74 (d, *J*=8.5, 2F), -132.44 (dd, *J*=20.7, 8.0, 2F), -163.07 (tt, *J*=20.7, 5.7, 1F).

$^{13}C$ NMR (101 MHz, $CDCl_3$) δ 165.22, 163.07, 162.99, 162.79, 161.79, 161.00, 160.91, 160.48, 160.41, 158.51, 158.46, 158.39, 155.77, 155.73, 134.73, 134.29, 114.96, 114.56, 109.20, 107.72, 107.61, 107.52, 107.12, 107.01, 106.52, 105.53, 99.41, 64.06, 56.03, 14.63.

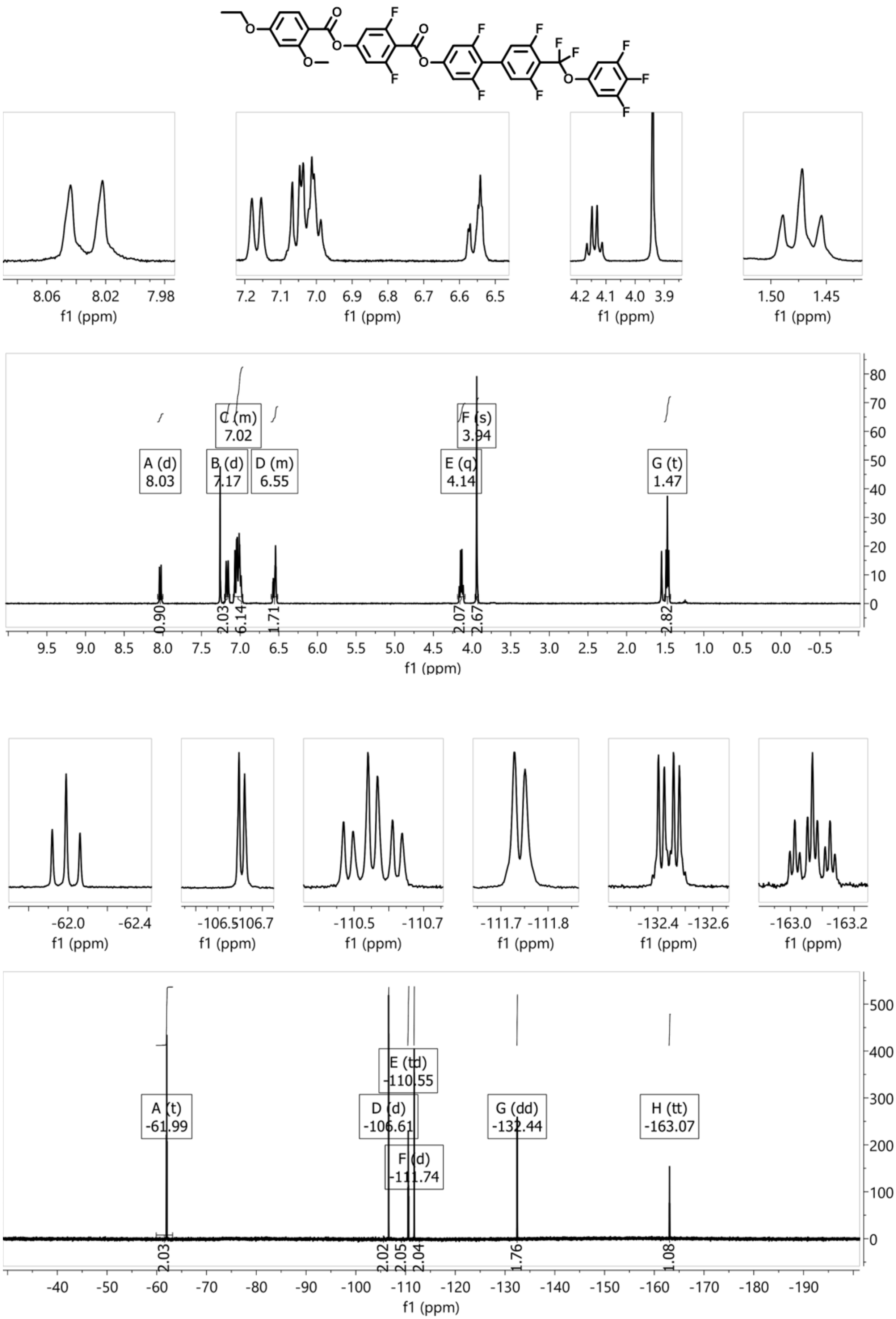

F
O
F
O
F
F
F
F
F
O
F
O
O
O
F
F
F
O
F
8.06
8.02
7.98
f1 (ppm)
7.2
7.1
7.0
6.9
6.8
6.7
6.6
6.5
f1 (ppm)
4.2
4.1
4.0
3.9
f1 (ppm)
1.50
1.45
f1 (ppm)
A (d)
8.03
B (d)
7.17
C (m)
7.02
D (m)
6.55
E (q)
4.14
F (s)
3.94
G (t)
1.47
0.90
2.03
6.14
1.71
2.07
2.67
2.82
9.5 9.0 8.5 8.0 7.5 7.0 6.5 6.0 5.5 5.0 4.5 4.0 3.5 3.0 2.5 2.0 1.5 1.0 0.5 0.0 -0.5
f1 (ppm)
80 70 60 50 40 30 20 10 0
-62.0
-62.4
f1 (ppm)
-106.5
-106.7
f1 (ppm)
-110.5
-110.7
f1 (ppm)
-111.7
-111.8
f1 (ppm)
-132.4
-132.6
f1 (ppm)
-163.0
-163.2
f1 (ppm)
A (t)
-61.99
D (d)
-106.61
E (td)
-110.55
F (d)
-111.74
G (dd)
-132.44
H (tt)
-163.07
2.03
2.02
2.05
2.04
1.76
1.08
-40 -50 -60 -70 -80 -90 -100 -110 -120 -130 -140 -150 -160 -170 -180 -190
f1 (ppm)
500 400 300 200 100 0

*Figure S 26. $^{1}H$ and $^{19}F$ NMR spectra of $T_2$-2-2.*

**$T_3$-2-2**

HRMS (ESI) m/z Calculated for $C_{37}H_{21}O_7F_{11}$:
$[M+H]^+$ theoretical mass: 787.1184, found 787.11838, difference -0.075 ppm.

$^{1}H$ NMR (400 MHz, cdcl$_3$) δ = 8.03 (d, *J*=8.7, 1H), 7.17 (d, *J*=10.5, 2H), 7.09 – 6.96 (m, 6H), 6.57 – 6.52 (m, 2H), 4.02 (t, *J*=6.3, 2H), 3.94 (s, 3H), 1.93 – 1.80 (m, 2H), 1.12 – 1.04 (m, 3H).

$^{19}F$ NMR (376 MHz, $CDCl_3$) δ = -61.99 (t, *J*=26.5, 2F), -106.62 (d, *J*=9.3, 2F), -110.55 (td, *J*=26.5, 9.9, 2F), -111.74 (d, *J*=8.3, 2F), -132.44 (dd, *J*=20.8, 7.9, 2F), -163.07 (tt, *J*=20.8, 5.8, 1F).

$^{13}C$ NMR (101 MHz, $CDCl_3$) δ 165.43, 163.08, 163.03, 162.81, 161.79, 161.05, 161.00, 160.86, 160.49, 160.45, 158.55, 158.51, 158.41, 155.69, 151.29, 134.70, 134.39, 114.88, 114.64, 108.04, 107.64, 107.40, 107.34, 107.31, 107.09, 107.05, 106.86, 106.84, 106.64, 106.58, 106.56, 105.60, 99.38, 69.99, 56.04, 22.43, 10.47.

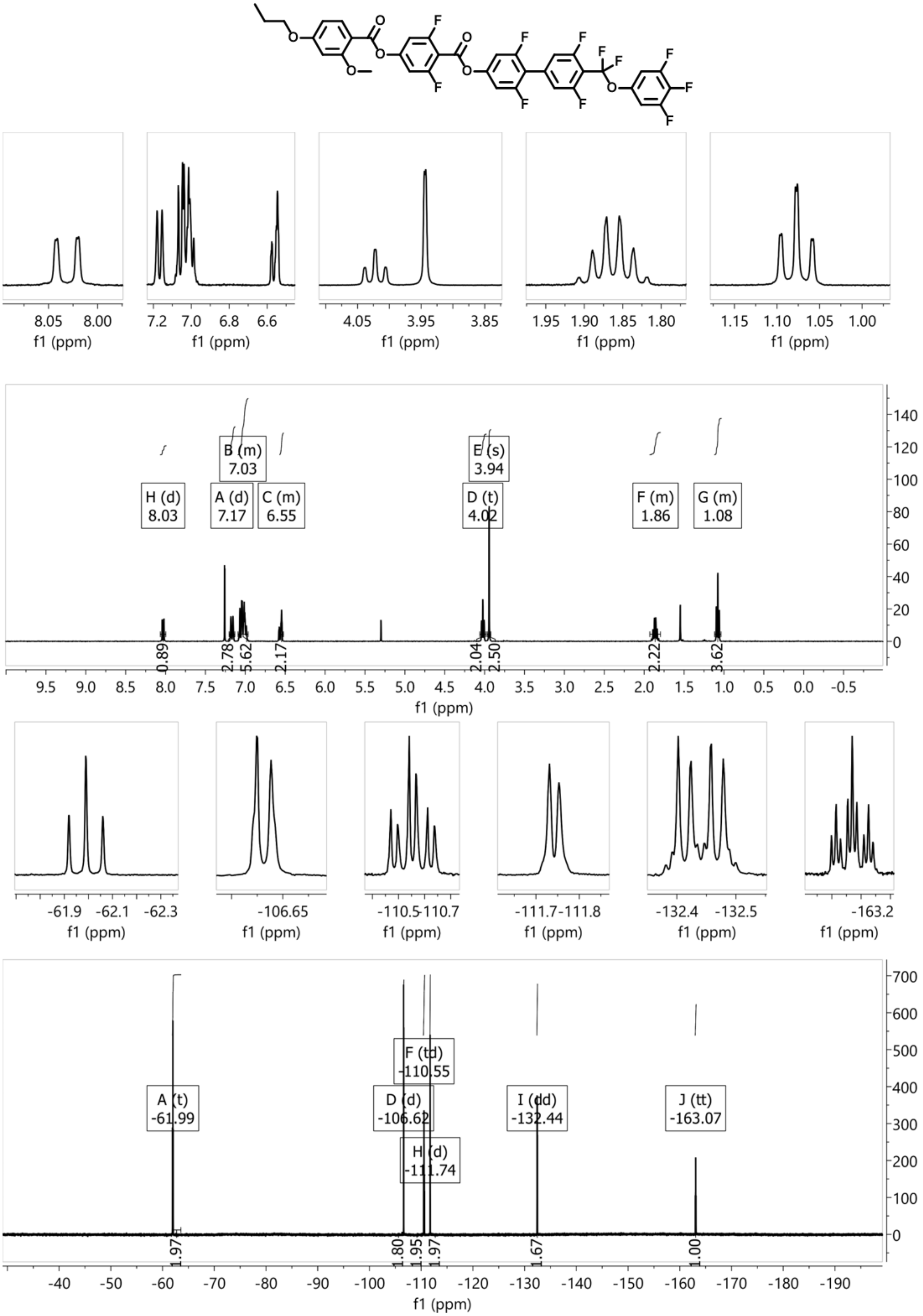

F
O
O
F
F
F
F
F
F
F
O
F
O
F
F
8.05 8.00
f1 (ppm)
7.2 7.0 6.8 6.6
f1 (ppm)
4.05 3.95 3.85
f1 (ppm)
1.95 1.90 1.85 1.80
f1 (ppm)
1.15 1.10 1.05 1.00
f1 (ppm)
H (d)
8.03
A (d)
7.17
B (m)
7.03
C (m)
6.55
D (t)
4.02
E (s)
3.94
F (m)
1.86
G (m)
1.08
0.89
2.78
5.62
2.17
2.04
2.50
2.22
3.62
9.5 9.0 8.5 8.0 7.5 7.0 6.5 6.0 5.5 5.0 4.5 4.0 3.5 3.0 2.5 2.0 1.5 1.0 0.5 0.0 -0.5
f1 (ppm)
140
120
100
80
60
40
20
0
-61.9 -62.1 -62.3
f1 (ppm)
-106.65
f1 (ppm)
-110.5 -110.7
f1 (ppm)
-111.7 -111.8
f1 (ppm)
-132.4 -132.5
f1 (ppm)
-163.2
f1 (ppm)
A (t)
-61.99
D (d)
-106.62
F (td)
-110.55
H (d)
-111.74
I (dd)
-132.44
J (tt)
-163.07
1.97
1.80
1.95
1.97
1.67
1.00
-40 -50 -60 -70 -80 -90 -100 -110 -120 -130 -140 -150 -160 -170 -180 -190
f1 (ppm)
700
600
500
400
300
200
100
0

*Figure S 27. $^{1}H$ and $^{19}F$ NMR spectra of $T_3$-2-2. $^{1}H$ spectrum shows trace DCM at 5.3 ppm.*

**$T_4$-2-2**

HRMS (ESI) m/z Calculated for $C_{38}H_{23}O_7F_{11}$:
$[M+H]^+$ theoretical mass: 801.1341, found 801.13426, difference 0.213 ppm.

$^{1}H$ NMR (400 MHz, $CDCl_3$) δ = 8.03 (d, *J*=8.7, 1H), 7.17 (d, *J*=10.6, 2H), 7.09 – 6.99 (m, 6H), 6.60 – 6.52 (m, 2H), 4.06 (t, *J*=6.4, 2H), 3.94 (s, 3H), 1.88 – 1.76 (m, 2H), 1.58 – 1.48 (m, 2H), 1.01 (t, *J*=7.4, 3H).
$^{19}F$ NMR (376 MHz, $CDCl_3$) δ = -61.99 (t, *J*=26.5, 2F), -106.61 (d, *J*=9.2, 2F), -110.45 – -110.76 (m, 2F), -111.74 (d, *J*=8.2, 2F), -132.44 (dd, *J*=20.9, 8.0, 2F), -162.87 – -163.16 (m, 1F).
$^{13}C$ NMR (101 MHz, $CDCl_3$) δ 165.44, 163.07, 162.81, 161.79, 161.03, 160.93, 160.43, 160.28, 158.57, 158.47, 158.43, 156.92, 155.76, 134.71, 114.88, 114.62, 109.06, 107.64, 107.59, 107.47, 107.40, 107.35, 107.32, 107.18, 107.10, 107.06, 106.87, 106.85, 106.65, 106.60, 106.57, 105.59, 99.37, 68.21, 56.04, 31.09, 19.18, 13.82.

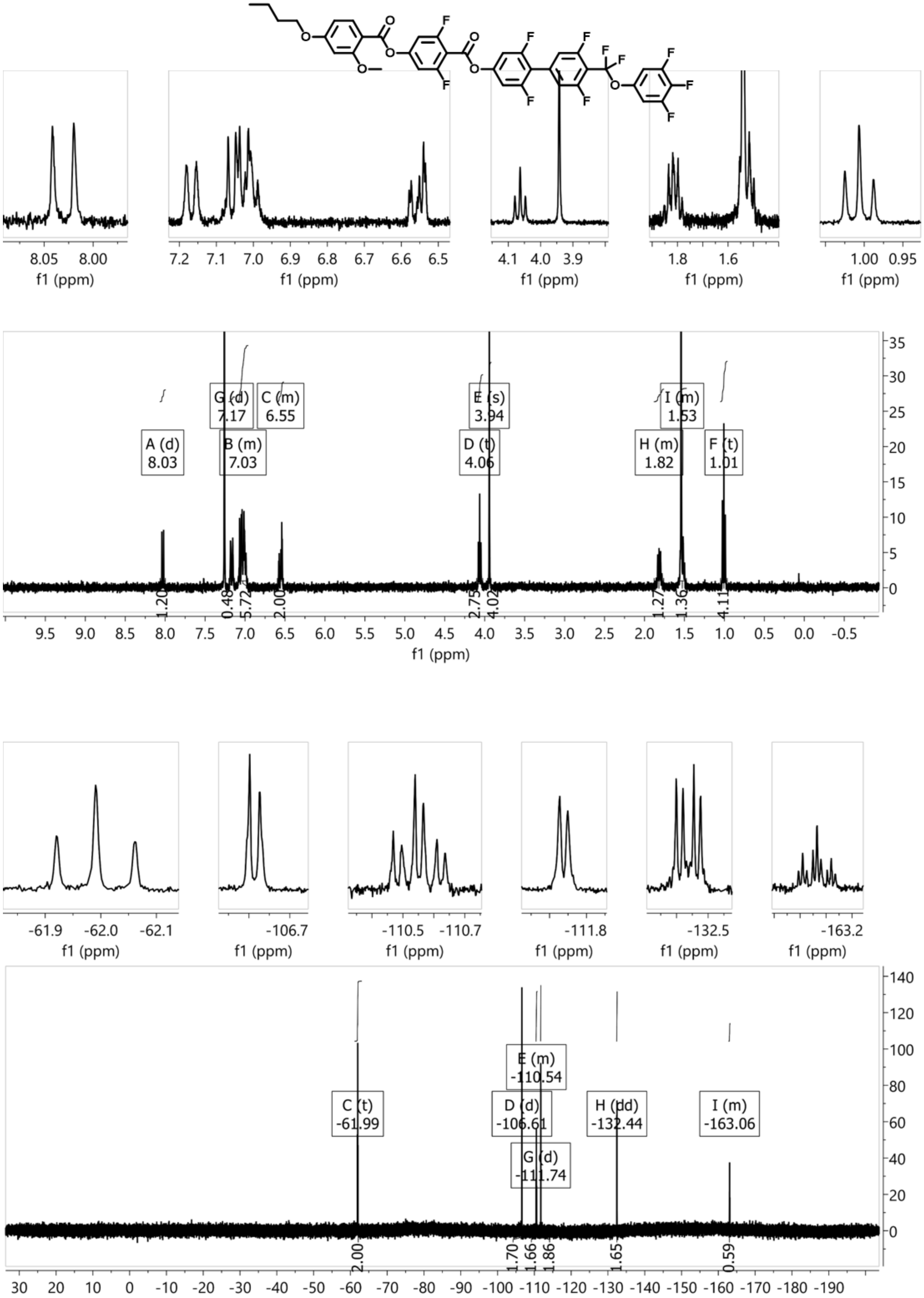

A (d)
8.03
G (d)
7.17
B (m)
7.03
C (m)
6.55
D (t)
4.06
E (s)
3.94
H (m)
1.82
I (m)
1.53
F (t)
1.01
f1 (ppm)
C (t)
-61.99
D (d)
-106.61
E (m)
-110.54
G (d)
-111.74
H (dd)
-132.44
I (m)
-163.06

*Figure S 28. $^{1}H$ and $^{19}F$ NMR spectra of $T_4$-2-2.*

**$T_5$-2-2**
HRMS (ESI) m/z Calculated for $C_{39}H_{25}O_7F_{11}$:
$[M+H]^+$ theoretical mass: 815.1497, found 815.14987, difference 0.160 ppm.

$^{1}H$ NMR (400 MHz, $CDCl_3$) δ = 8.03 (d, *J*=8.7, 1H), 7.17 (d, *J*=10.4, 2H), 7.10 – 6.96 (m, 6H), 6.60 – 6.51 (m, 2H), 4.05 (t, *J*=6.5, 2H), 3.94 (s, 3H), 1.88 – 1.78 (m, 2H), 1.50 – 1.38 (m, 4H), 0.96 (t, *J*=7.2, 3H).

$^{19}F$ NMR (376 MHz, $CDCl_3$) δ = -61.99 (t, *J*=26.5, 2F), -106.61 (d, *J*=9.3, 2F), -110.55 (td, *J*=26.6, 9.9, 2F), -111.74 (d, *J*=8.2, 2F), -132.44 (dd, *J*=20.8, 7.9, 2F), -163.06 (tt, *J*=20.8, 6.1, 1F).

$^{13}C$ NMR (101 MHz, $CDCl_3$) δ 165.44, 163.07, 163.00, 162.81, 161.79, 160.98, 160.90, 160.49, 160.41, 158.53, 158.47, 158.43, 158.39, 155.70, 155.70, 151.43, 151.29, 149.83, 134.70, 114.88, 114.30, 109.08, 107.64, 107.55, 107.47, 107.40, 107.34, 107.30, 107.09, 107.05, 106.86, 106.84, 106.59, 106.56, 105.59, 99.37, 68.52, 56.03, 28.76, 28.11, 22.42, 13.99.

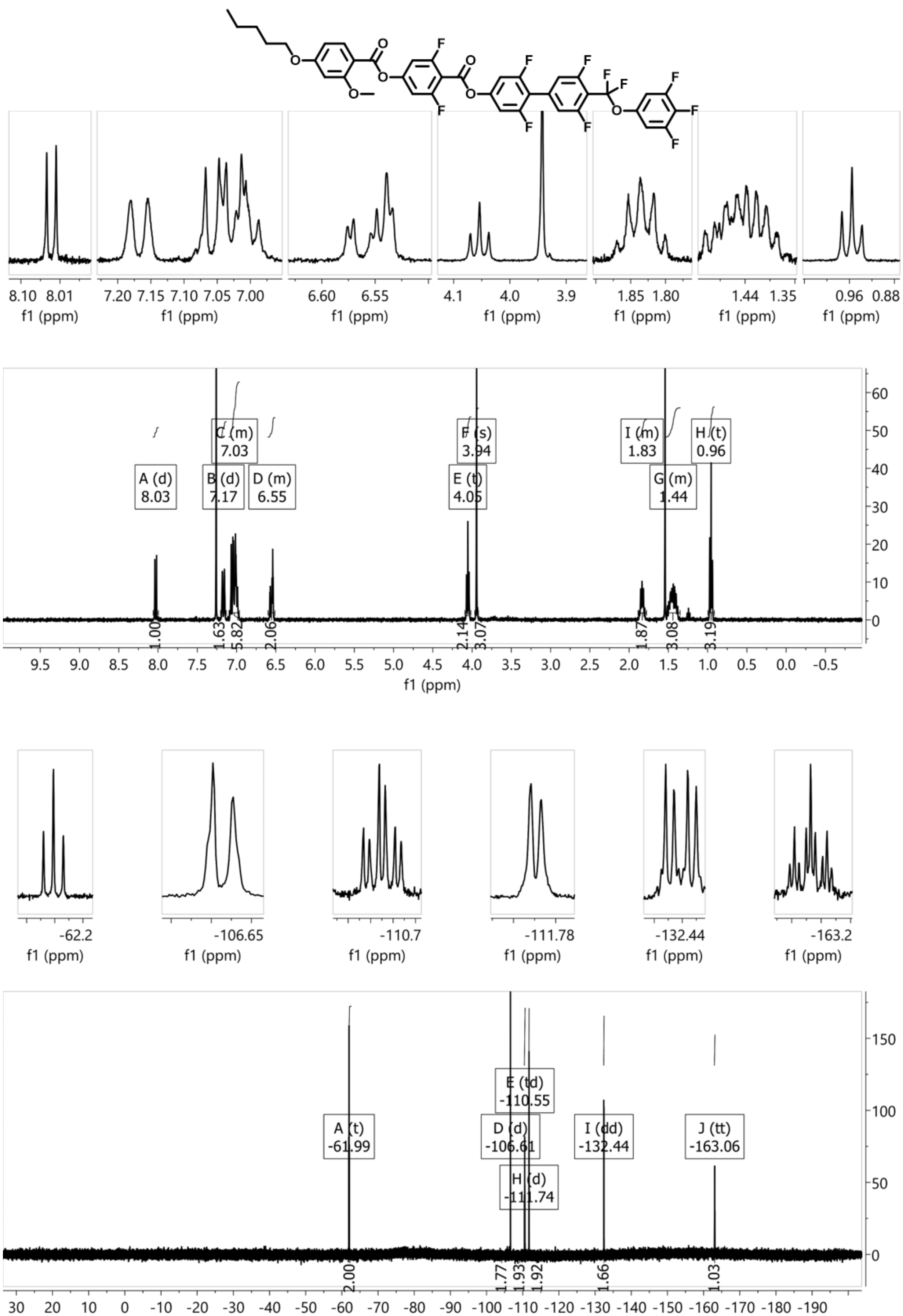

8.10 8.01
f1 (ppm)
7.20 7.15 7.10 7.05 7.00
f1 (ppm)
6.60 6.55
f1 (ppm)
4.1 4.0 3.9
f1 (ppm)
1.85 1.80
f1 (ppm)
1.44 1.35
f1 (ppm)
0.96 0.88
f1 (ppm)
A (d)
8.03
B (d)
7.17
C (m)
7.03
D (m)
6.55
E (t)
4.05
F (s)
3.94
I (m)
1.83
G (m)
1.44
H (t)
0.96
1.00
1.63
5.82
2.06
2.14
3.07
1.87
3.08
3.19
9.5 9.0 8.5 8.0 7.5 7.0 6.5 6.0 5.5 5.0 4.5 4.0 3.5 3.0 2.5 2.0 1.5 1.0 0.5 0.0 -0.5
f1 (ppm)
-62.2
f1 (ppm)
-106.65
f1 (ppm)
-110.7
f1 (ppm)
-111.78
f1 (ppm)
-132.44
f1 (ppm)
-163.2
f1 (ppm)
A (t)
-61.99
D (d)
-106.61
E (td)
-110.55
H (d)
-111.74
I (dd)
-132.44
J (tt)
-163.06
2.00
1.77
1.93
1.92
1.66
1.03
30 20 10 0 -10 -20 -30 -40 -50 -60 -70 -80 -90 -100 -110 -120 -130 -140 -150 -160 -170 -180 -190
f1 (ppm)

*Figure S 29. $^{1}H$ and $^{19}F$ NMR spectra of $T_5$-2-2.*

**$T_6$-2-2**

HRMS (ESI) m/z Calculated for $C_{40}H_{27}O_7F_{11}$:
$[M+H]^+$ theoretical mass: 829.1654, found 829.16421, difference -1.422 ppm.
$[M+Na]^+$ theoretical mass: 851.1473, found 851.14637, difference -1.132 ppm.

$^{1}H$ NMR (400 MHz, $CDCl_3$) δ = 8.03 (d, *J*=8.7, 1H), 7.17 (d, *J*=10.5, 2H), 7.10 – 6.94 (m, 6H), 6.60 – 6.51 (m, 2H), 4.05 (t, *J*=6.5, 2H), 3.94 (s, 3H), 1.83 (p, *J*=6.7, 2H), 1.55 – 1.43 (m, 2H), 1.43 – 1.32 (m, 4H), 0.96 – 0.88 (m, 3H).
$^{19}F$ NMR (376 MHz, $CDCl_3$) δ = -61.99 (t, *J*=26.5, 2F), -106.62 (d, *J*=9.7, 2F), -110.56 (td, *J*=26.5, 10.4, 2F), -111.75 (d, *J*=8.5, 2F), -132.37 – -132.53 (m, 2F), -163.08 (tt, *J*=20.9, 5.8, 1F).
$^{13}C$ NMR (101 MHz, $CDCl_3$) δ 165.44, 163.07, 162.99, 162.80, 161.79, 161.04, 160.98, 160.89, 160.49, 160.41, 158.52, 158.50, 158.47, 158.44, 158.39, 155.85, 155.70, 155.55, 152.33, 152.28, 151.44, 151.29, 151.16, 149.84, 149.78, 149.75, 149.63, 144.53, 139.73, 137.24, 137.10, 134.81, 134.70, 134.59, 134.39, 134.27, 134.17, 133.99, 114.94, 114.77, 114.76, 114.53, 109.07, 107.60, 107.30, 106.97, 106.75, 106.72, 106.03, 105.63, 105.57, 99.41, 99.34, 68.53, 56.11, 31.52, 29.03, 25.64, 22.57, 14.01.

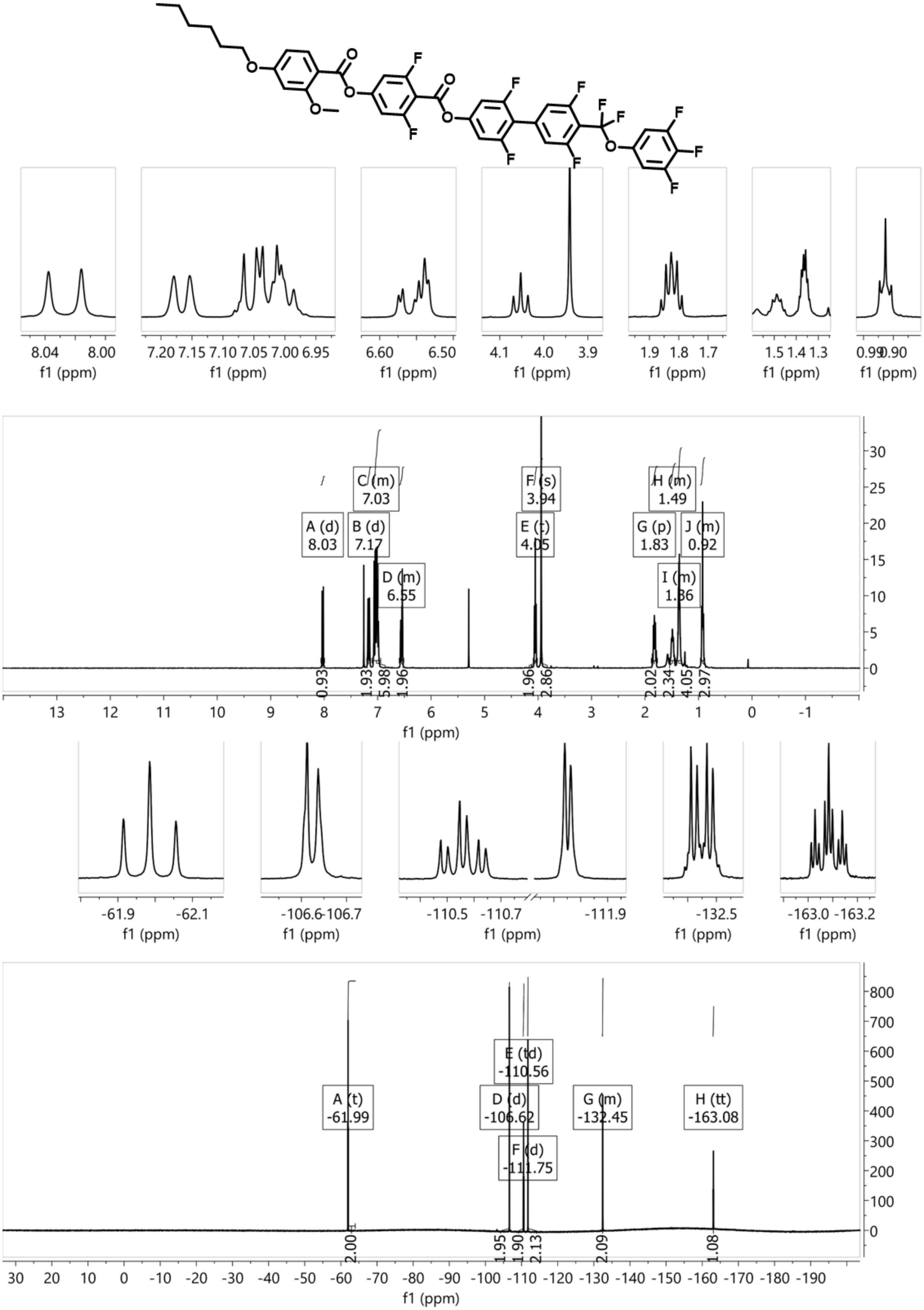

*Figure S 30. $^{1}H$ and $^{19}F$ NMR spectra of $T_6$-2-2.*

**$T_7$-2-2**
HRMS (ESI) m/z Calculated for $C_{41}H_{29}O_7F_{11}$:
$[M+H]^+$ theoretical mass: 843.1810, found 843.17953, difference -1.790 ppm.
$[M+Na]^+$ theoretical mass: 865.1630, found 865.16179, difference -1.380 ppm.

$^{1}H$ NMR (400 MHz, $CDCl_3$) δ = 8.03 (d, *J*=8.7, 1H), 7.17 (d, *J*=10.6, 2H), 7.10 – 6.96 (m, 6H), 6.60 – 6.51 (m, 2H), 4.05 (t, *J*=6.5, 2H), 3.94 (s, 2H), 1.83 (p, *J*=6.8, 2H), 1.54 – 1.42 (m, 2H), 1.42 – 1.29 (m, 6H), 0.94 – 0.87 (m, 3H).

$^{19}F$ NMR (376 MHz, $CDCl_3$) δ = -61.99 (t, *J*=26.5, 2F), -106.62 (d, *J*=9.6, 2F), -110.56 (td, *J*=26.5, 10.4, 2F), -111.75 (d, *J*=8.6, 2F), -132.36 – -132.53 (m, 2F), -163.07 (tt, *J*=20.9, 5.8, 1F).

$^{13}C$ NMR (101 MHz, $CDCl_3$) δ 165.44, 163.07, 162.99, 162.81, 161.79, 160.98, 160.89, 160.48, 160.41, 158.54, 158.53, 158.47, 158.38, 155.70, 152.32, 152.16, 151.47, 151.31, 151.14, 149.81, 149.65, 144.66, 144.55, 134.70, 134.27, 134.21, 114.77, 114.56, 109.07, 107.56, 107.10, 106.67, 105.64, 105.57, 99.41, 99.34, 68.54, 31.74, 29.07, 25.93, 22.60, 14.08.

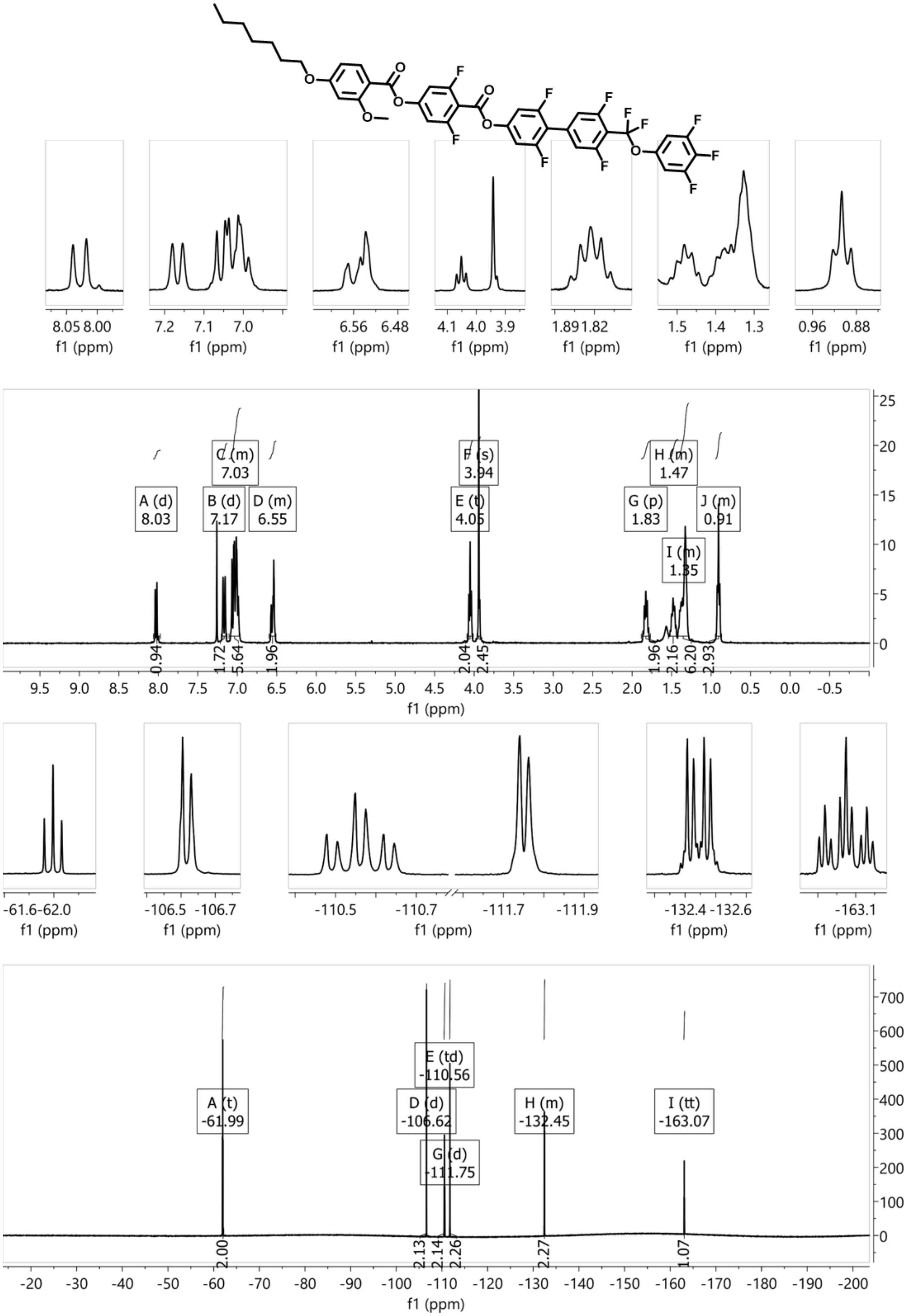

A (d)
8.03
B (d)
7.17
C (m)
7.03
D (m)
6.55
E (t)
4.05
F (s)
3.94
G (p)
1.83
H (m)
1.47
I (m)
1.35
J (m)
0.91
f1 (ppm)
A (t)
-61.99
D (d)
-106.62
E (td)
-110.56
G (d)
-111.75
H (m)
-132.45
I (tt)
-163.07
f1 (ppm)

*Figure S 31. $^{1}H$ and $^{19}F$ NMR spectra of $T_7$-2-2.*

**$T_8$-2-2**

HRMS (ESI) m/z Calculated for $C_{42}H_{31}O_7F_{11}$:
$[M+H]^+$ theoretical mass: 857.1967, found 857.19551, difference -1.376 ppm.
$[M+Na]^+$ theoretical mass: 879.1786, found 879.17742, difference -1.381 ppm.

$^{1}H$ NMR (400 MHz, $CDCl_3$) δ = 8.03 (d, *J*=8.7, 1H), 7.17 (d, *J*=10.6, 2H), 7.13 – 6.96 (m, 6H), 6.60 – 6.51 (m, 2H), 4.05 (t, *J*=6.5, 2H), 3.94 (s, 3H), 1.83 (p, *J*=6.7, 2H), 1.47 (q, *J*=7.2, 2H), 1.40 – 1.27 (m, 8H), 0.94 – 0.86 (m, 3H).

$^{19}F$ NMR (376 MHz, $CDCl_3$) δ = -61.99 (t, *J*=26.6, 2F), -106.61 (d, *J*=9.6, 2F), -110.55 (td, *J*=26.6, 10.7, 2F), -111.74 (d, *J*=8.8, 2F), -132.44 (dd, *J*=20.9, 8.2, 2F), -162.98 – -163.15 (m, 1F).

$^{13}C$ NMR (101 MHz, $CDCl_3$) δ 165.44, 162.98, 162.80, 161.79, 161.03, 160.98, 160.89, 160.49, 160.39, 158.51, 158.47, 158.40, 155.70, 152.29, 152.18, 151.29, 151.14, 134.72, 134.68, 134.25, 114.99, 114.51, 109.07, 107.85, 107.19, 107.02, 106.70, 106.05, 105.56, 105.55, 99.42, 99.33, 68.54, 56.14, 55.93, 31.79, 29.21, 29.07, 25.97, 22.65, 14.13.

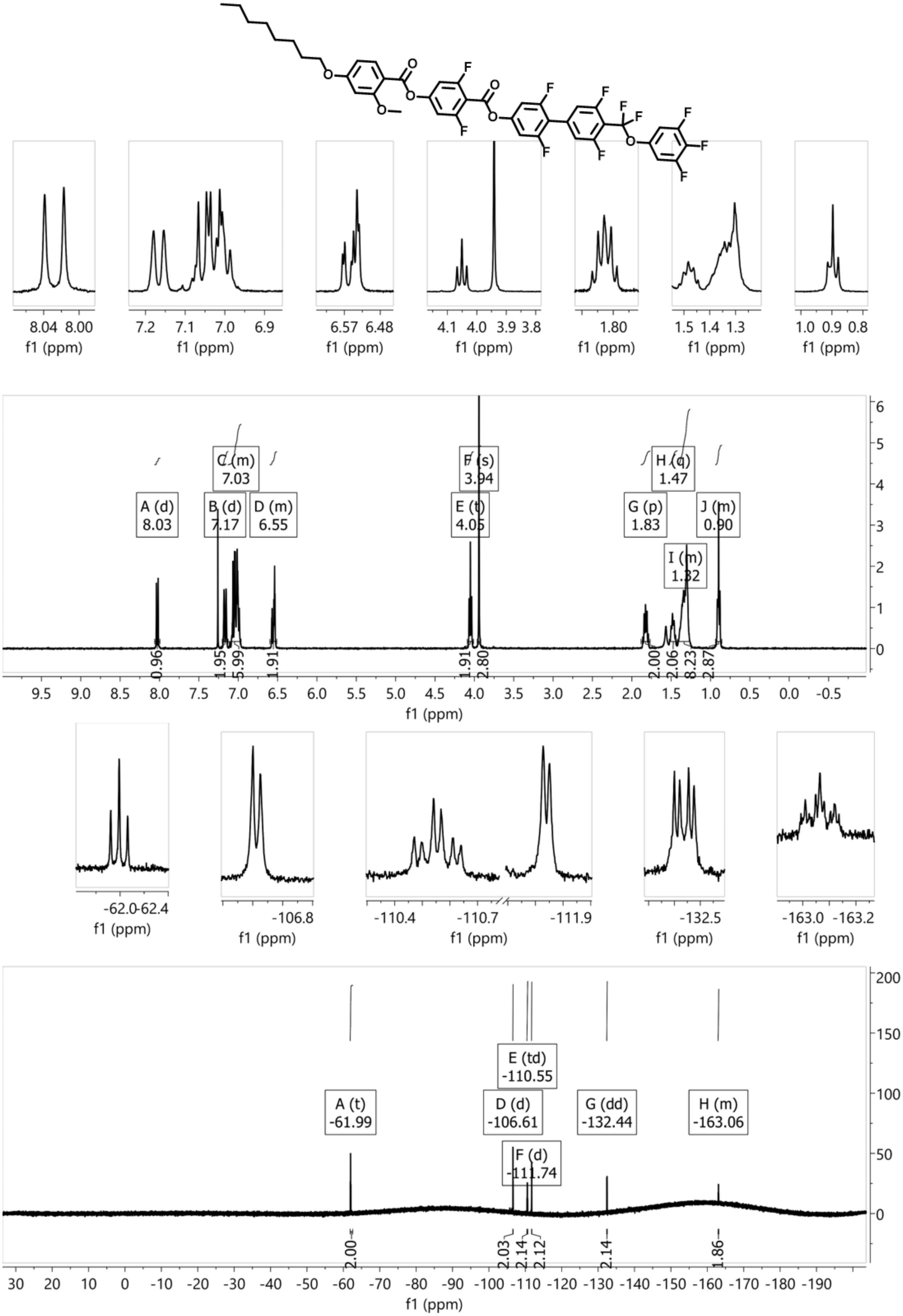

A (d)
8.03
B (d)
7.17
C (m)
7.03
D (m)
6.55
E (t)
4.05
F (s)
3.94
G (p)
1.83
H (q)
1.47
I (m)
1.32
J (m)
0.90
f1 (ppm)
A (t)
-61.99
D (d)
-106.61
E (td)
-110.55
F (d)
-111.74
G (dd)
-132.44
H (m)
-163.06

*Figure S 32. $^{1}H$ and $^{19}F$ NMR spectra of $T_8$-2-2.*

**$T_9$-2-2**

HRMS (ESI) m/z Calculated for $C_{43}H_{33}O_7F_{11}$:
$[M+H]^+$ theoretical mass: 871.21234, found 871.21010, difference -2.571 ppm.
$[M+Na]^+$ theoretical mass: 893.19428, found 893.19213, difference -2.411 ppm.

$^{1}H$ NMR (400 MHz, $CDCl_3$) δ = 8.03 (d, *J*=8.7, 1H), 7.17 (d, *J*=10.5, 2H), 7.09 – 6.96 (m, 6H), 6.60 – 6.51 (m, 2H), 4.05 (t, *J*=6.5, 2H), 3.94 (s, 3H), 1.87 – 1.78 (m, 2H), 1.53 – 1.43 (m, 2H), 1.41 – 1.24 (m, 10H), 0.93 – 0.85 (m, 3H).

$^{19}F$ NMR (376 MHz, $CDCl_3$) δ = -61.98 (t, *J*=26.4, 2F), -106.63 (d, *J*=10.2, 2F), -110.38 – -110.71 (m, 2F), -111.75 (d, *J*=8.8, 2F), -132.45 (dd, *J*=20.8, 8.7, 2F), -163.08 (t, *J*=20.8, 1F).

$^{13}C$ NMR (101 MHz, $CDCl_3$) δ 165.44, 163.07, 162.99, 162.80, 161.79, 160.98, 160.89, 160.49, 160.42, 158.48, 158.40, 155.84, 155.70, 152.28, 151.45, 151.30, 149.84, 134.69, 134.27, 120.00, 114.86, 114.62, 109.09, 107.63, 107.56, 107.44, 107.39, 107.33, 107.29, 107.08, 107.04, 106.85, 106.83, 106.78, 106.64, 106.57, 106.56, 106.20, 106.04, 105.88, 105.61, 99.38, 68.54, 56.02, 31.86, 29.50, 29.35, 29.24, 29.07, 25.96, 22.66, 14.10.

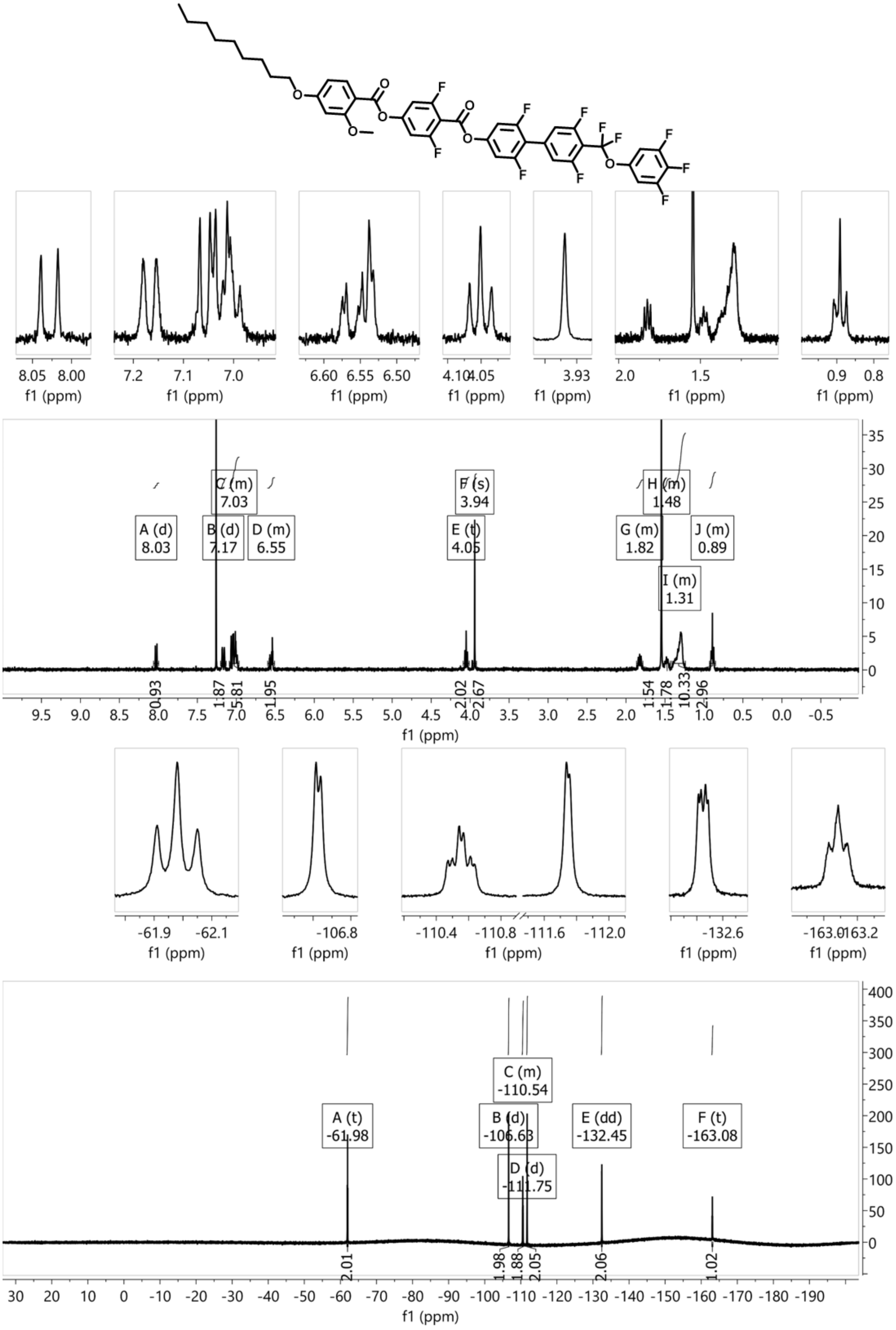

A (d)
8.03
B (d)
7.17
C (m)
7.03
D (m)
6.55
E (t)
4.05
F (s)
3.94
G (m)
1.82
H (m)
1.48
I (m)
1.31
J (m)
0.89
f1 (ppm)
A (t)
-61.98
B (d)
-106.63
C (m)
-110.54
D (d)
-111.75
E (dd)
-132.45
F (t)
-163.08
f1 (ppm)

*Figure S 33. $^{1}H$ and $^{19}F$ NMR spectra of $T_9$-2-2.*

**$T_{10}$-2-2**

HRMS (ESI) m/z Calculated for $C_{44}H_{35}O_7F_{11}$:
$[M+H]^+$ theoretical mass: 885.2280, found 885.22484, difference -3.558 ppm.
$[M+Na]^+$ theoretical mass: 907.2099, found 907.20685, difference -3.399 ppm.

$^{1}H$ NMR (400 MHz, $CDCl_3$) δ = 8.03 (d, *J*=8.7, 1H), 7.17 (d, *J*=10.6, 2H), 7.09 – 6.95 (m, 6H), 6.60 – 6.51 (m, 2H), 4.05 (t, *J*=6.5, 2H), 3.94 (s, 3H), 1.82 (p, *J*=6.8, 2H), 1.50 – 1.46 (m, 2H), 1.41 – 1.26 (m, 12H), 0.93 – 0.85 (m, 3H).

$^{19}F$ NMR (376 MHz, $CDCl_3$) δ = -61.88 (t, *J*=26.3, 2F), -106.50 (d, *J*=11.8, 2F), -110.45 (t, *J*=26.3, 2F), -111.51 – -111.75 (m, 2F), -132.18 – -132.48 (m, 2F), -162.78 – -163.13 (m, 1F).

$^{13}C$ NMR (101 MHz, $CDCl_3$) δ 165.44, 162.99, 162.81, 161.80, 161.71, 161.07, 160.98, 160.90, 160.49, 158.48, 158.39, 155.73, 155.55, 151.31, 146.10, 143.12, 142.85, 142.71, 134.70, 134.28, 114.87, 114.82, 114.65, 114.62, 109.08, 107.63, 107.56, 107.45, 107.44, 107.39, 107.34, 107.30, 107.30, 107.09, 107.05, 106.86, 106.83, 106.64, 106.59, 106.56, 105.61, 99.38, 68.54, 56.03, 31.89, 29.54, 29.34, 29.31, 29.07, 25.96, 22.67, 14.11.

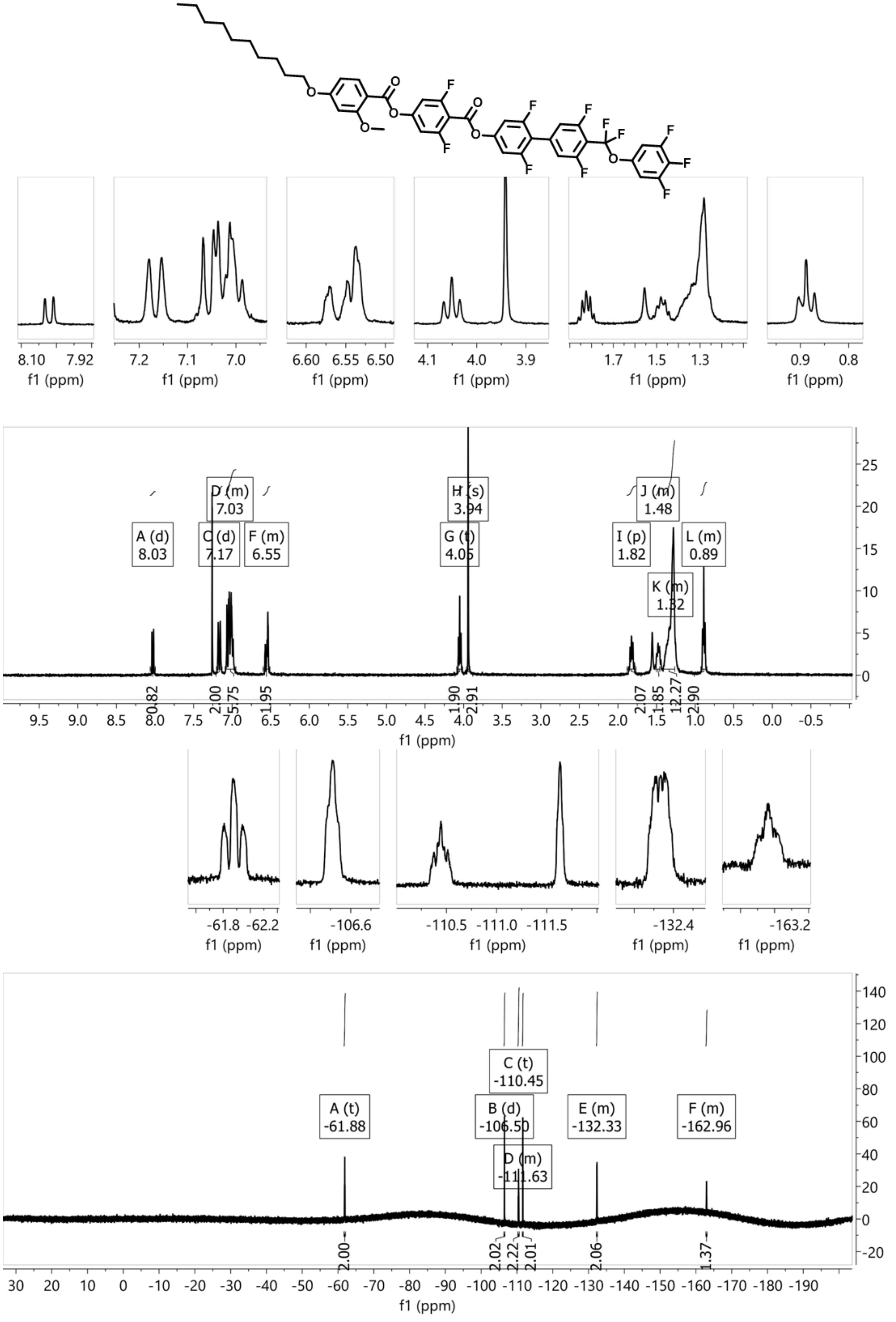

8.10 7.92 f1 (ppm)
7.2 7.1 7.0 f1 (ppm)
6.60 6.55 6.50 f1 (ppm)
4.1 4.0 3.9 f1 (ppm)
1.7 1.5 1.3 f1 (ppm)
0.9 0.8 f1 (ppm)
A (d) 8.03
C (d) 7.17
D (m) 7.03
F (m) 6.55
G (t) 4.05
H (s) 3.94
I (p) 1.82
J (m) 1.48
K (m) 1.32
L (m) 0.89
0.82 2.00 5.75 1.95 1.90 2.91 2.07 1.85 12.27 2.90
f1 (ppm)
-61.8 -62.2 f1 (ppm)
-106.6 f1 (ppm)
-110.5 -111.0 -111.5 f1 (ppm)
-132.4 f1 (ppm)
-163.2 f1 (ppm)
A (t) -61.88
B (d) -106.50
C (t) -110.45
D (m) -111.63
E (m) -132.33
F (m) -162.96
2.00 2.02 2.22 2.01 2.06 1.37
f1 (ppm)

*Figure S 34. $^{1}$H and $^{19}$F NMR spectra of $T_{10}$-2-2.*

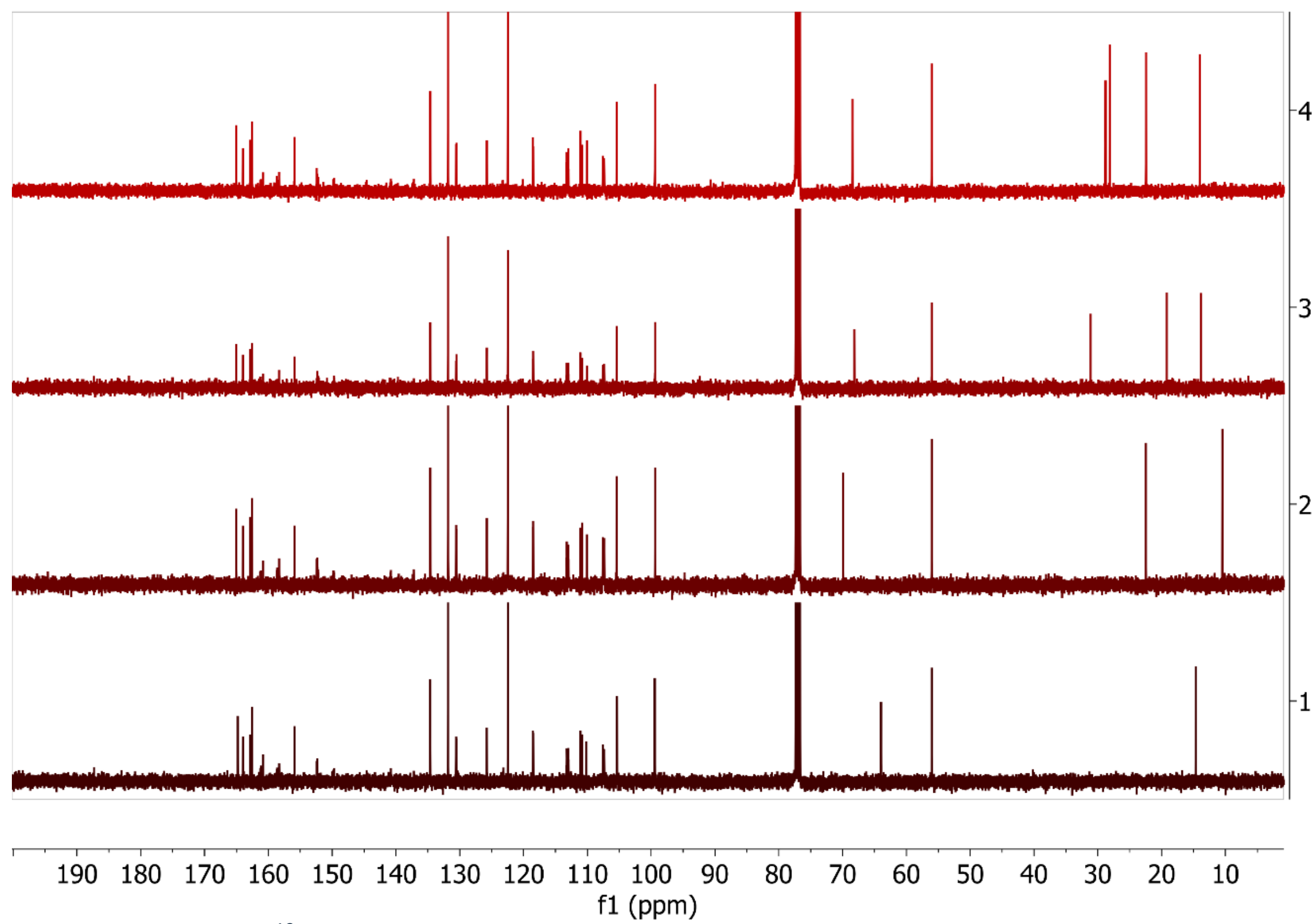

*Figure S 35. Stacked $^{13}$C NMR spectra of (1) $T_2$-0-1, (2) $T_3$-0-1, (3) $T_4$-0-1, (4) $T_5$-0-1.*

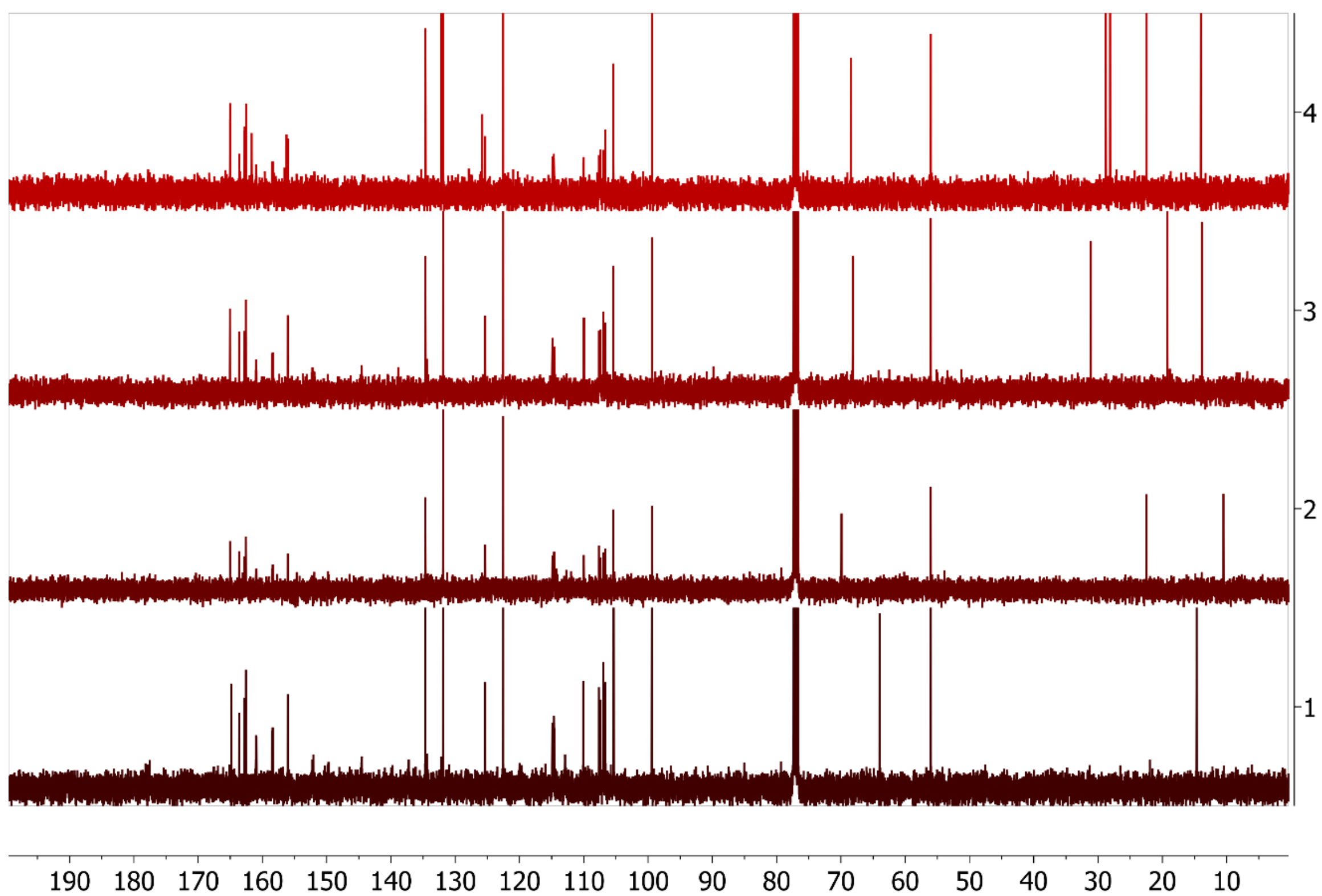

*Figure S 36. Stacked $^{13}C$ NMR spectra of (1) $T_2$-1-1, (2) $T_3$-1-1, (3) $T_4$-1-1, (4) $T_5$-1-1.*

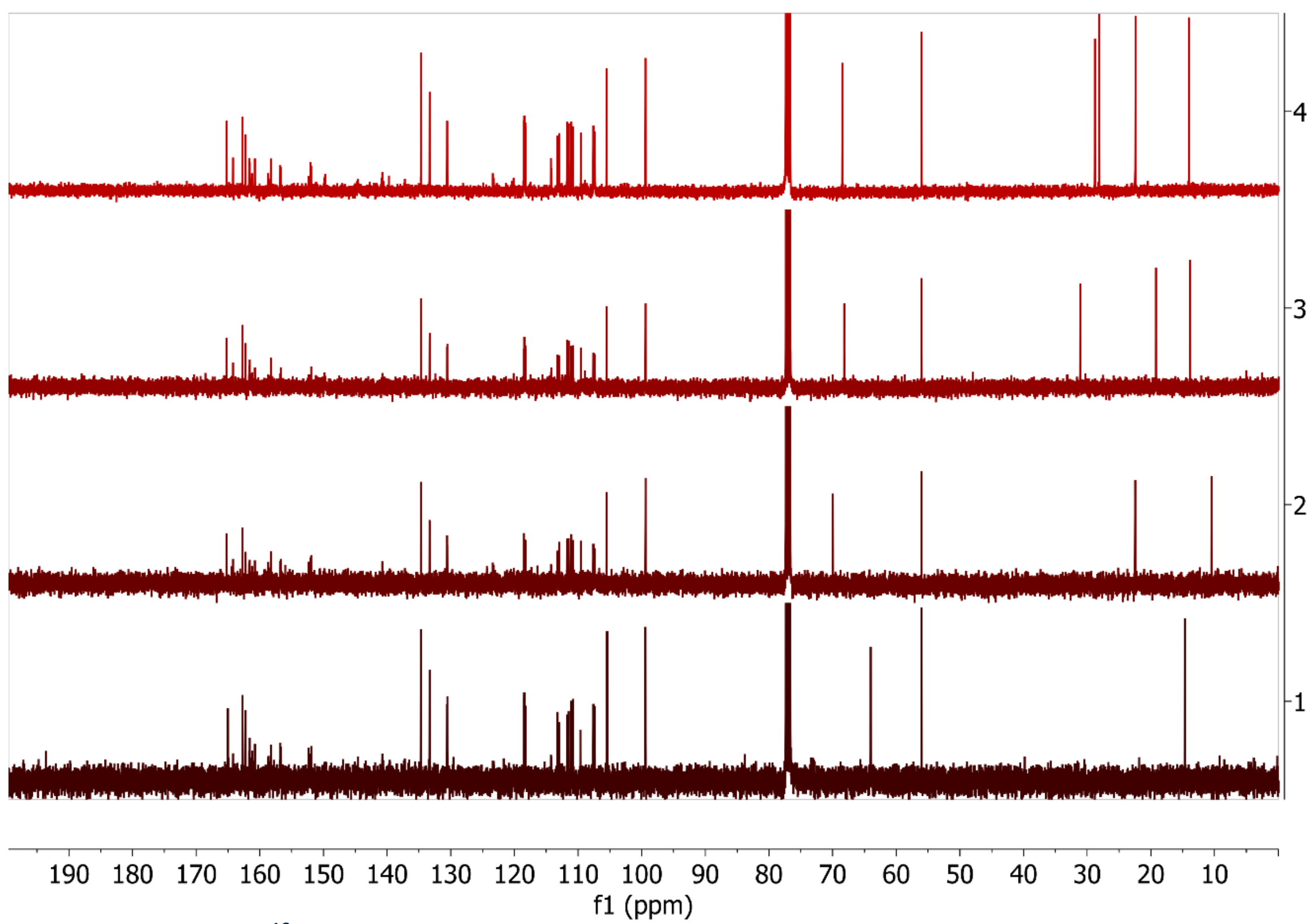

*Figure S 37. Stacked $^{13}C$ NMR spectra of (1) $T_2$-0-2, (2) $T_3$-0-2, (3) $T_4$-0-2, (4) $T_5$-0-2.*

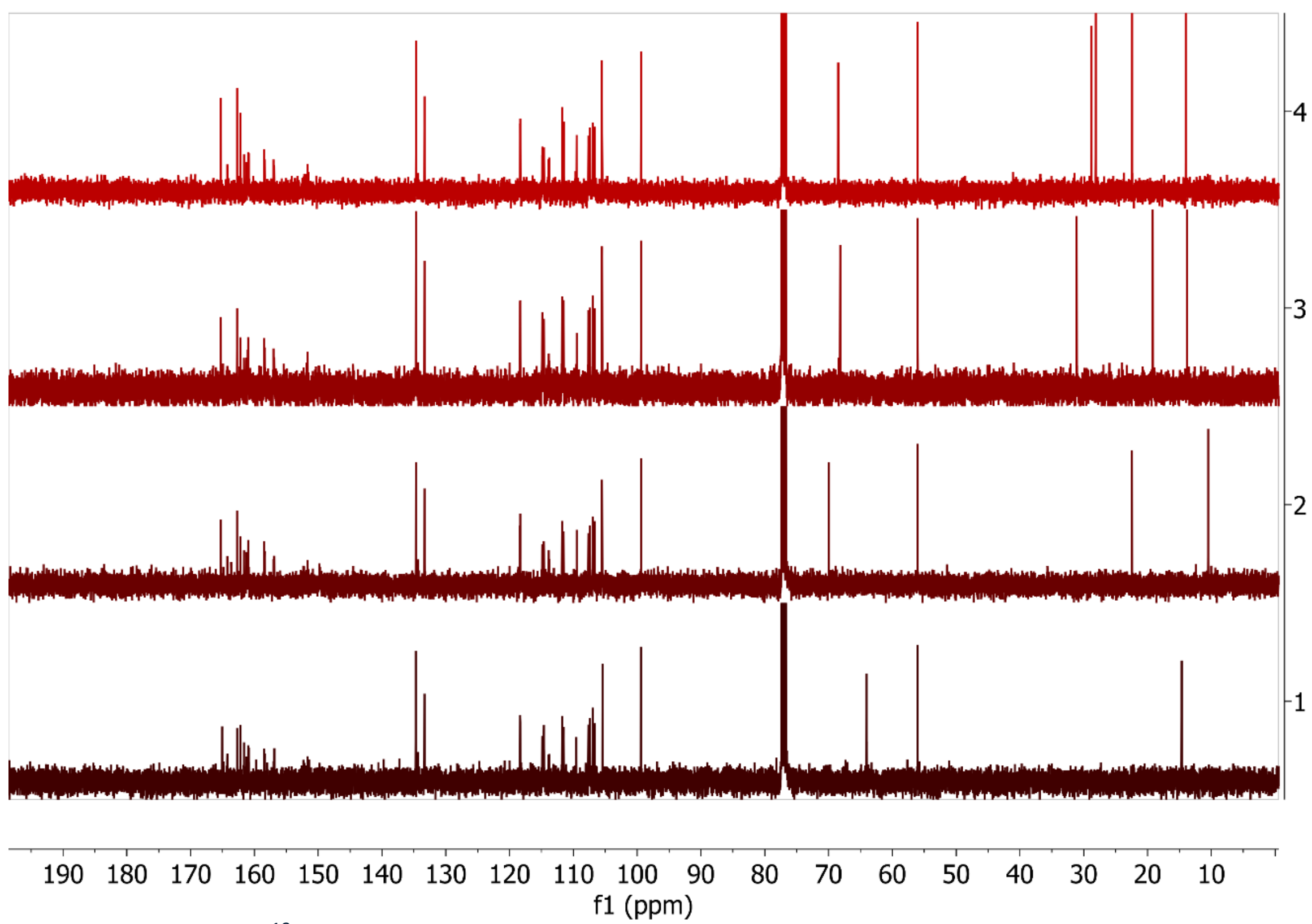

*Figure S 38. Stacked $^{13}C$ NMR spectra of (1) $T_2$-1-2, (2) $T_3$-1-2, (3) $T_4$-1-2, (4) $T_5$-1-2.*

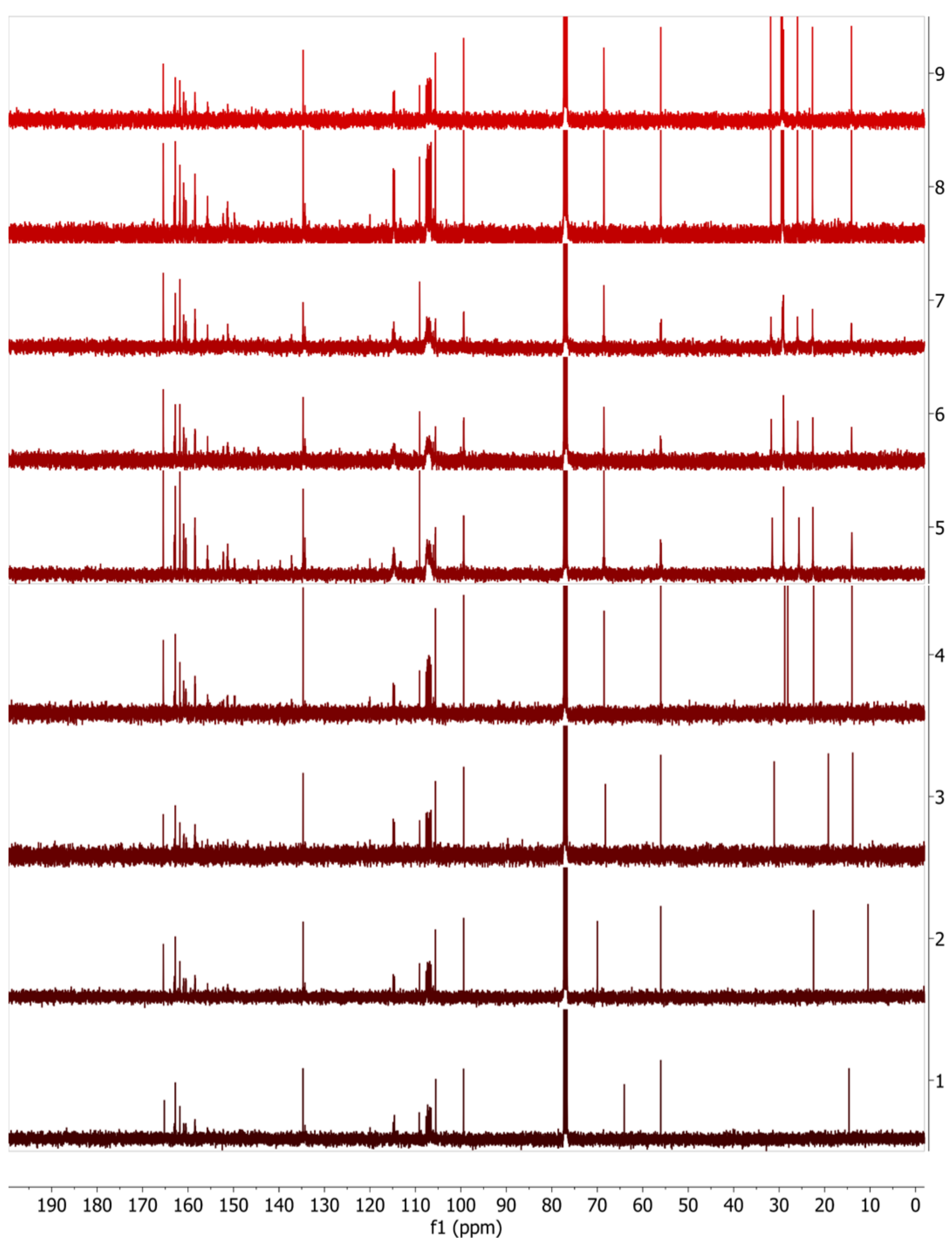

*Figure S 39. Stacked $^{13}C$ NMR spectra of (1) $T_2$-2-2, (2) $T_3$-2-2, (3) $T_4$-2-2, (4) $T_5$-2-2, (5) $T_6$-2-2, (6) $T_7$-2-2, (7) $T_8$-2-2, (8) $T_9$-2-2, (9) $T_{10}$-2-2.*